\documentclass[traditabstract]{aa} 
\usepackage{graphicx}
\usepackage{stfloats}
\usepackage[varg]{txfonts}
\usepackage{natbib}
\usepackage{lscape}
\usepackage[normalem]{ulem}
\usepackage{subfig}
\usepackage{caption}
\usepackage{longtable}
\bibpunct{(}{)}{;}{a}{}{,} 

\begin{document}

   \title{Granulation properties of giants, dwarfs, and white dwarfs from the
     CIFIST 3D model atmosphere grid}

   \author{P.-E. Tremblay\inst{1}
          \and
          H.-G. Ludwig\inst{1}
          \and
          B. Freytag\inst{2}
          \and
          M. Steffen\inst{3}
          \and
          E. Caffau\inst{1,4}
          }

   \institute{Zentrum f\"ur Astronomie der Universit\"at Heidelberg,
     Landessternwarte, K\"onigstuhl 12, 69117
     Heidelberg
     \\ \email{ptremblay@lsw.uni-heidelberg.de,hludwig@lsw.uni-heidelberg.de,ecaffau@lsw.uni-heidelberg.de}
     \and Centre de
     Recherche Astrophysique de Lyon, UMR 5574: CNRS, Universit\'e de Lyon,
     \'Ecole Normale Sup\'erieure de Lyon, 46 all\'ee d'Italie, F-69364 Lyon
     Cedex 07, France \\ 
     \email{Bernd.Freytag@ens-lyon.fr} 
     \and Leibniz-Institut f\"ur Astrophysik Potsdam, An der Sternwarte 16,
     D-14482 Potsdam, Germany
     \\ \email{msteffen@aip.de}
     \and GEPI, Observatoire de Paris, CNRS, Univ. Paris Diderot, Place Jules Janssen, 92190, Meudon, France}

   \date{Received ..; accepted ..}
 
  \abstract {3D model atmospheres for giants, dwarfs, and white dwarfs,
    computed with the CO5BOLD code and part of the CIFIST grid, have been
      used for spectroscopic and asteroseismic studies. Unlike existing
    plane-parallel 1D structures, these simulations predict the spatially
      and temporally resolved emergent intensity so that granulation
    can be analysed, which provides insights on how convective energy transfer
    operates in stars. The wide range of atmospheric parameters of the CIFIST
    3D simulations ($3600 < T_{\rm eff}$ (K) $< 13,000$ and $1 < \log g < 9$)
    allows the comparison of convective processes in significantly different
    environments. We show that the relative intensity contrast is correlated
    with both the Mach and P\'eclet numbers in the photosphere. The
      horizontal size of granules varies between 3 and 10 times the local
      pressure scale height, with a tight correlation between the factor and
      the Mach number of the flow. Given that convective giants, dwarfs, and
    white dwarfs cover the same range of Mach and P\'eclet numbers, we
    conclude that photospheric convection operates in a very similar way in
    those objects.}{}{}{}{}


   \titlerunning{Granulation properties in giants, dwarfs, and white dwarfs}
   \authorrunning{Tremblay et al.}
   \maketitle

\section{Introduction}

In late-type stars, giants, and cool white dwarfs, surface convection is
responsible for the granulation pattern directly observed in high-resolution
images of the Sun. Granulation is more difficult to observe directly in other
stars, although it has been possible to constrain the lifetime and size of
granules through asteroseismic studies of stochastically excited p-modes
\citep{asterod,astero2,mathur11,samadi13} and interferometry in the case of the red supergiant
Betelgeuse \citep{inter,inter2}.

Early numerical hydrodynamical experiments \citep{num2,num1} have provided
predictions for the size and temperature contrast of granulation as a function
of characteristic flow numbers. The radiation-hydrodynamics (RHD) 3D
simulations, starting from the work of \citet{nordlund82}, have improved
predictions for the solar granulation with more realistic stratifications. For
many years, the observed values of intensity contrast between dark and bright
granules in the Sun were significantly lower than those predicted from RHD
simulations.  However, it was shown recently that if the instrumental image
degradation is taken into account properly, the simulations can reproduce the
observed values very well \citep{irms1}.

For 3D simulations of solar-like stars, computed with the CO$^5$BOLD code
\citep{freytag12} and the \citet{stein} code, the relative intensity
contrast and size of granules relative to the local pressure scale height
($H_{\rm p}$) appear to be changing significantly with $T_{\rm eff}$
  \citep{trampe}. Furthermore, for hotter stars and supergiants, a
bifurcation is observed from a regular pattern of bright cells and dark lanes
towards a regime with small shallow cells and large granules with deep
reaching downdrafts \citep{freytag12}. However, for objects with different
gravities, granulation properties are
in general fairly similar \citep{magic13}, albeit with a shift in $T_{\rm eff}$. Surface
convection is still significant in white dwarfs at $T_{\rm eff}$ up to
15,000~K for pure-H atmospheres \citep{tremblay13}, and up to 30,000~K for pure-He atmospheres
\citep{bergeron11}.

Granulation properties can not be derived from existing plane-parallel 1D
structures. Furthermore, the convective energy transfer, usually modelled with
the mixing-length theory \citep[][hereafter MLT]{MLT}, is parameterised in a
different way in giants, stars, and white dwarfs
\citep{f93,bergeron95,ludwig12}. As a consequence, it is difficult to
compare how convection works in stars with different surface gravities and
temperatures.

\begin{figure*}[b!]
\captionsetup[subfigure]{labelformat=empty}
\begin{center}
\subfloat[]{
\includegraphics[bb=70pt 10pt 532pt 775pt,width=2.0in,angle=90]{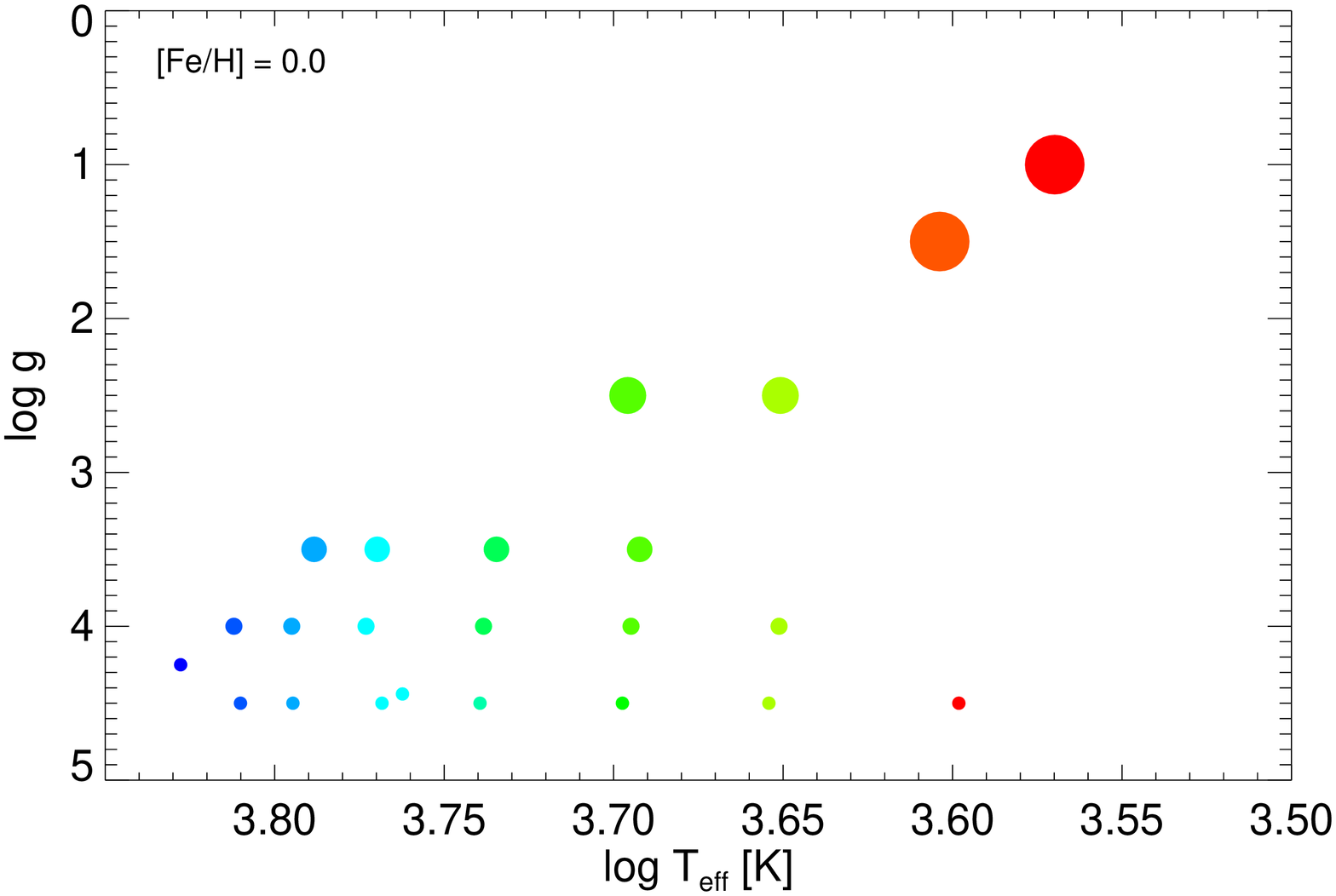}}
\subfloat[]{
\includegraphics[bb=70pt 10pt 532pt 775pt,width=2.0in,angle=90]{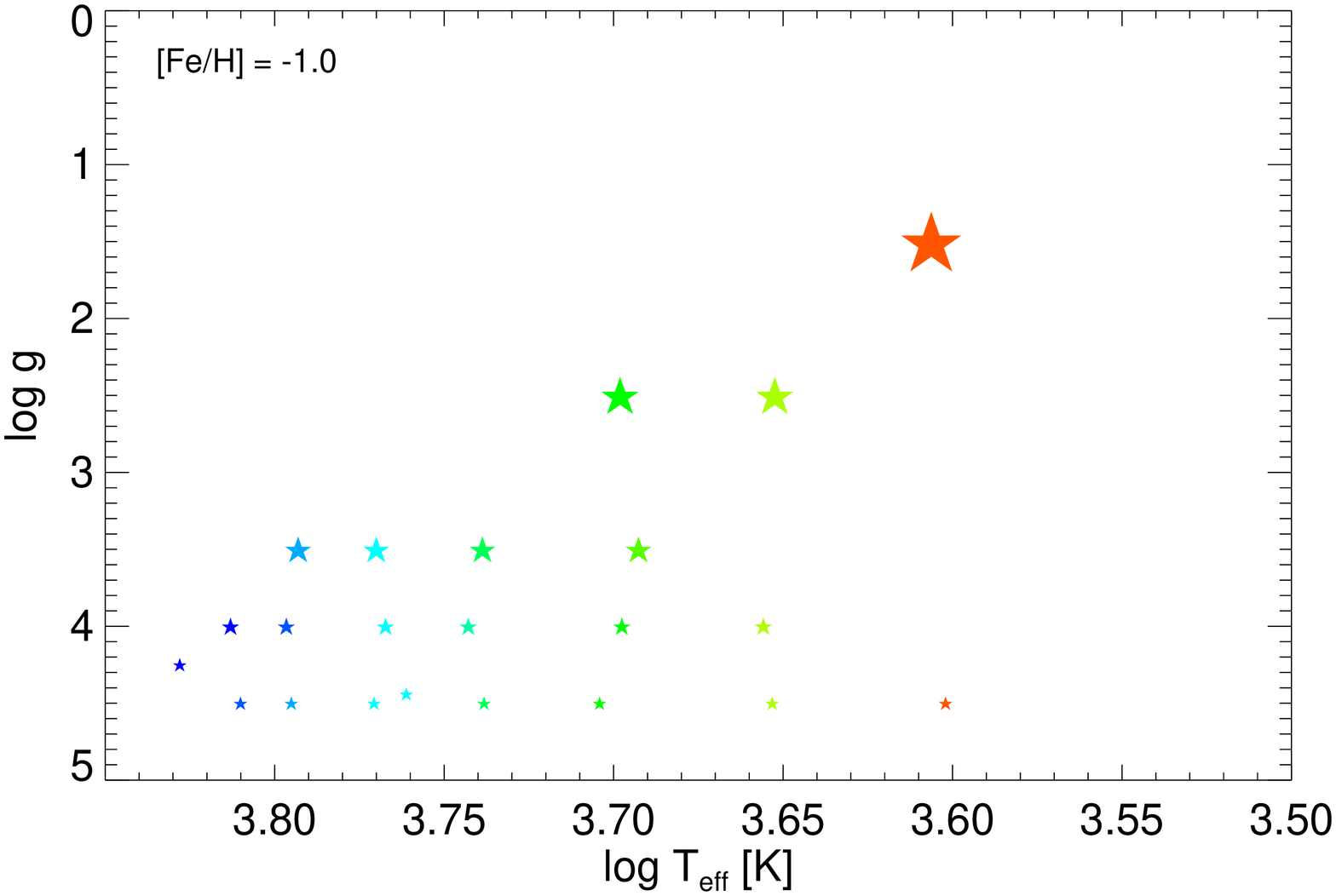}}\\
\subfloat[]{
\includegraphics[bb=70pt 10pt 532pt 775pt,width=2.0in,angle=90]{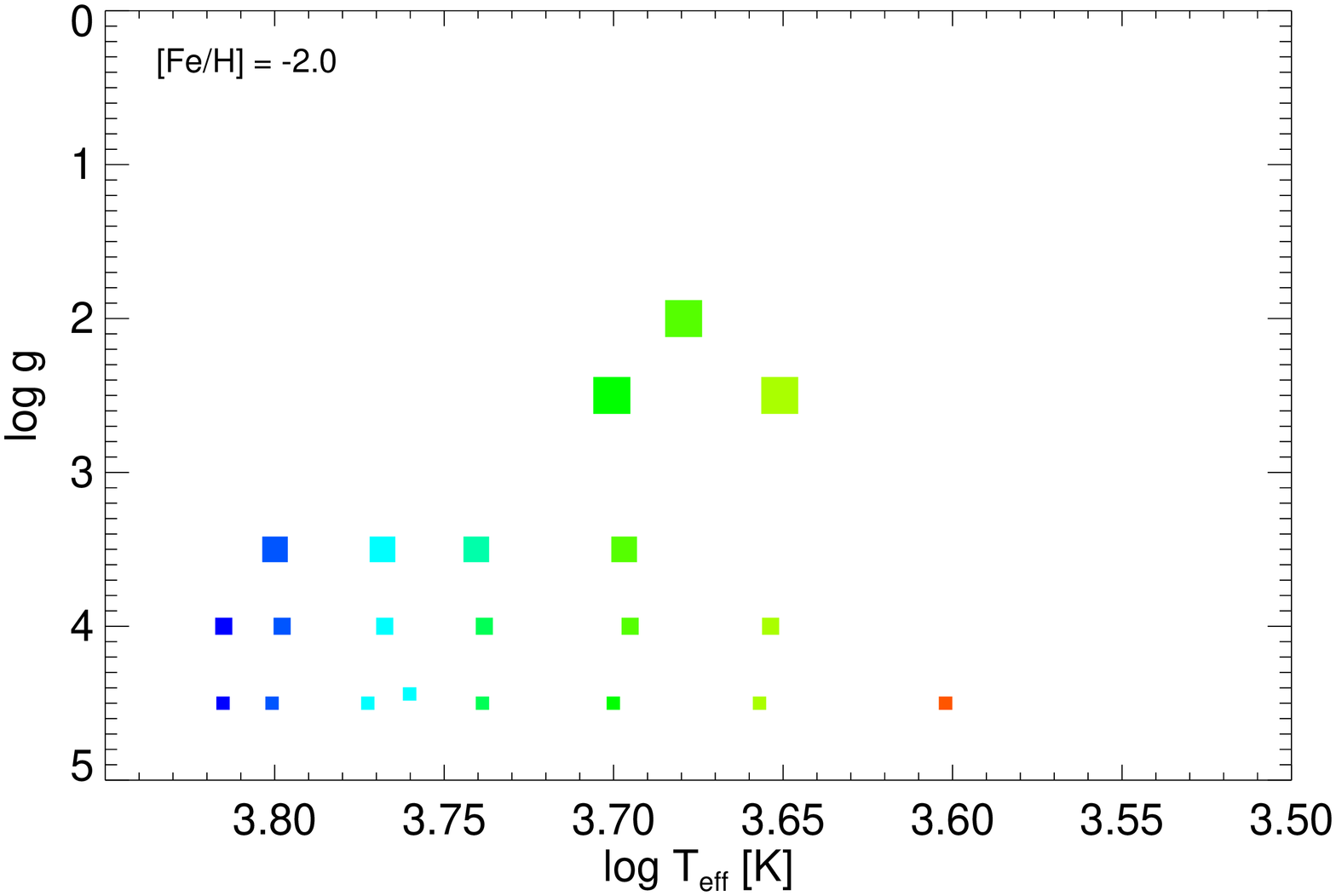}}
\subfloat[]{
\includegraphics[bb=70pt 10pt 532pt 775pt,width=2.0in,angle=90]{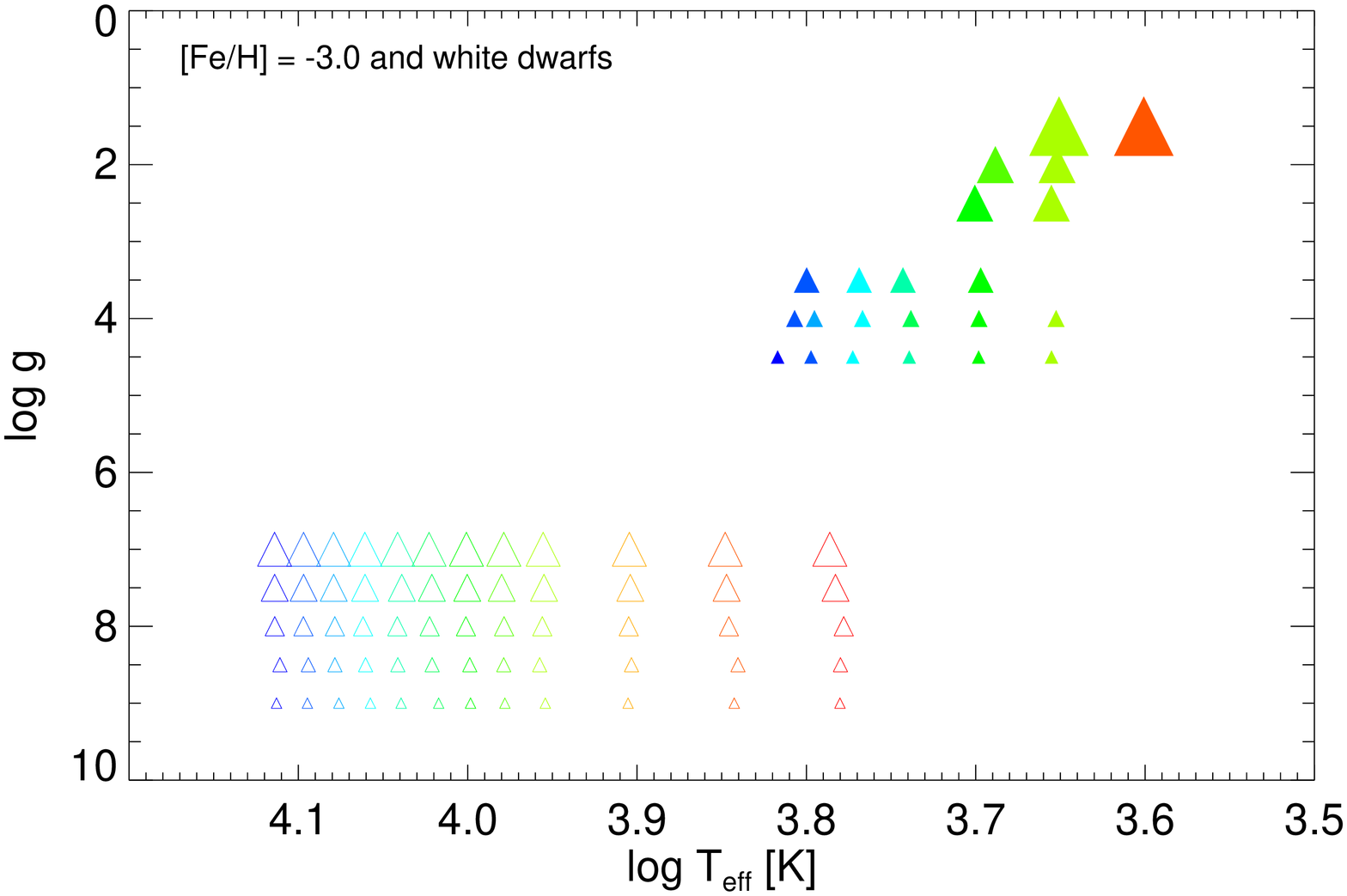}}
\caption{Surface gravity and mean $T_{\rm eff}$ for the 3D model atmospheres
  in the CIFIST grid. The top-left (circles), top-right (stars), bottom-left
  (squares) and bottom-right (triangles) panels represent different
  metallicities, with [Fe/H] = $0, -1, -2$ and $-3$, respectively. For
  simplicity, pure-hydrogen white dwarfs (open triangles) are represented
  along the [Fe/H] = $-3$ stars (note the different scale for this
  panel). $T_{\rm eff}$ values are colour-coded from red (cool) to blue (hot)
  and $\log g$ values (cgs units) are size-coded from small (high gravity) to
  large (low gravity) symbols. The x- and y-axis can be used as legends for the following figures. There
  is no link between white dwarfs and stars with the same sizes and colours.
\label{fg:f1}}
\end{center}
\end{figure*}

In this work, we study how convective energy transfer behaves in the
photosphere of stars by relying on 148 CO$^5$BOLD 3D simulations for dwarfs,
white dwarfs, and in a few cases, giants. In particular, we derive the
relative intensity contrast of surface granulation, which is a measure of the
deviation from a 1D plane-parallel approximation. This is an indication of the
strength of 3D effects but in practice, the 3D effects on the mean structure
and line formation are a complex function of local inhomogeneities
\citep{pmodes,paper1,ku13} and do not scale well with the intensity
contrast. Our 3D computations can also predict the characteristic size and
lifetime for the granulation. These quantities are of interest for the
understanding of the granulation background of asteroseismological
observations. Compared to the study of \citet{trampe}, we rely on a different
  code and include white dwarfs, sub-solar metallicity dwarfs, and giants.

We aim at defining how the granulation properties can be formulated as a
function of the physical conditions in the photosphere. These relations are of
interest, first of all, to set up new 3D simulations. Furthermore, such
parameterisations can suggest how the observed properties of convection in the
Sun are expected to be scaled to other objects. For instance, it is an open
question whether small scale magnetic fields generated by a turbulent dynamo
can exist in white dwarfs like it is observed in the Sun
\citep{mhd1,mhd2}. Finally, we hope that our analysis can provide insights on
how to improve globally the current 1D models and the mixing-length theory.

We describe in Sect.~2 the 3D model atmospheres that we rely on for this
study. We follow in Sect.~3 with a presentation of the characteristic
granulation properties as derived from the grid of 3D simulations. We discuss
these results in Sect.~4 and conclude in Sect.~5.

\section{3D model atmospheres}

\subsection{Giants and dwarfs}

We rely on 88 simulations of giants and dwarfs computed with CO$^{5}$BOLD and
part of the CIFIST grid \citep{abun3,caffau11}. These are non-gray RHD models
with the surface gravity and entropy flux at the bottom of the atmosphere as
input parameters. The implementation of the boundary conditions is described
in detail in \citet[][see Sect. 3.2]{freytag12}. In brief, the lateral
boundaries are periodic, and the top boundary is open to material flows and
radiation. The bottom layer is open to convective flows where a zero total
mass flux is enforced. The $T_{\rm eff}$ is not an input parameter, although
we specify the entropy of the ascending material to obtain approximately the
desired value, which is derived from the temporal and spatial average of the
emergent stellar flux.

The models typically have a resolution of 140$\times$140$\times$150 ($x$$\times
$$y$$\times$$z$) grid points. The vertical extent of the simulations, from the
photosphere at Rosseland optical depth ($\tau_{\rm R}$) unity to the bottom boundary, is always several orders of magnitude
(three or more) of pressure scale height, which ensures that convective eddies
in the photosphere are not impacted by boundary conditions. The upper boundary
extends beyond a mean $\tau_{\rm R}$ of
10$^{-6}$. The horizontal extent of the simulations was chosen so that at
least of the order of ten granules are resolved at the surface.

We rely on band-averaged opacities to describe the band-integrated radiative
transfer, based on the procedure laid out in \citet{nordlund82,ludwig94} and
\citet{voegler04}. A total number of 5 bins was used for models of solar
metallicity, while 6 bins were employed for models of sub-solar
metallicity\footnote{The solar model and the model at $T_{\rm eff} \sim 6250$
  K, $\log g = 4.0$ and [Fe/H] = $-$3.0 were computed with 12 opacity
  bins.}. This setup should be more than sufficient for our study of
wavelength-integrated (white light) intensity maps, which is much less
sensitive to the binning procedure than detailed spectral synthesis.

We assume four different metallicities in the range $ -3 <$ [Fe/H] $< 0$. The
equation of state takes into account H and He ionization and H$_2$
formation. The opacities are calculated with the Uppsala package
\citep{opac_cifist}. Chemical abundances are from \citet{abun1}, with the
exception of CNO, which are updated following \citet{abun2}. More details are
provided in \citet{abun3}. In Table~1 at the end of this work, the
atmospheric parameters and composition of our different simulations are given.

\begin{figure}[!h]
\captionsetup[subfigure]{labelformat=empty}
\begin{center}
\includegraphics[bb=40 40 552 765,width=2.3in,angle=90]{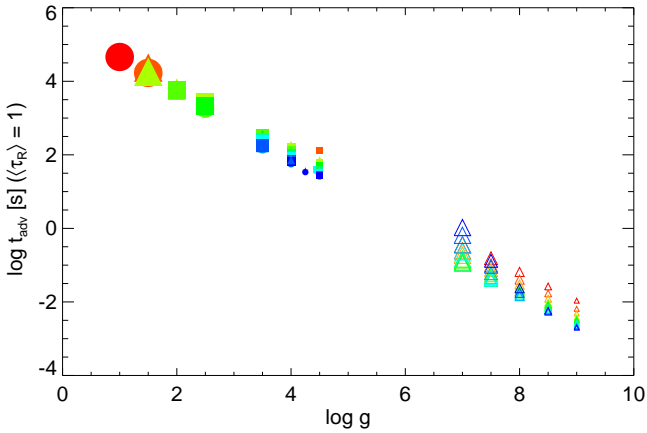}
\caption{Logarithm of the advective timescale (evaluated at $\langle\tau_\mathrm{R}\rangle = 1$) as
  a function of the surface gravity.
\label{fg:f15}}
\end{center}
\end{figure}

\begin{figure}[!h]
\captionsetup[subfigure]{labelformat=empty}
\begin{center}
\includegraphics[bb=40 40 552 765,width=2.3in,angle=90]{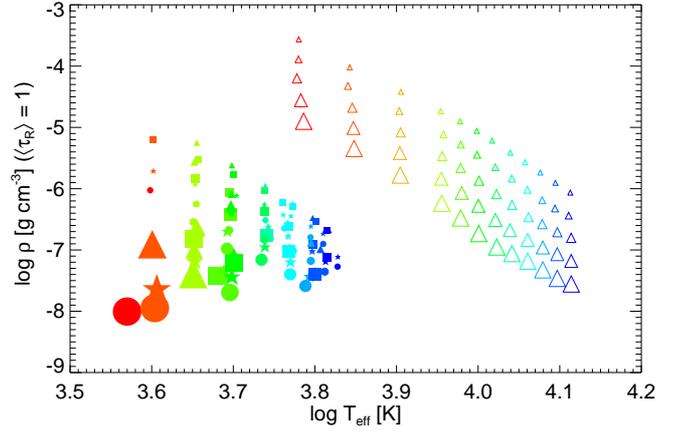}
\caption{Logarithm of the density (evaluated at $\langle\tau_\mathrm{R}\rangle = 1$) as
  a function of $\log T_{\rm eff}$.
\label{fg:f2}}
\end{center}
\end{figure}

\begin{figure}[!h]
\captionsetup[subfigure]{labelformat=empty}
\begin{center}
\includegraphics[bb=40 40 552 765,width=2.3in,angle=90]{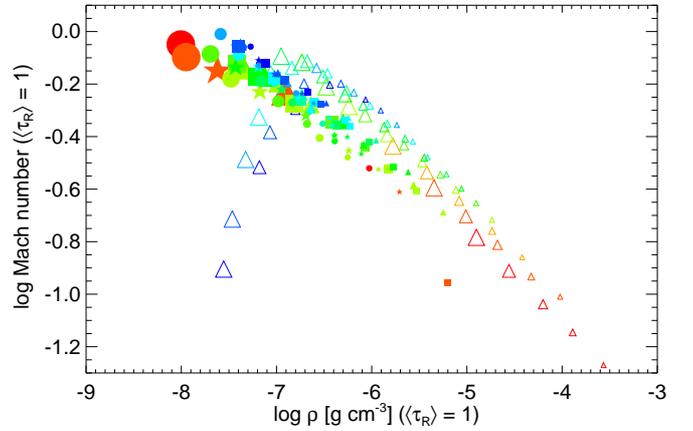}
\caption{Logarithm of the Mach number as a function of the logarithm of
  the density (both quantities evaluated at $\langle\tau_\mathrm{R}\rangle = 1$).
\label{fg:f3}}
\end{center}
\end{figure}

\begin{figure}[!h]
\captionsetup[subfigure]{labelformat=empty}
\begin{center}
\includegraphics[bb=40 40 552 765,width=2.3in,angle=90]{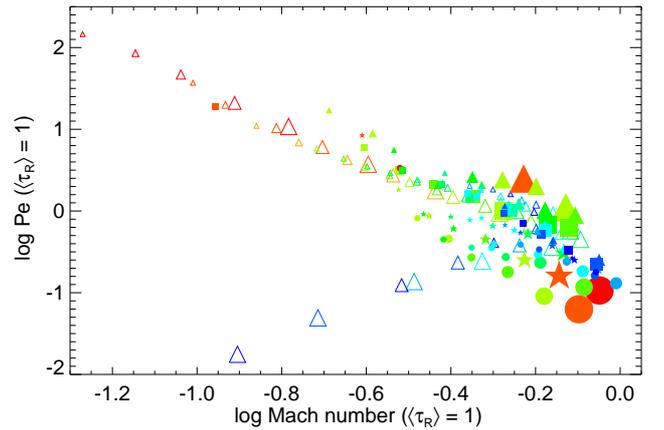}
\caption{Logarithm of the P\'eclet number as a function of Mach number (both
  quantities evaluated at $\langle\tau_\mathrm{R}\rangle = 1$).
\label{fg:f4}}
\end{center}
\end{figure}

\begin{figure*}[!]
\captionsetup[subfigure]{labelformat=empty}
\begin{center}
\subfloat[]{
\includegraphics[bb=130 400 442 705,width=2.20in]{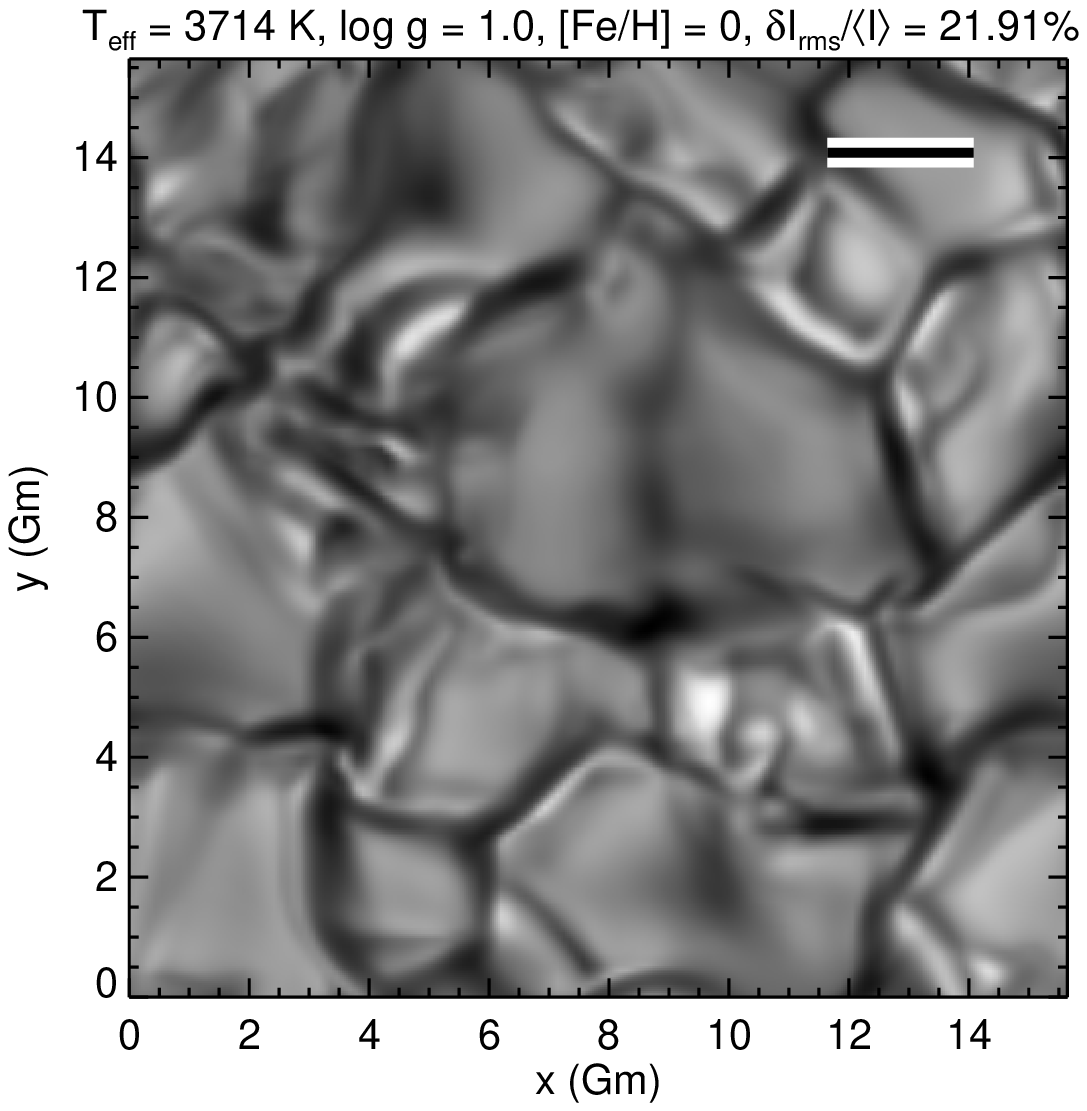}}
\subfloat[]{
\includegraphics[bb=130 400 442 705,width=2.20in]{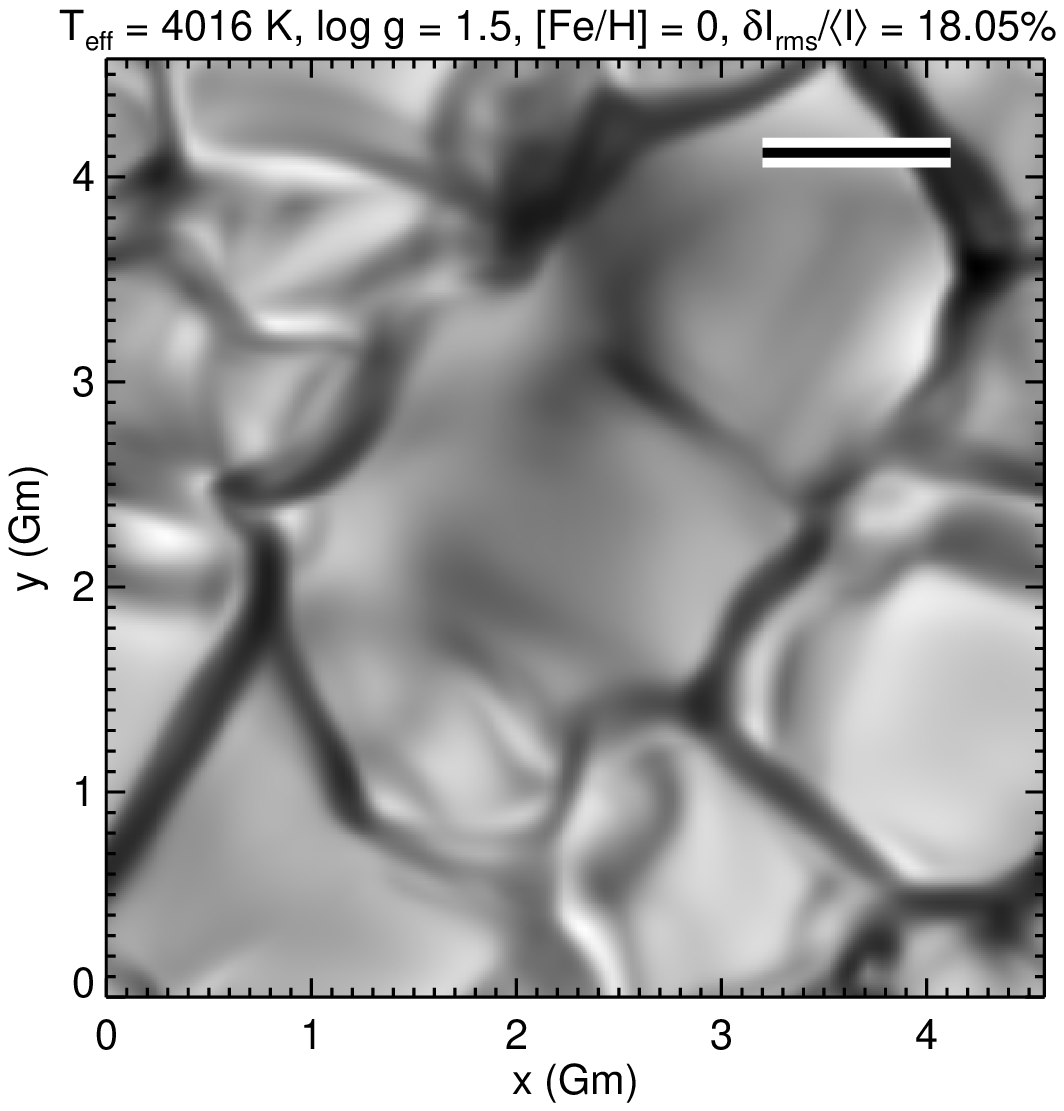}}
\subfloat[]{
\includegraphics[bb=130 400 442 705,width=2.20in]{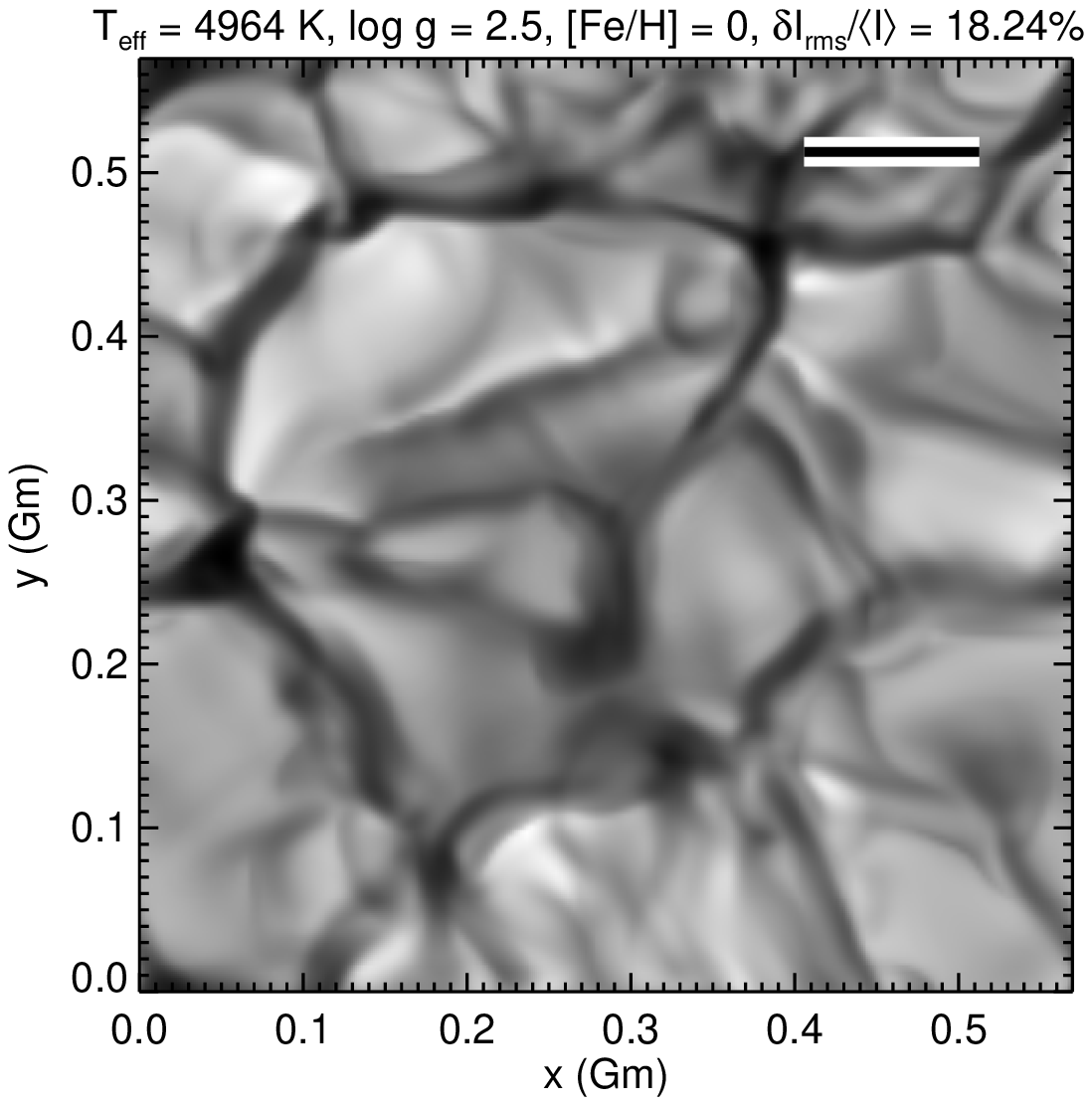}}
\caption{Emergent bolometric intensity at the top of the horizontal $xy$ plane
  for giant simulations at solar metallicity. The atmospheric parameters and
  the rms intensity contrast with respect to the mean intensity are given
  above the snapshots. The length of the upper bar in the top right is 10
  times the pressure scale height at $\langle\tau_\mathrm{R}\rangle = 1$.
\label{fg:f5}}
\end{center}
\end{figure*}

\begin{figure*}[!]
\captionsetup[subfigure]{labelformat=empty}
\begin{center}
\subfloat[]{
\includegraphics[bb=130 400 442 705,width=2.20in]{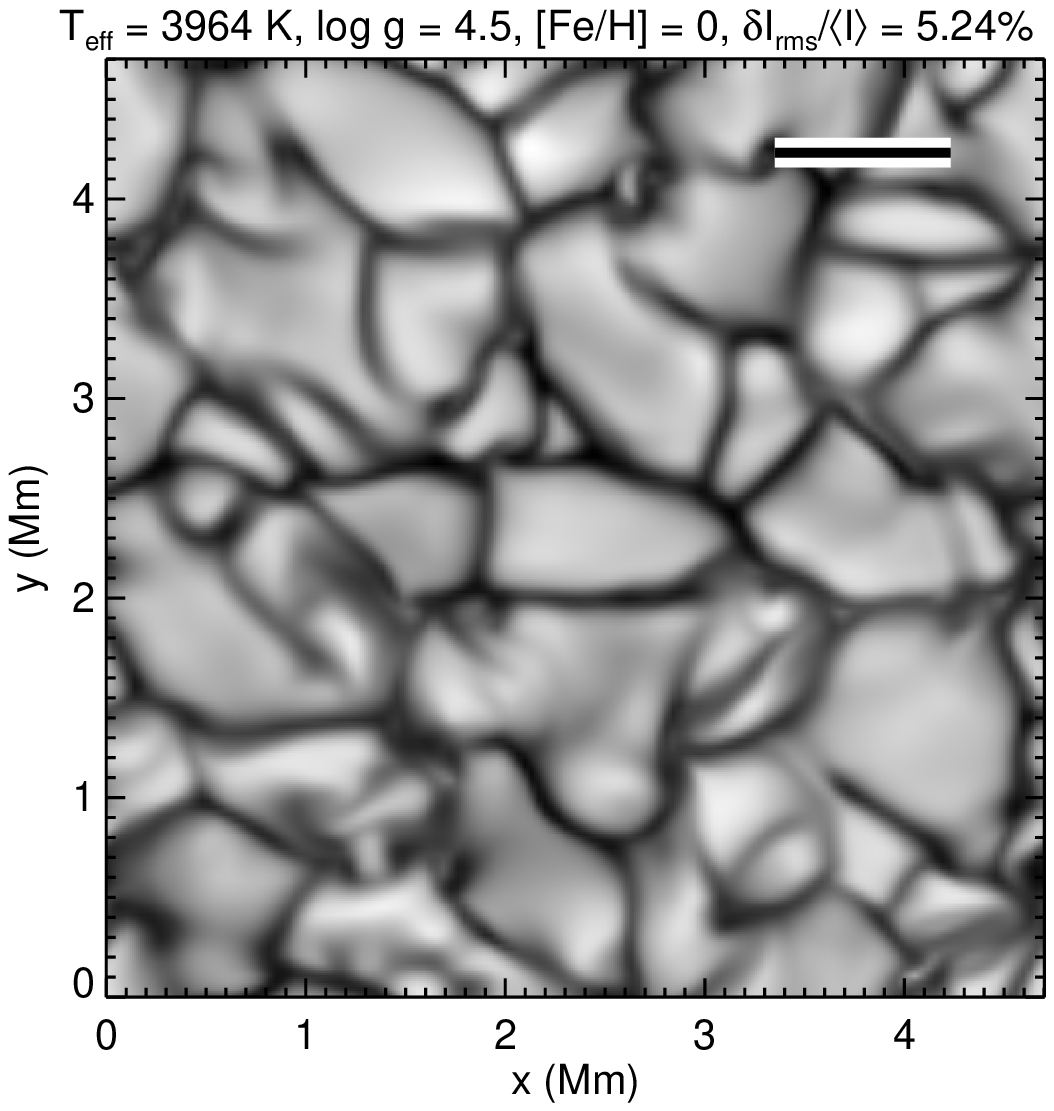}}
\subfloat[]{
\includegraphics[bb=130 400 442 705,width=2.20in]{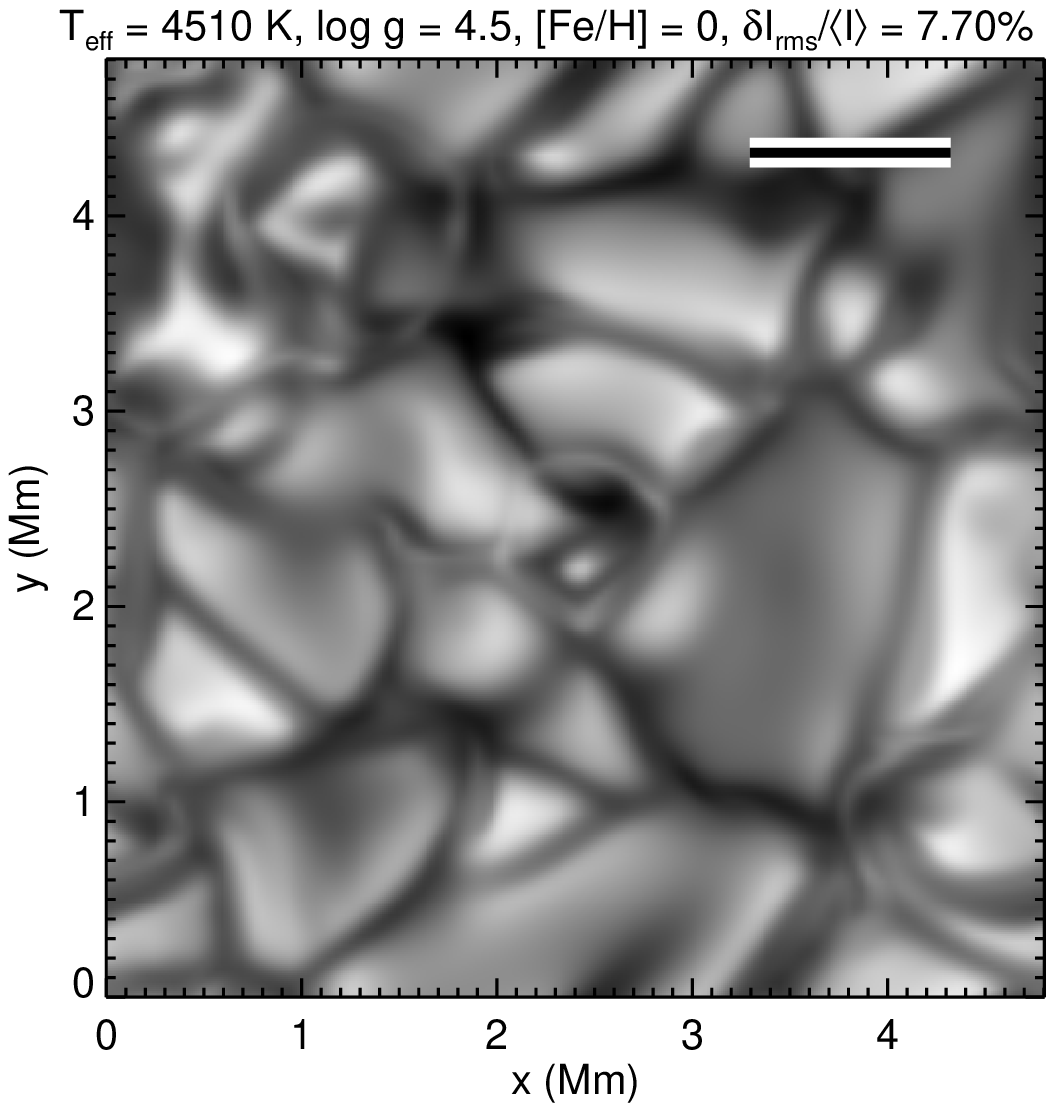}}
\subfloat[]{
\includegraphics[bb=130 400 442 705,width=2.20in]{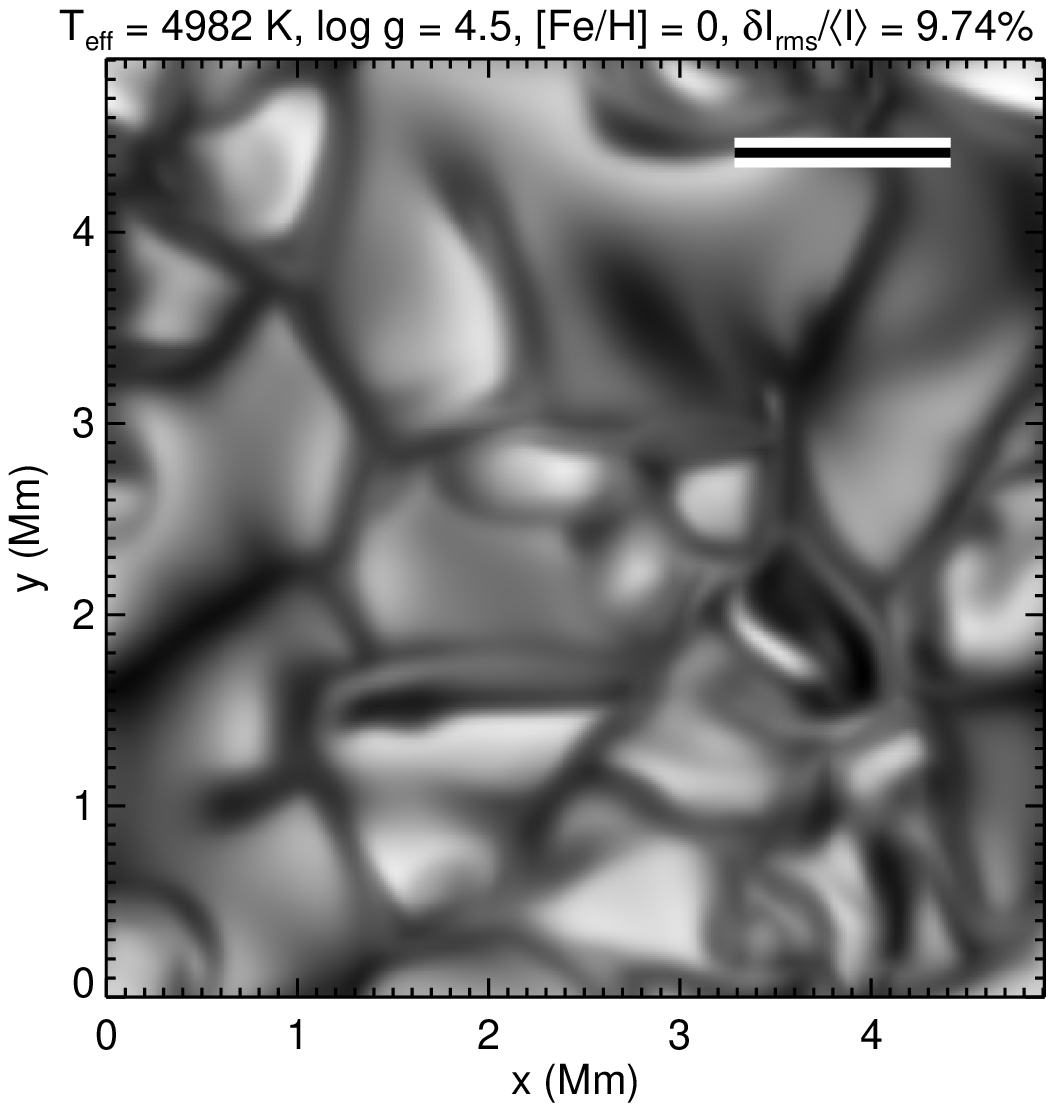}}\\
\subfloat[]{
\includegraphics[bb=130 400 442 675,width=2.20in]{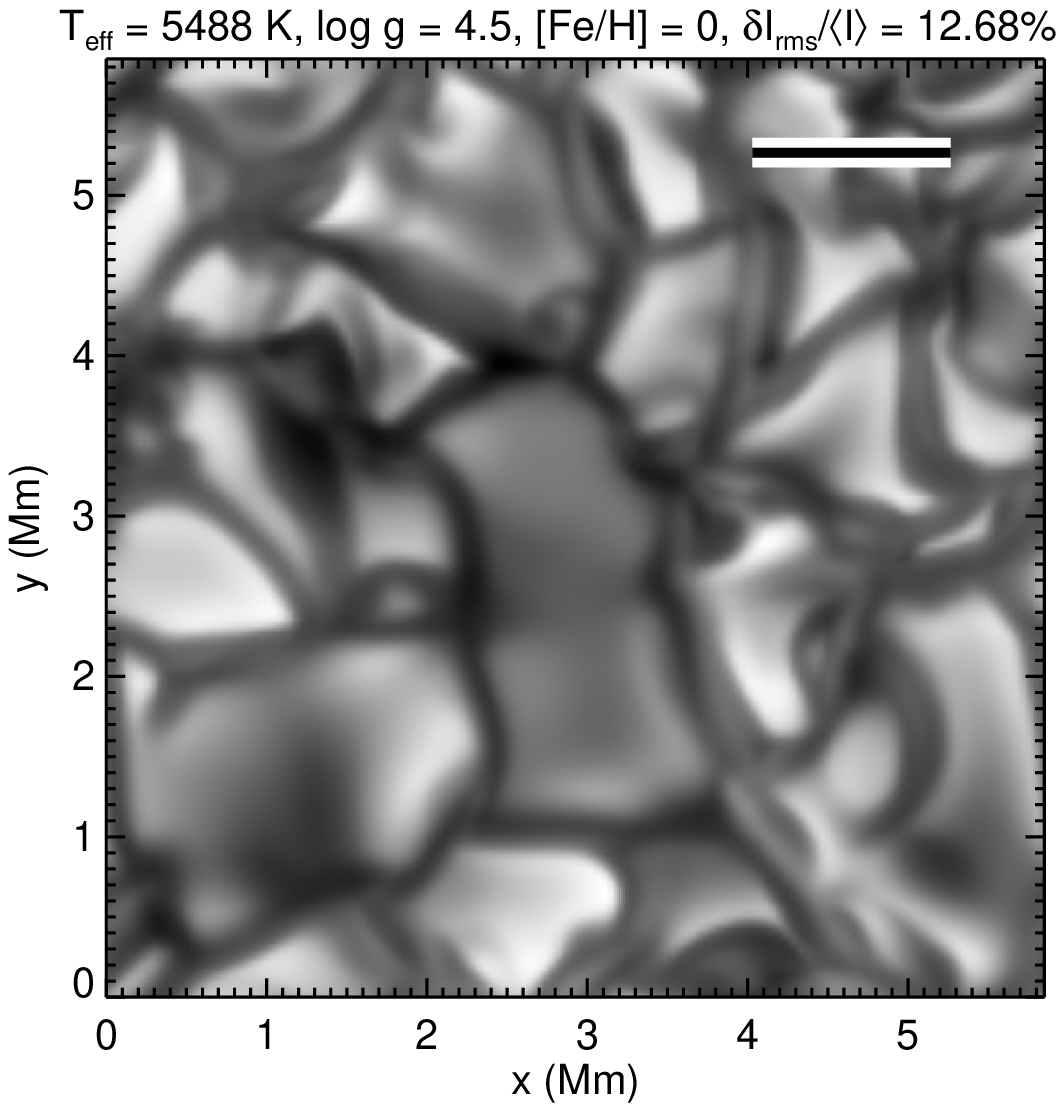}}
\subfloat[]{
\includegraphics[bb=130 400 442 675,width=2.20in]{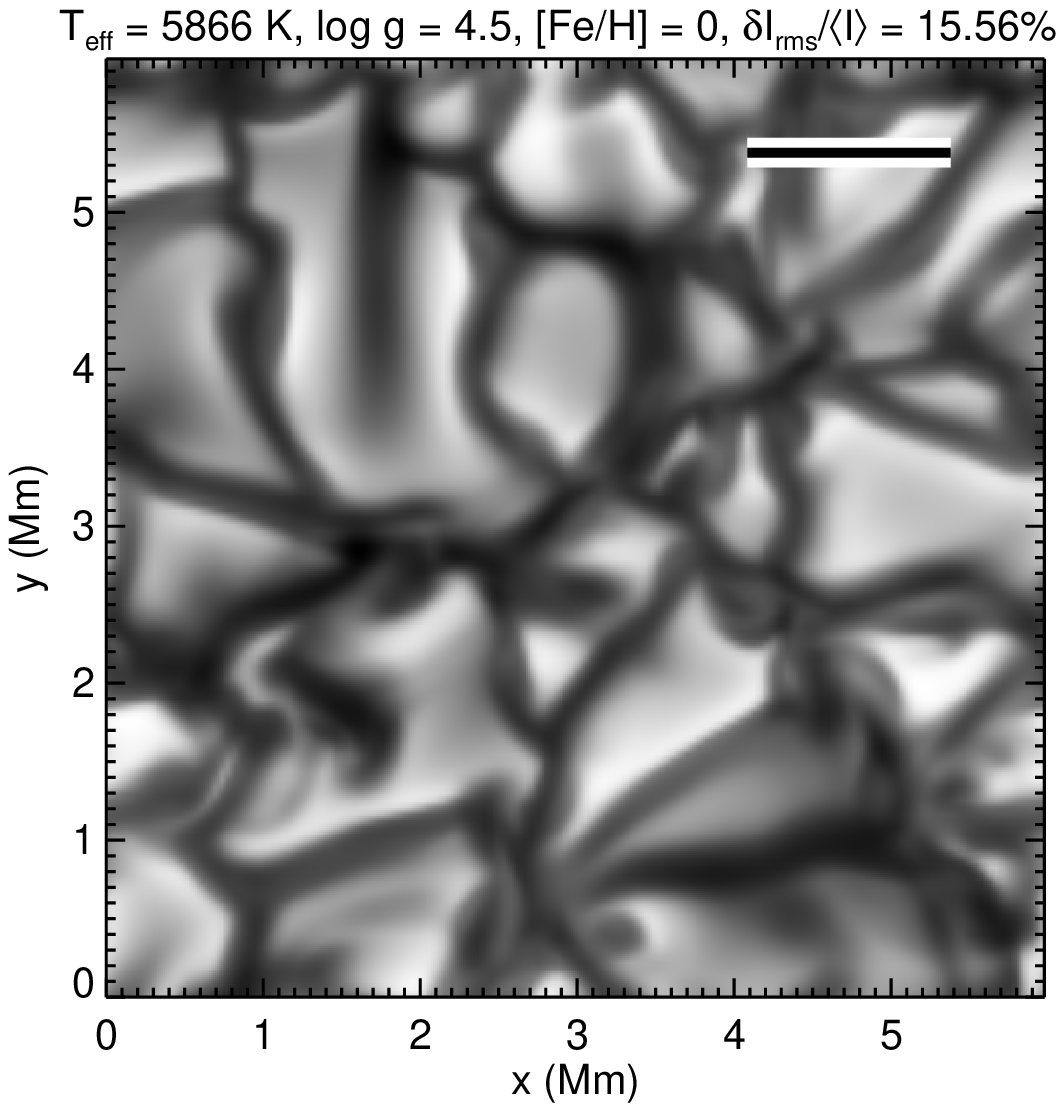}}
\subfloat[]{
\includegraphics[bb=130 400 442 675,width=2.20in]{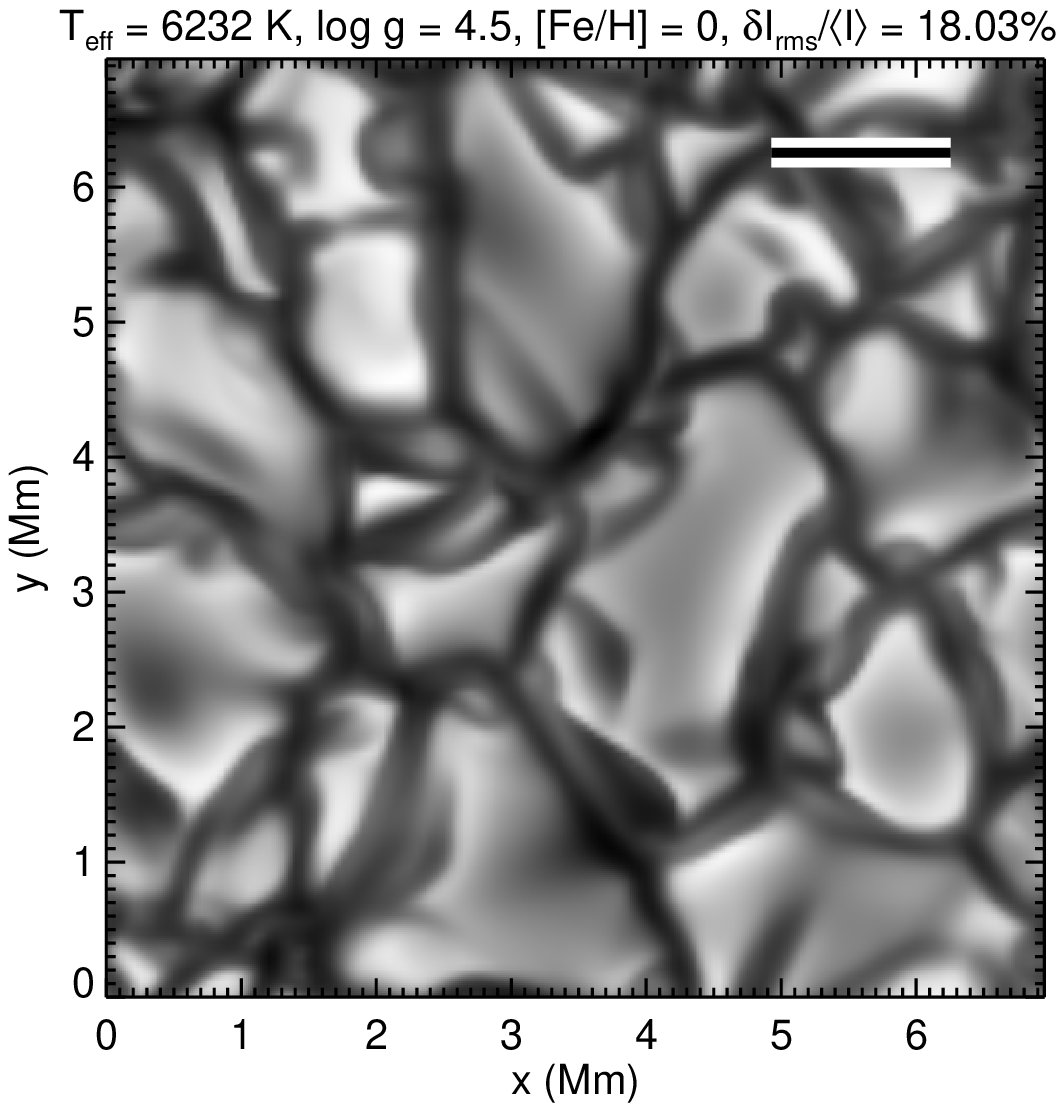}}\\
\subfloat[]{
\includegraphics[bb=130 400 442 675,width=2.20in]{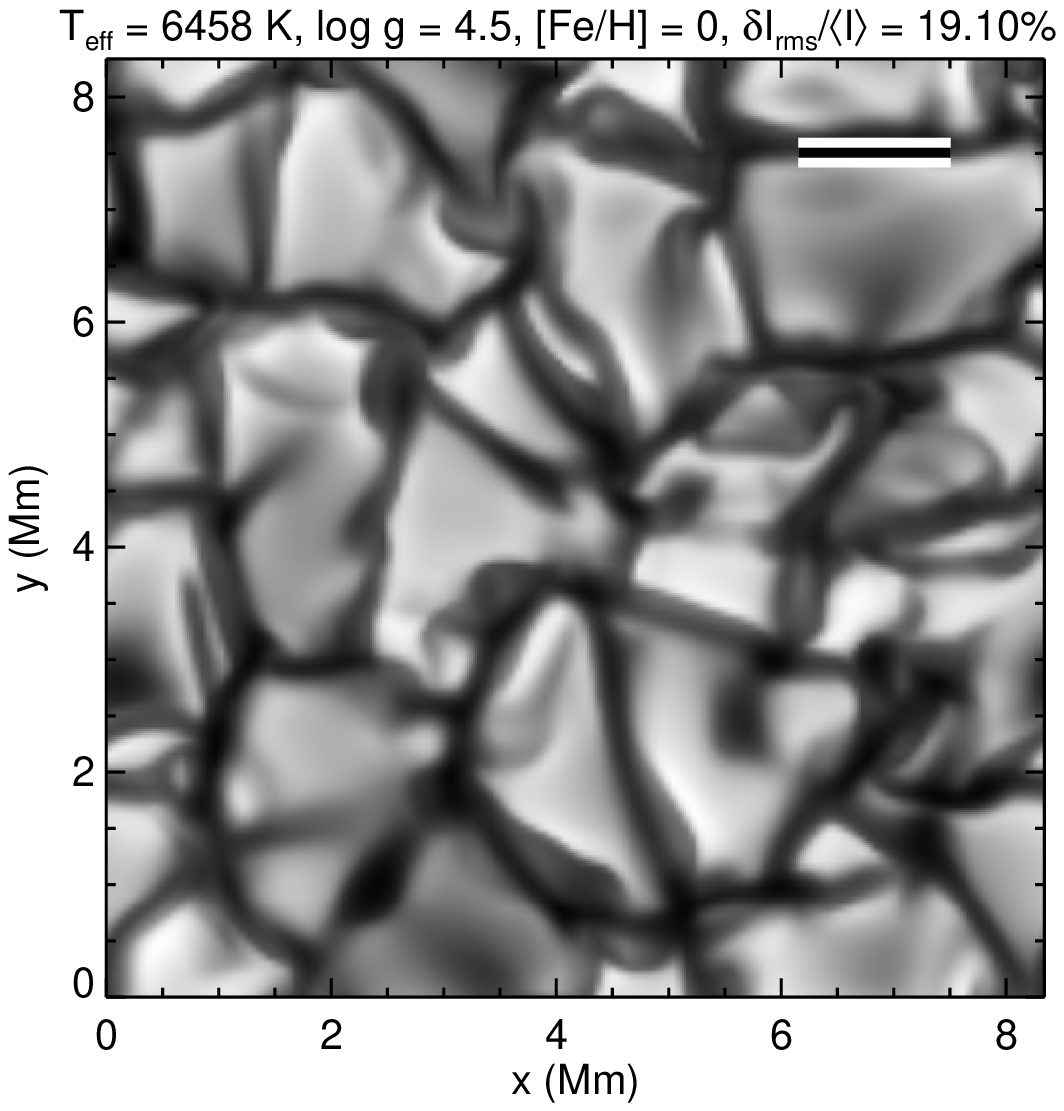}}
\subfloat[]{
\includegraphics[bb=130 400 442 675,width=2.20in]{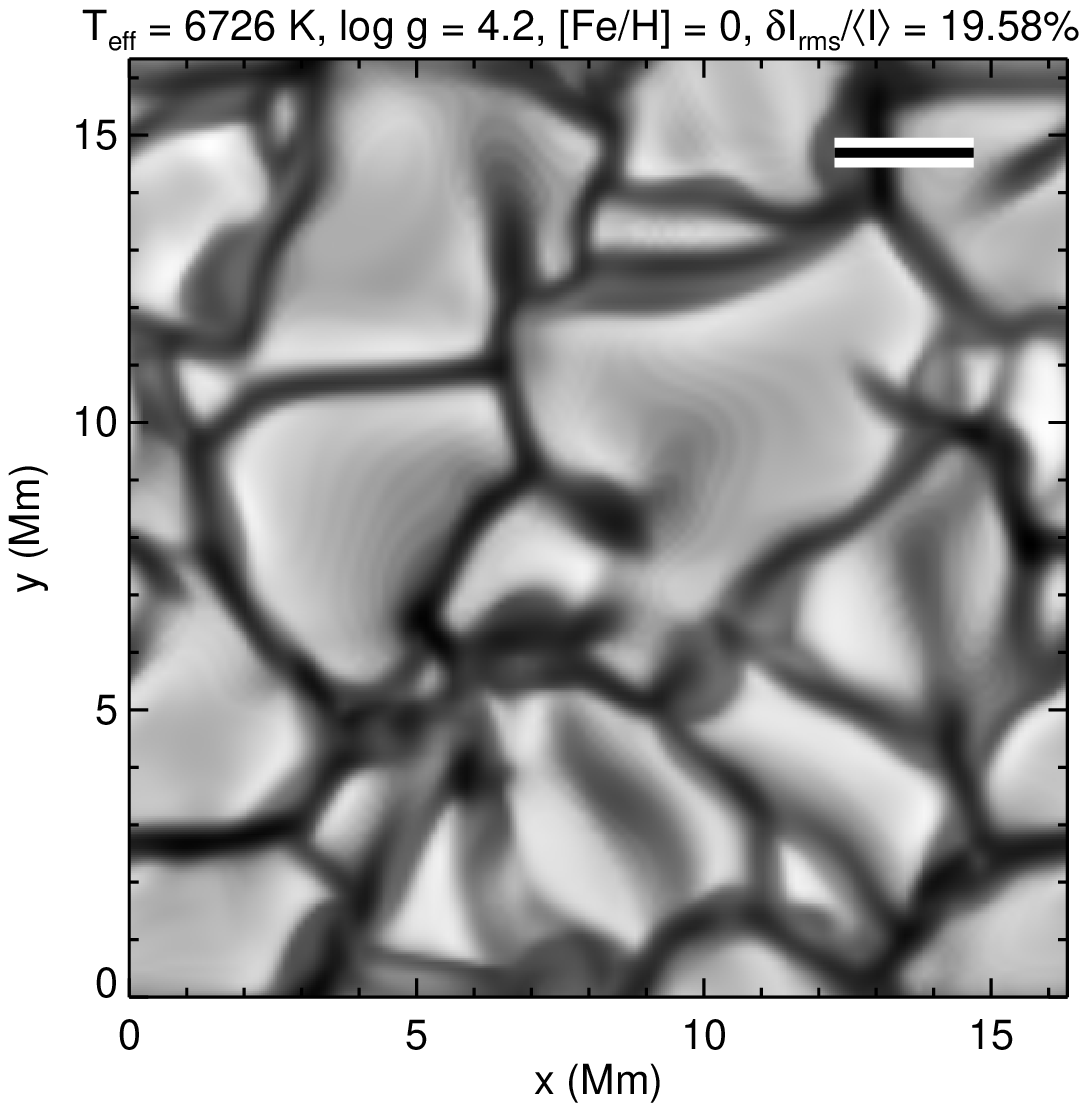}}
\caption{Emergent bolometric intensity for dwarf simulations at solar
  metallicity in the range 4000 $< T_{\rm eff}$ (K) $<$ 6750 with the
  atmospheric parameters given above the snapshots.
\label{fg:f6}}
\end{center}
\end{figure*}

\begin{figure*}[!]
\captionsetup[subfigure]{labelformat=empty}
\begin{center}
\subfloat[]{
\includegraphics[bb=130 400 442 705,width=2.20in]{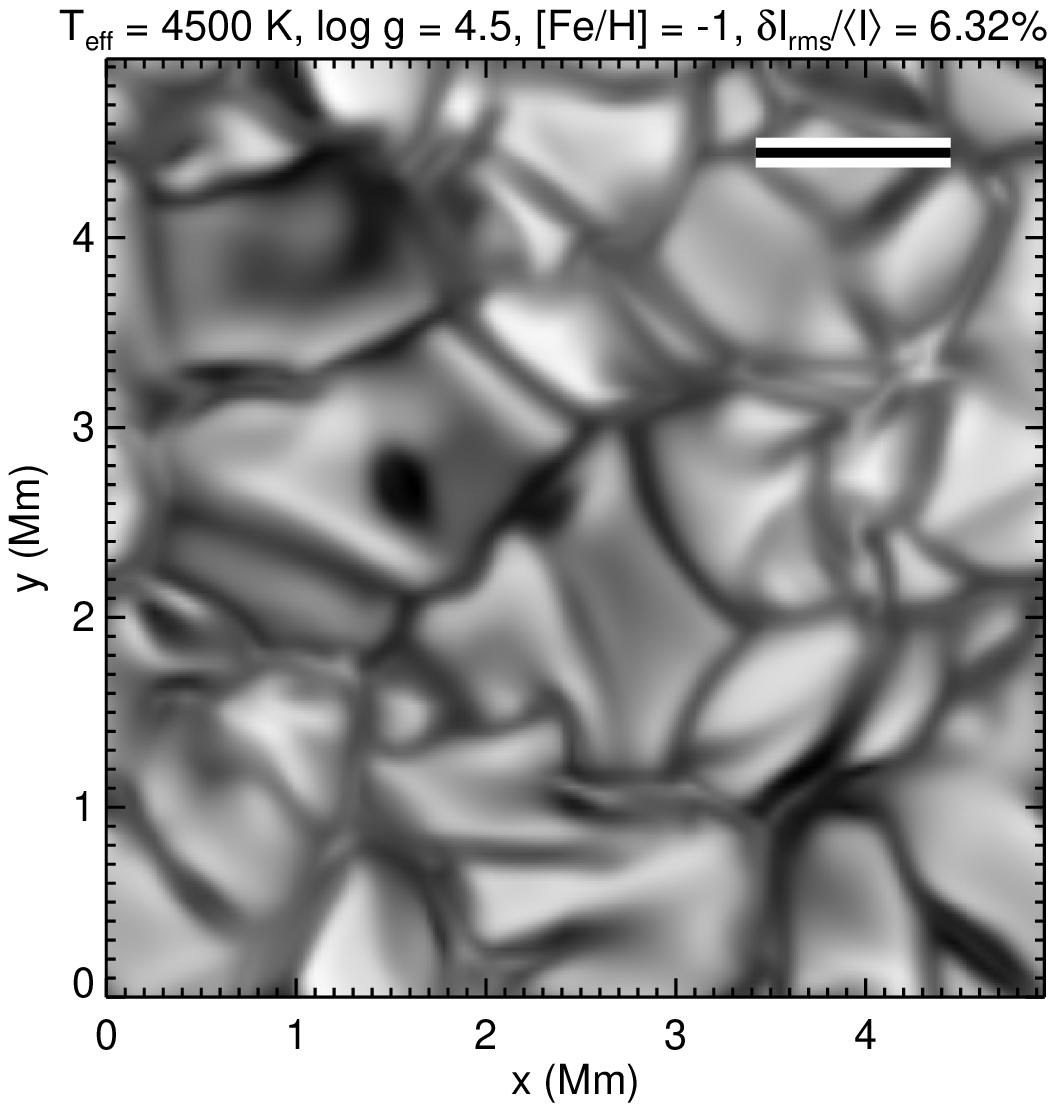}}
\subfloat[]{
\includegraphics[bb=130 400 442 705,width=2.20in]{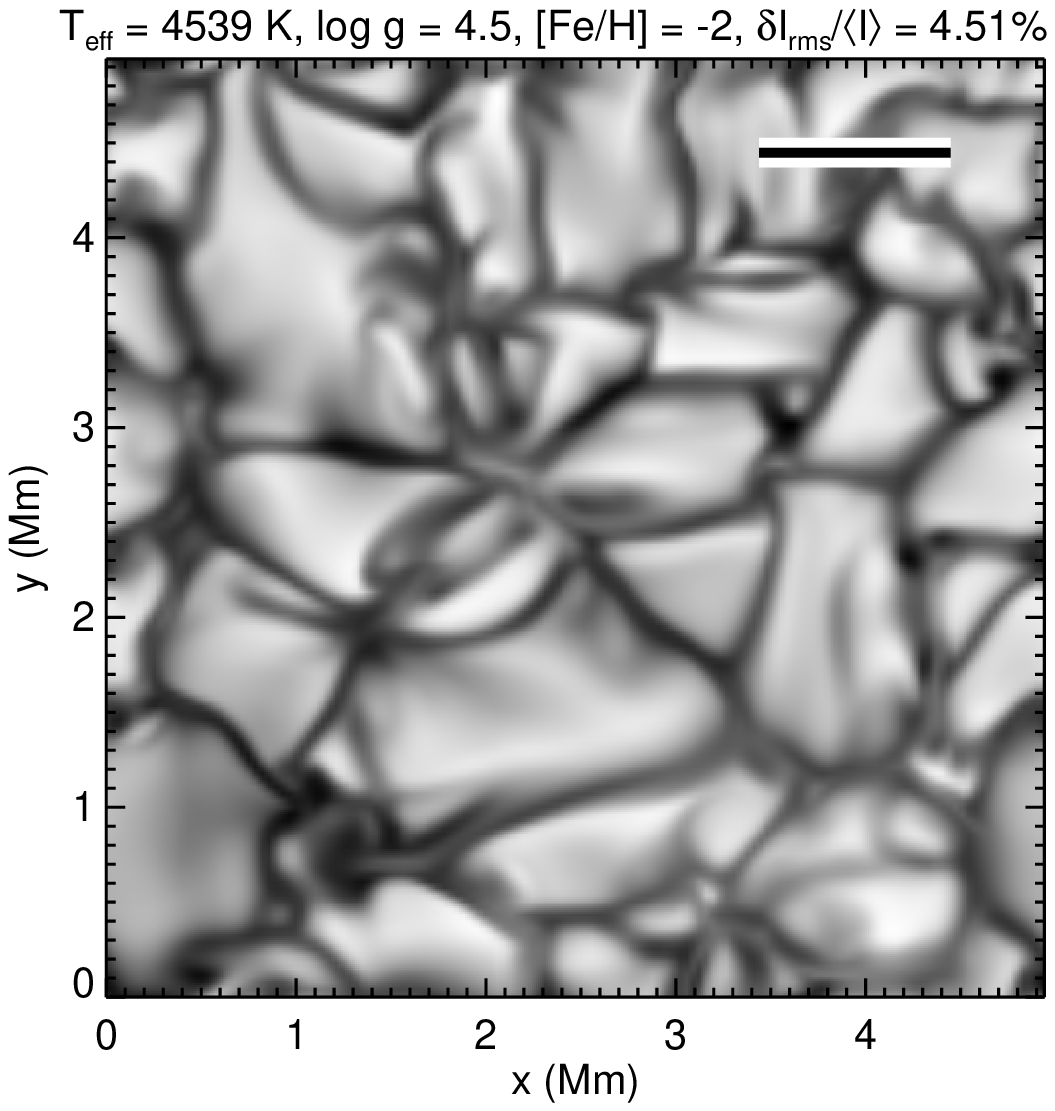}}
\subfloat[]{
\includegraphics[bb=130 400 442 705,width=2.20in]{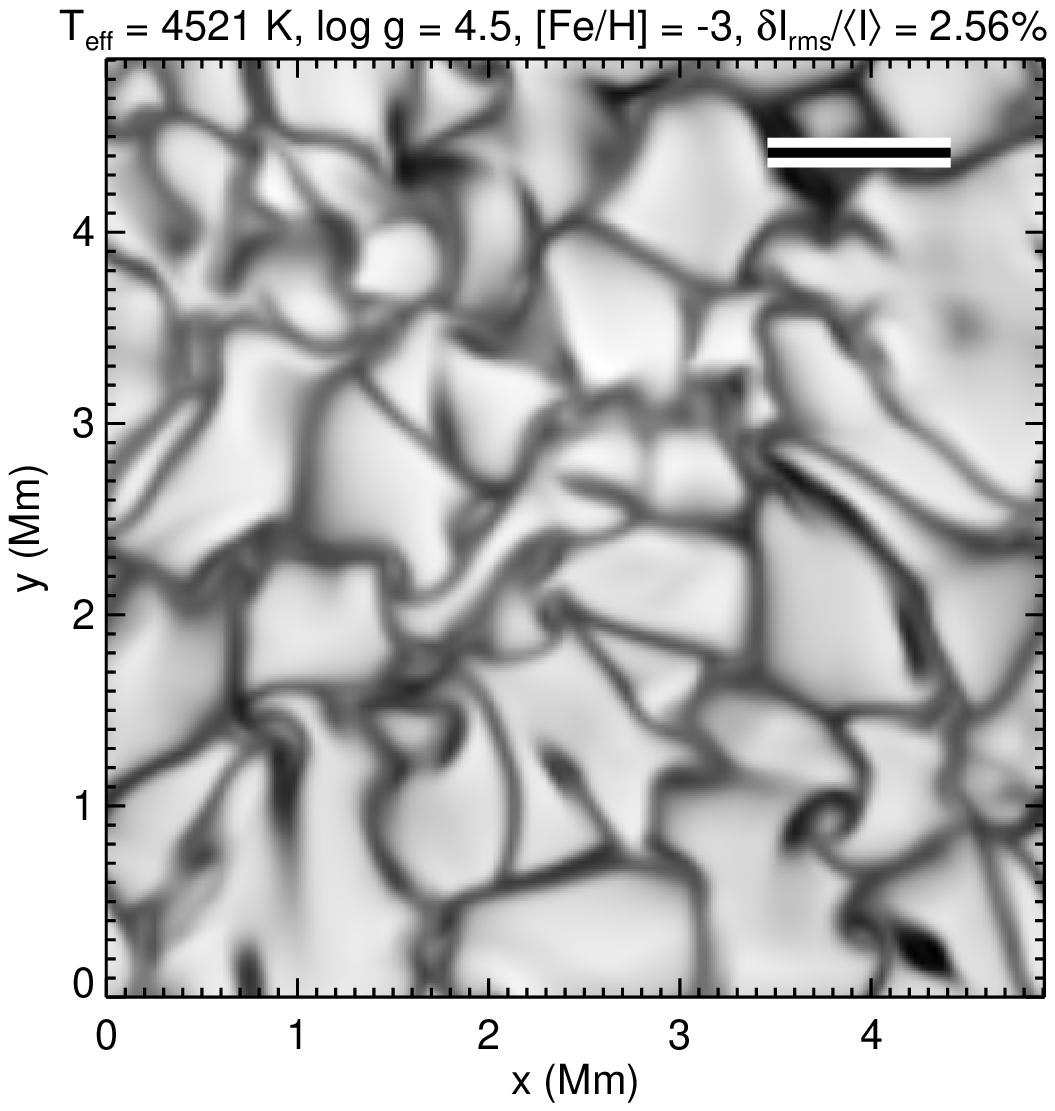}}\\
\subfloat[]{
\includegraphics[bb=130 400 442 675,width=2.20in]{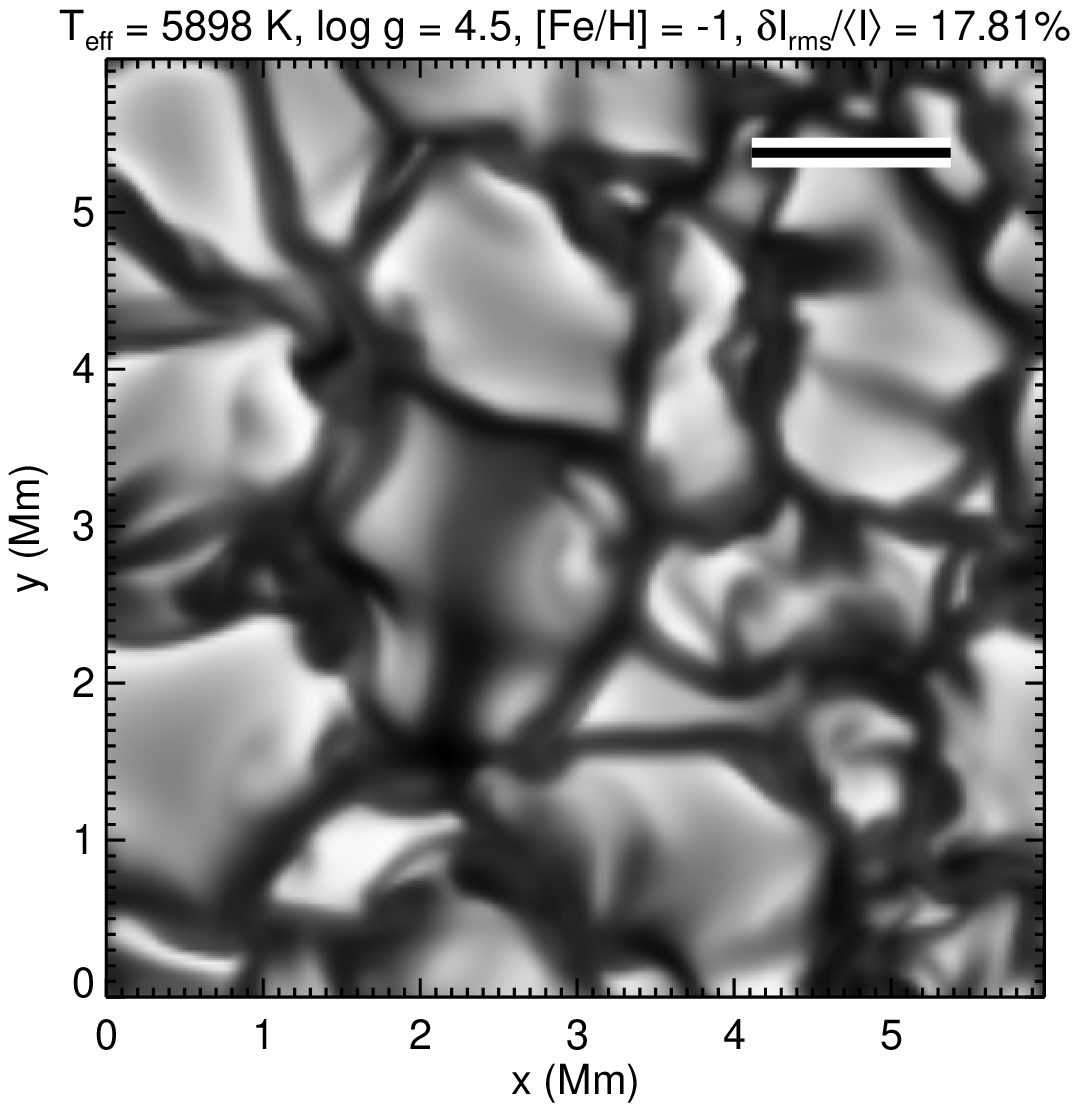}}
\subfloat[]{
\includegraphics[bb=130 400 442 675,width=2.20in]{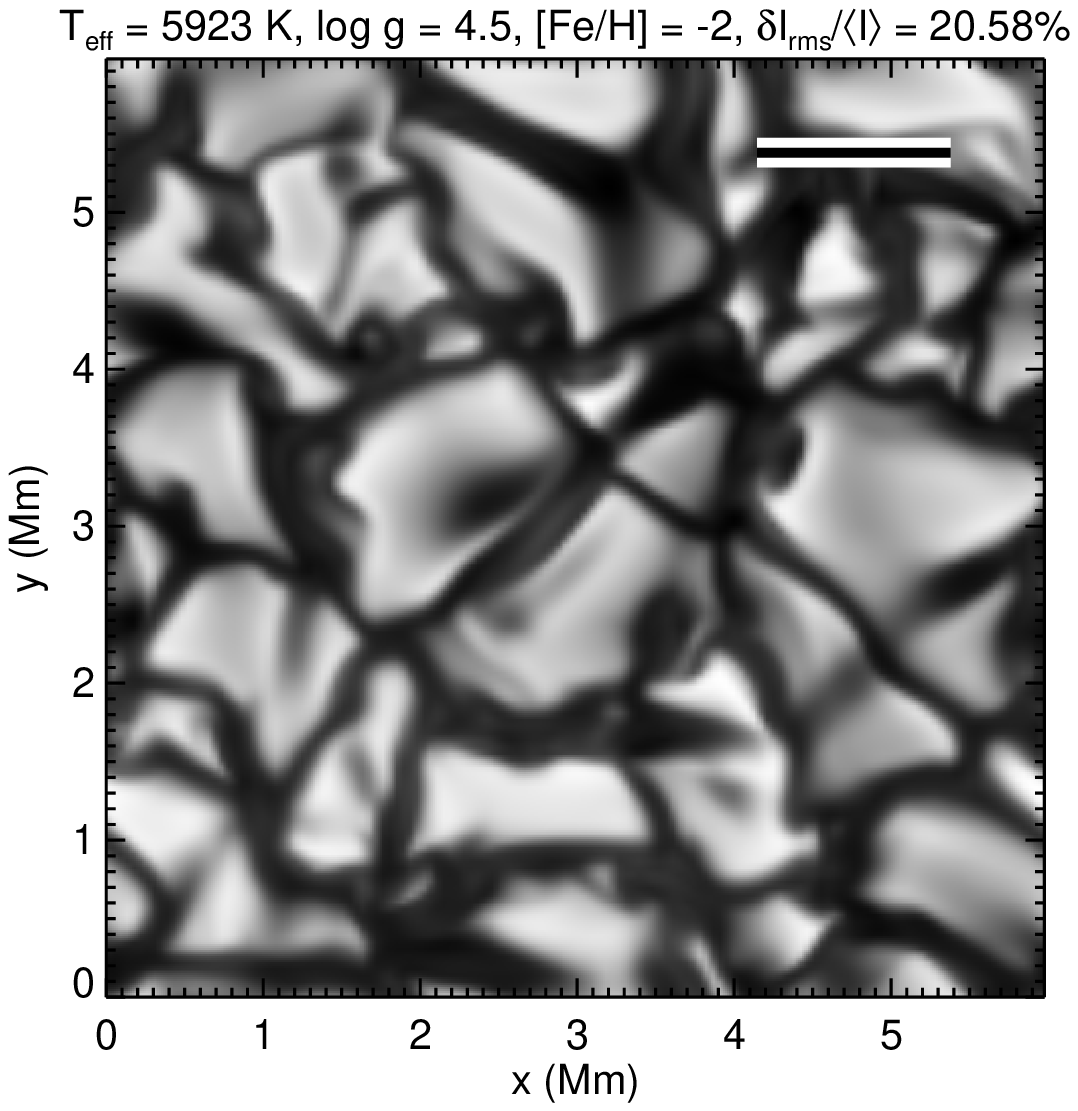}}
\subfloat[]{
\includegraphics[bb=130 400 442 675,width=2.20in]{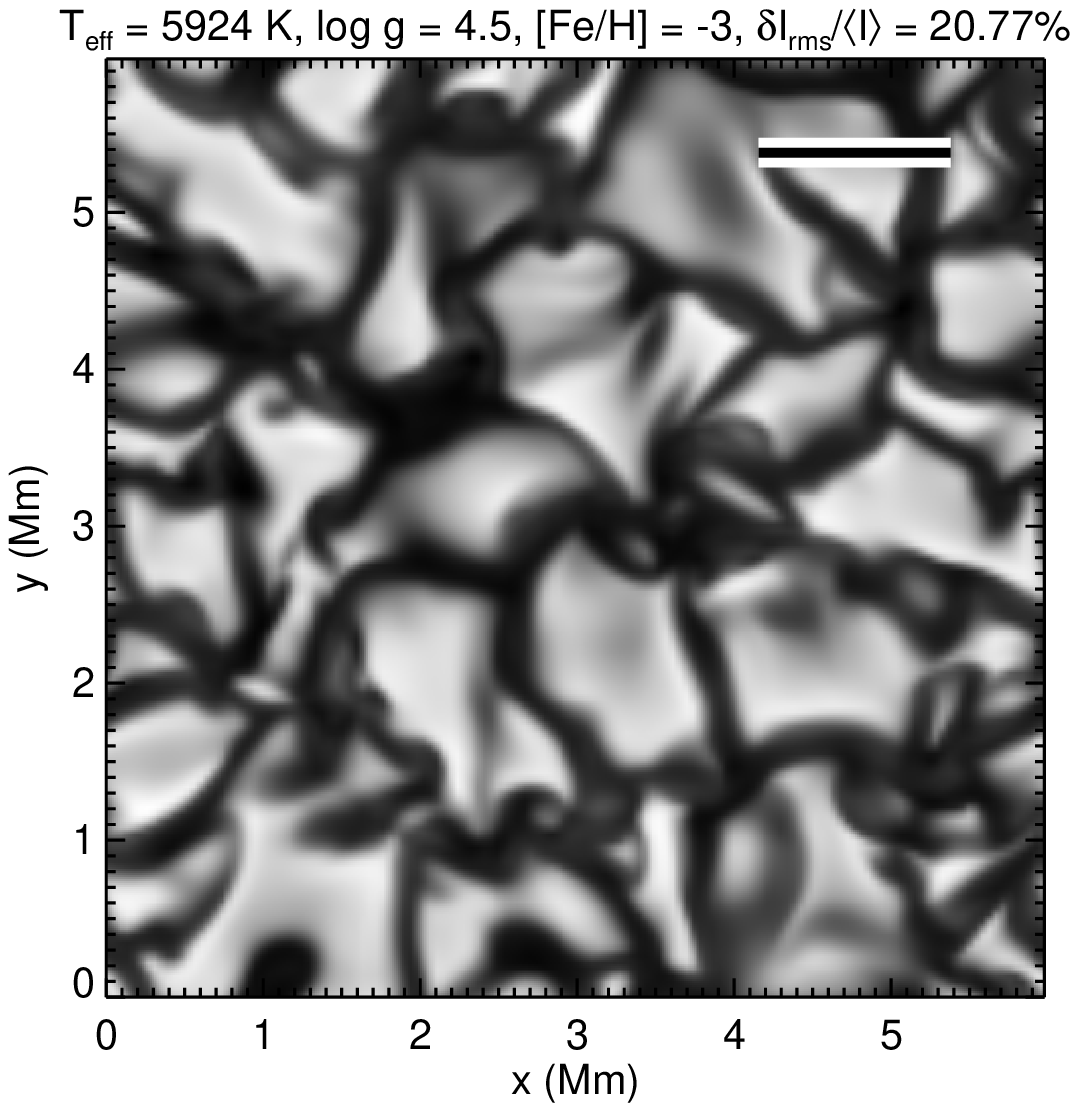}}
\caption{Emergent bolometric intensity for dwarfs at different
  metallicities. In the first row, $T_{\rm eff} =$ 4500~K, $\log g =$ 4.5, and
  [Fe/H] = $-1, -2$ and $-3$ (from left to right). In the second row, $T_{\rm
    eff} =$ 5900~K, $\log g =$ 4.5, and [Fe/H] varies again from $-1$ to $-3$.
\label{fg:f7}}
\end{center}
\end{figure*}

\begin{figure*}[!]
\captionsetup[subfigure]{labelformat=empty}
\begin{center}
\subfloat[]{
\includegraphics[bb=130 400 442 705,width=2.20in]{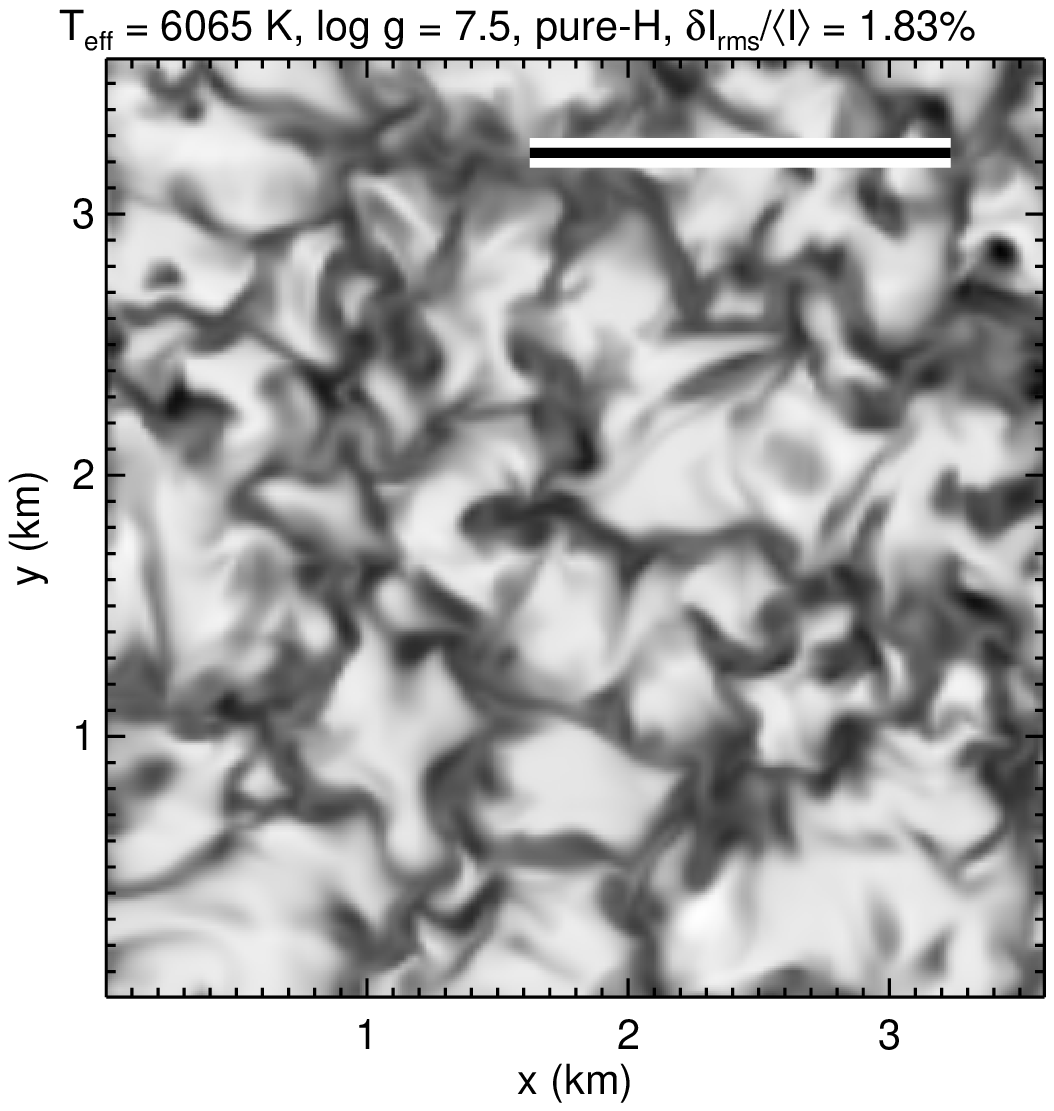}}
\subfloat[]{
\includegraphics[bb=130 400 442 705,width=2.20in]{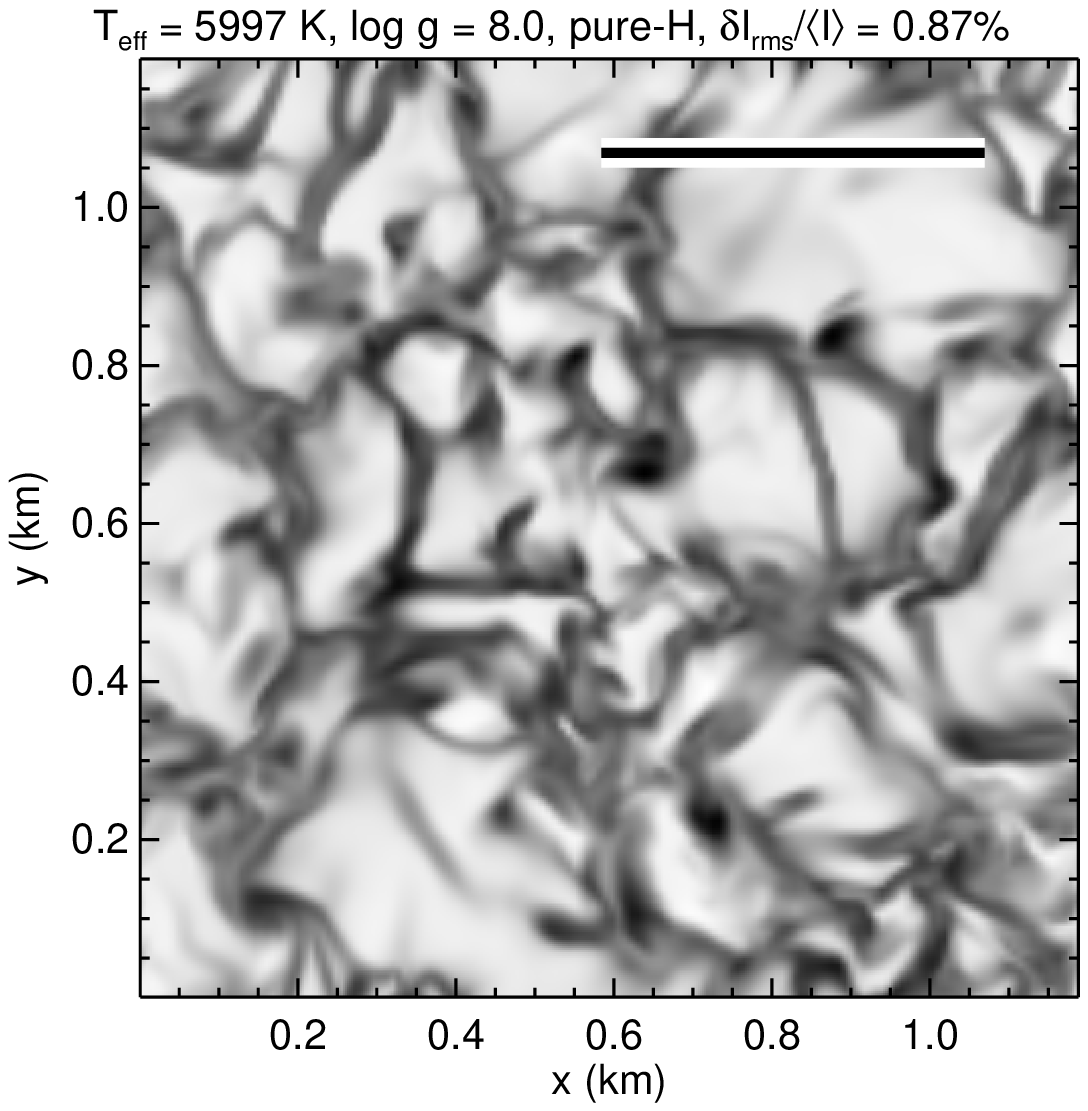}}
\subfloat[]{
\includegraphics[bb=130 400 442 705,width=2.20in]{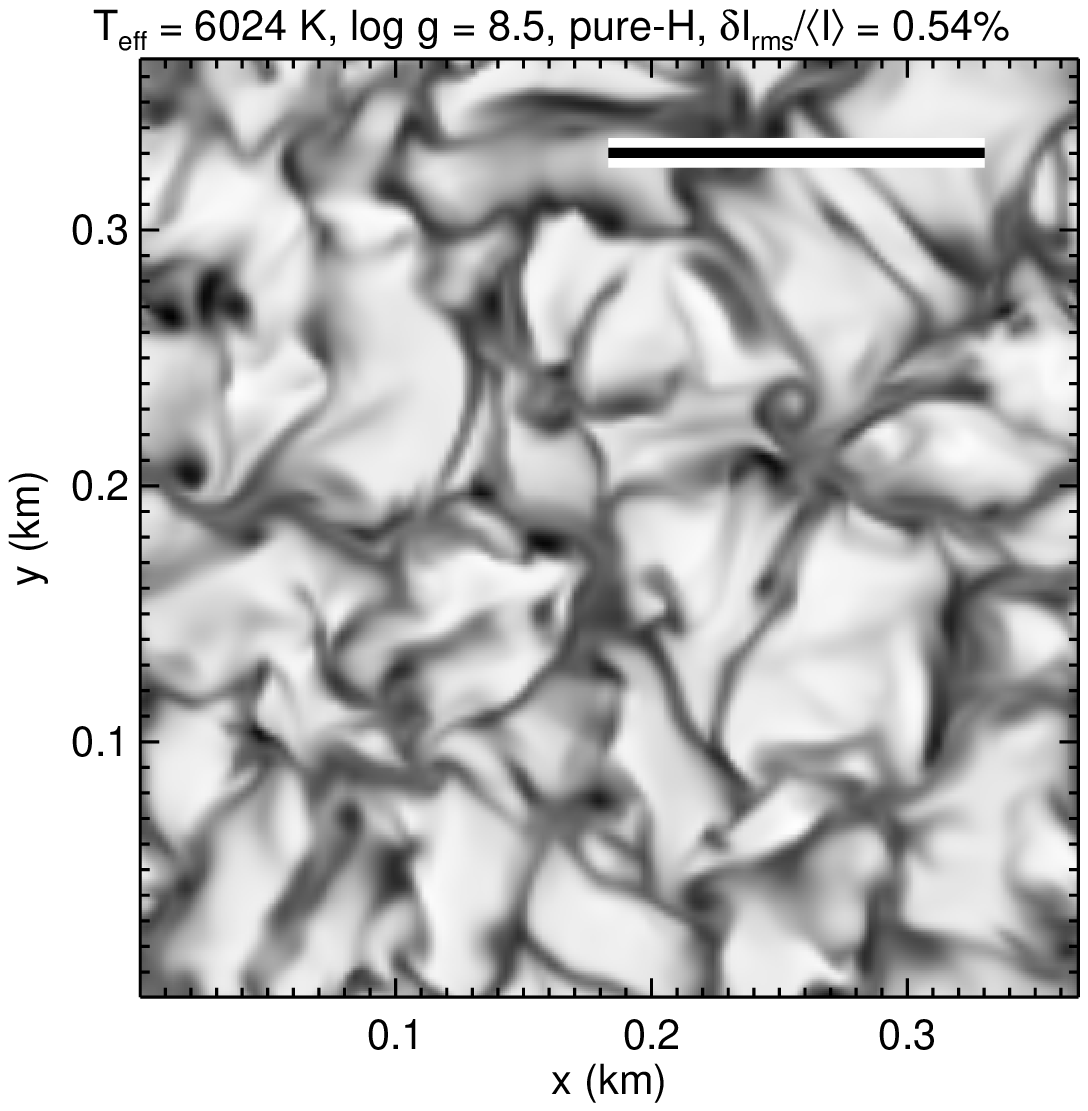}}\\
\subfloat[]{
\includegraphics[bb=130 400 442 675,width=2.20in]{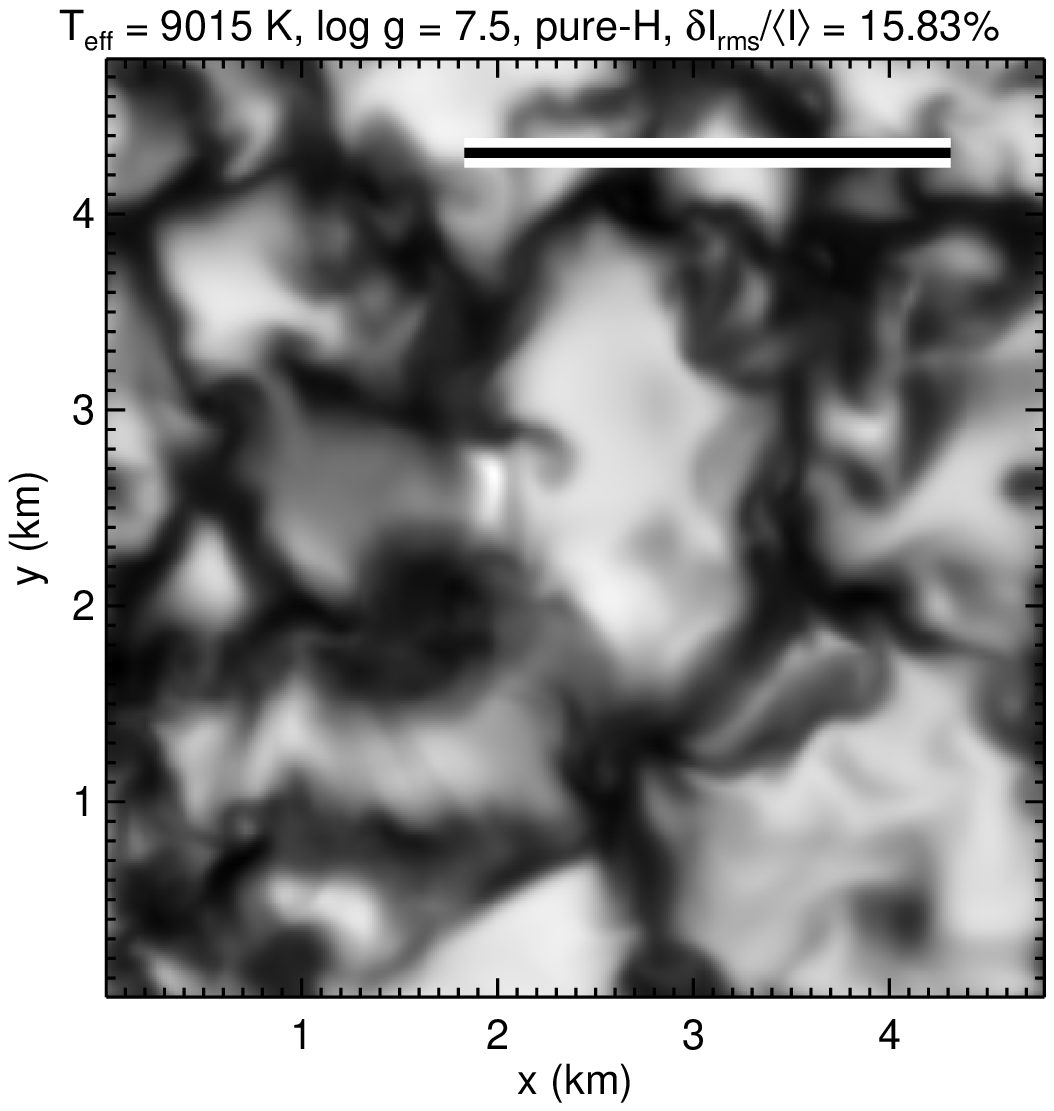}}
\subfloat[]{
\includegraphics[bb=130 400 442 675,width=2.20in]{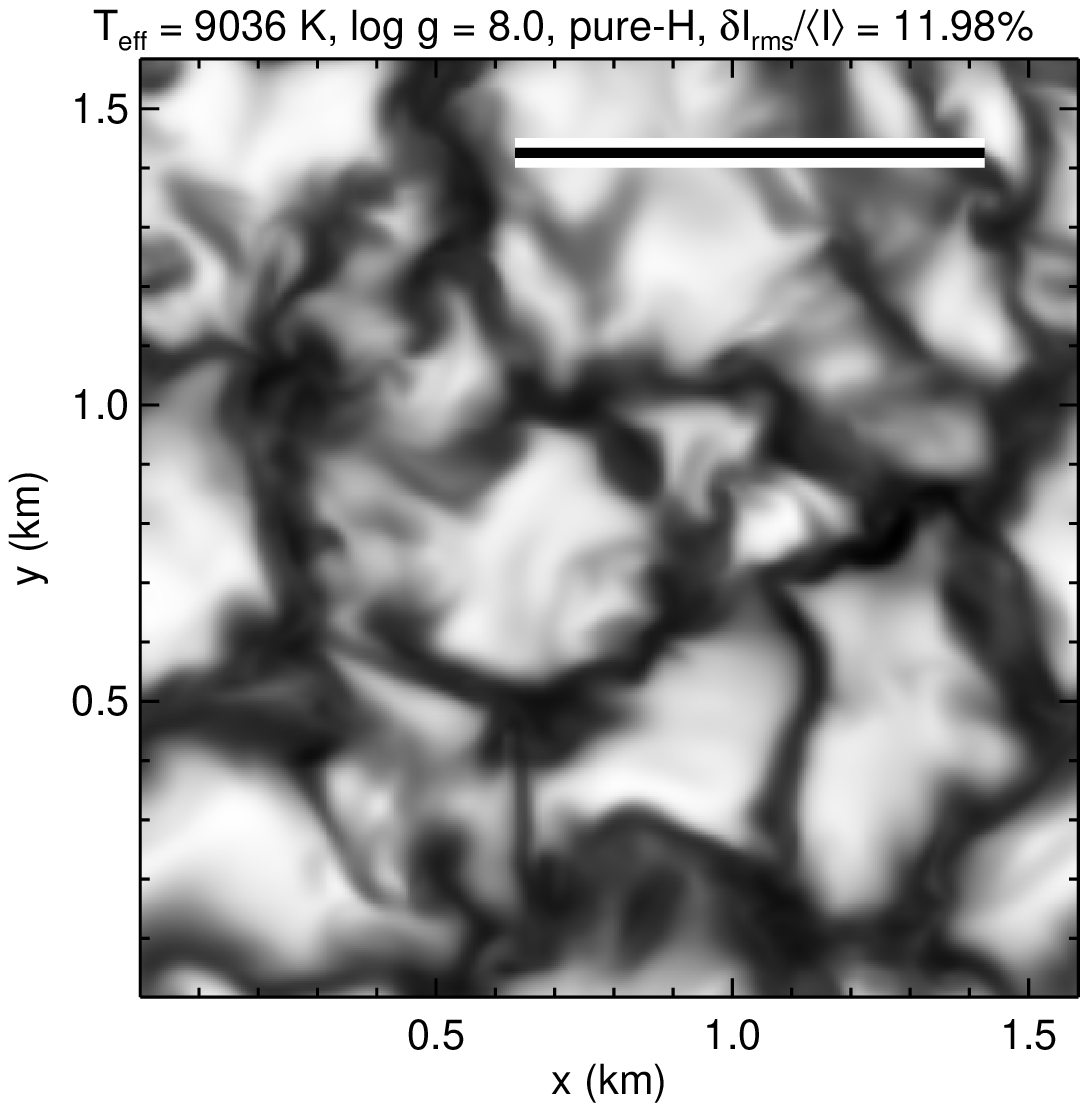}}
\subfloat[]{
\includegraphics[bb=130 400 442 675,width=2.20in]{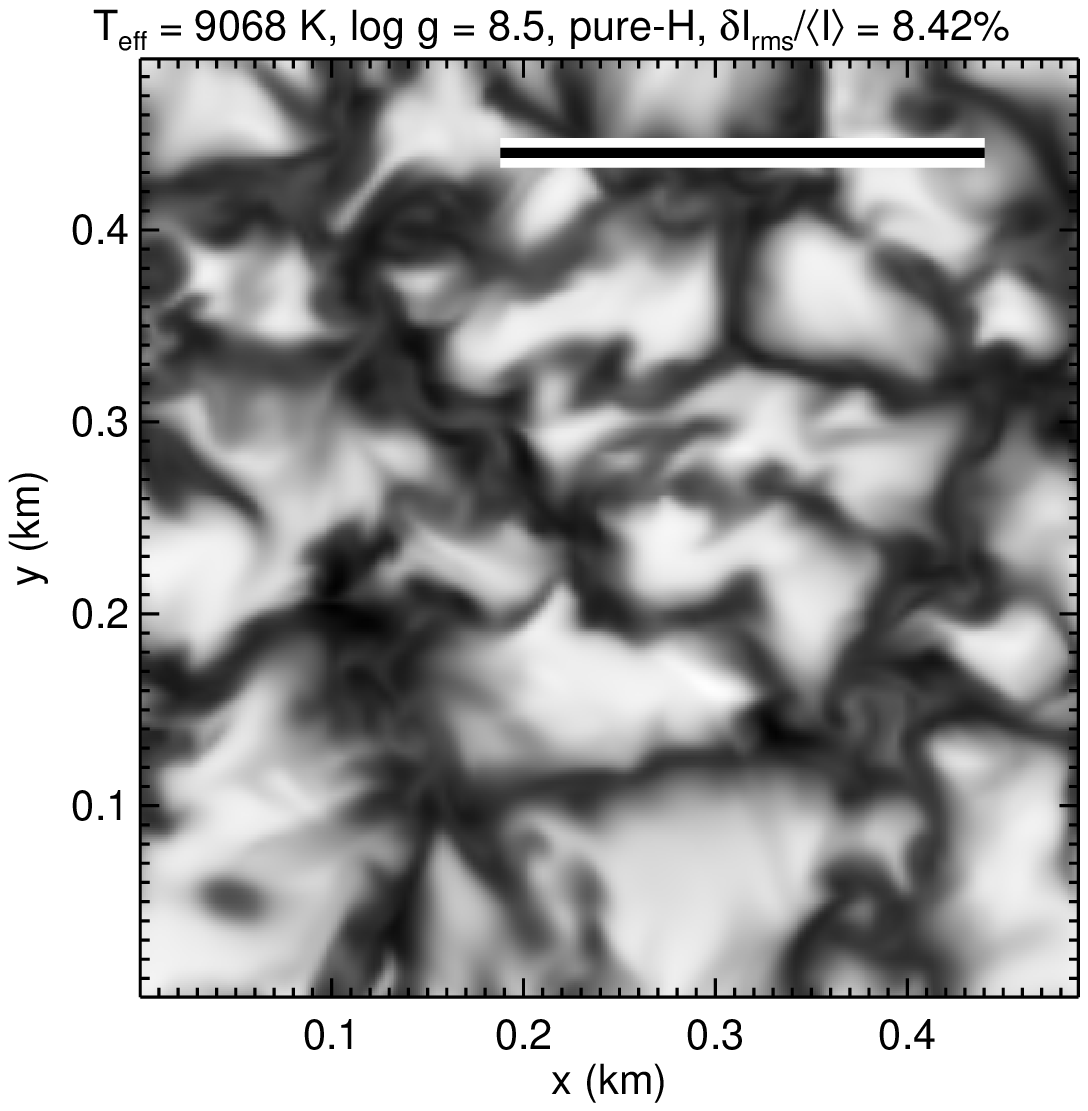}}\\
\subfloat[]{
\includegraphics[bb=130 400 442 675,width=2.20in]{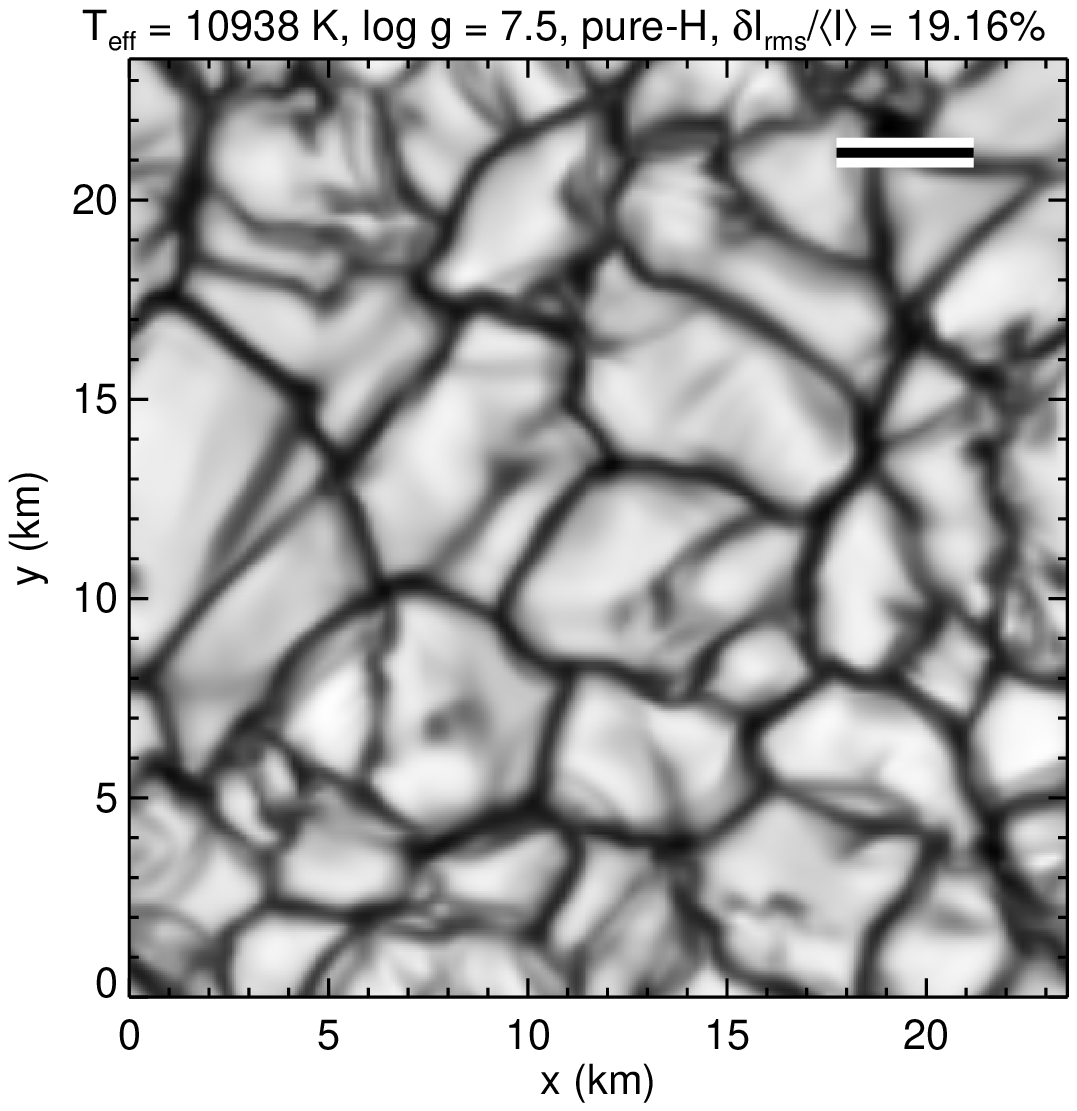}}
\subfloat[]{
\includegraphics[bb=130 400 442 675,width=2.20in]{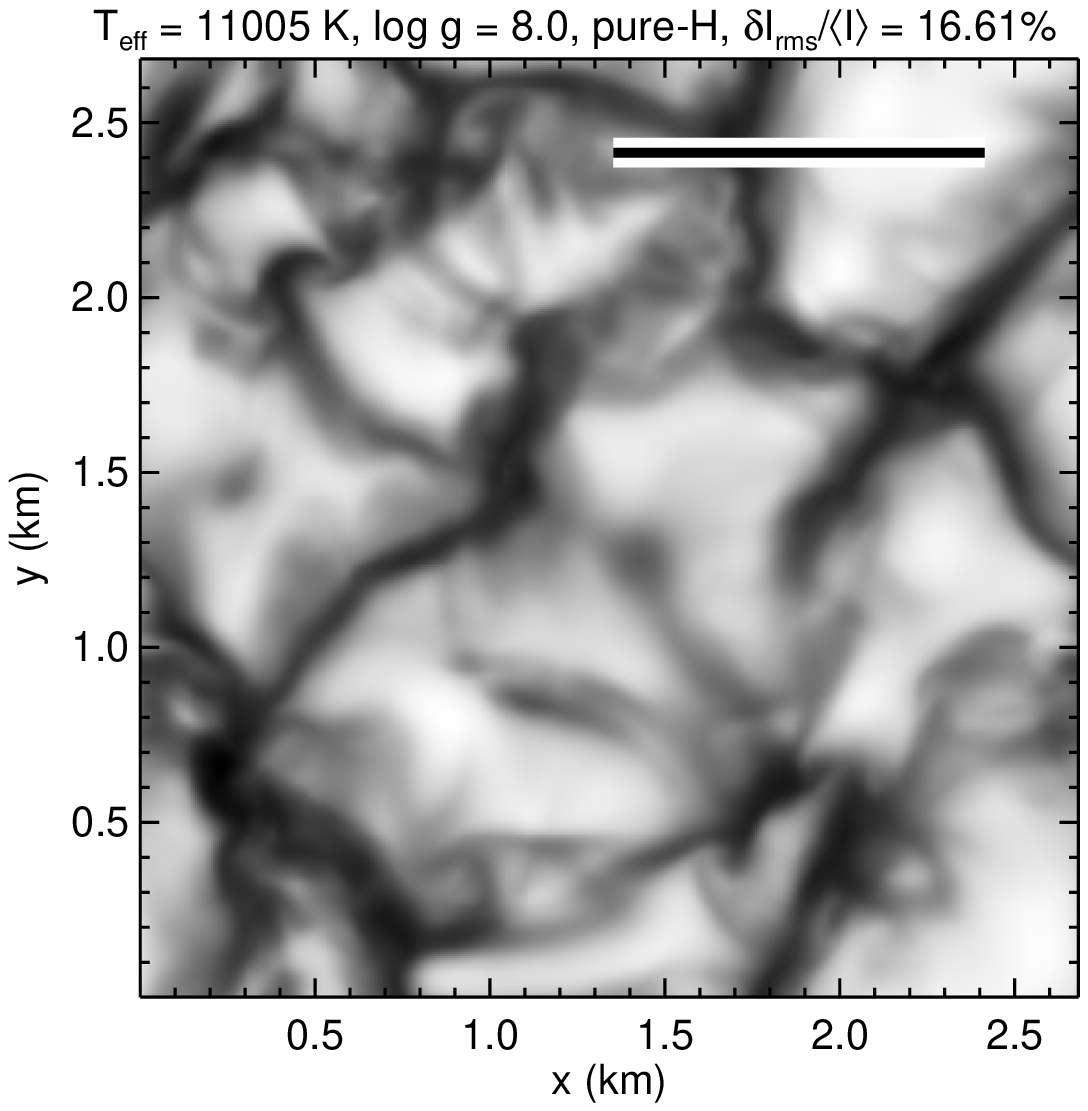}}
\subfloat[]{
\includegraphics[bb=130 400 442 675,width=2.20in]{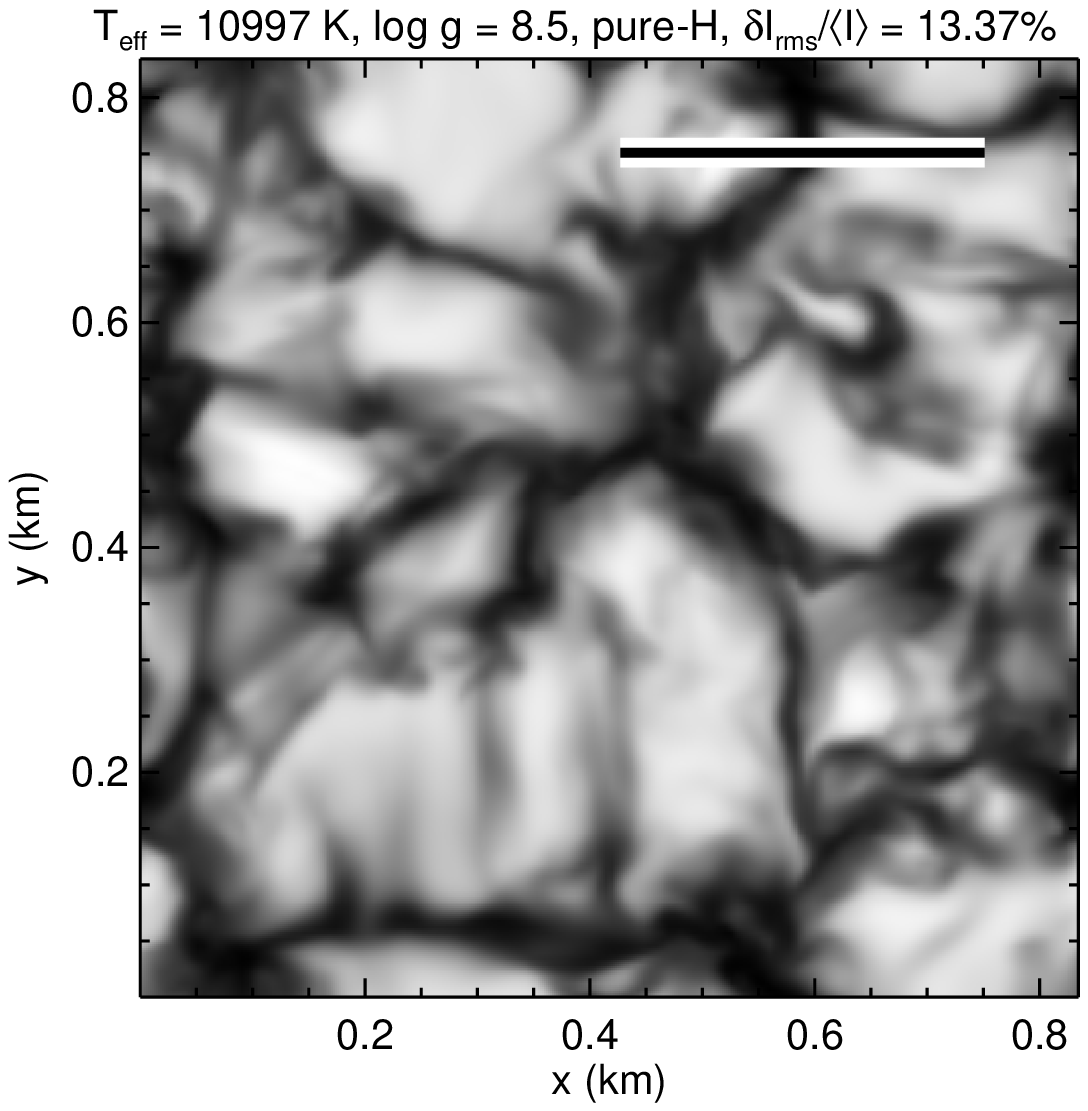}}\\
\subfloat[]{
\includegraphics[bb=130 400 442 675,width=2.20in]{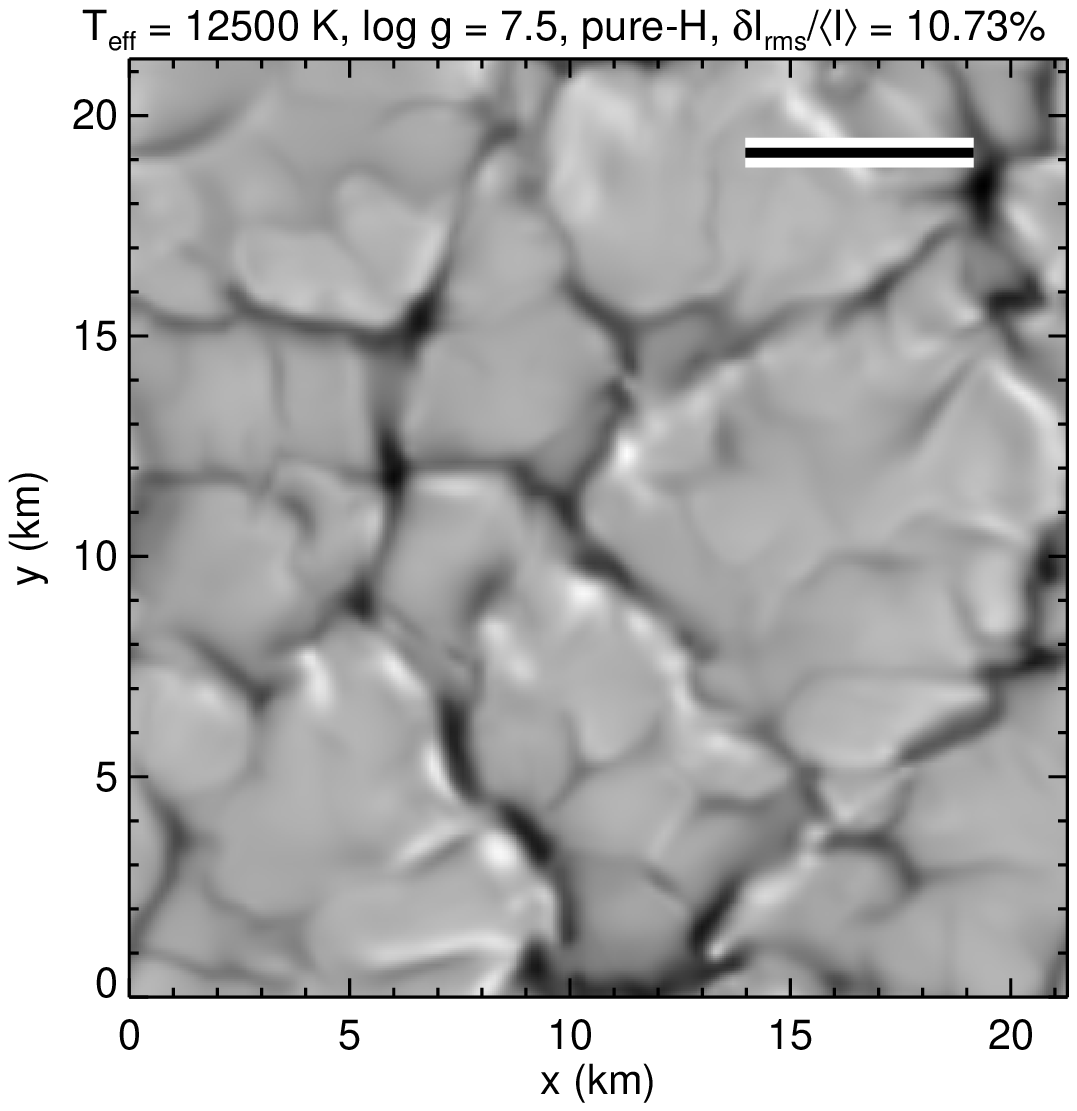}}
\subfloat[]{
\includegraphics[bb=130 400 442 675,width=2.20in]{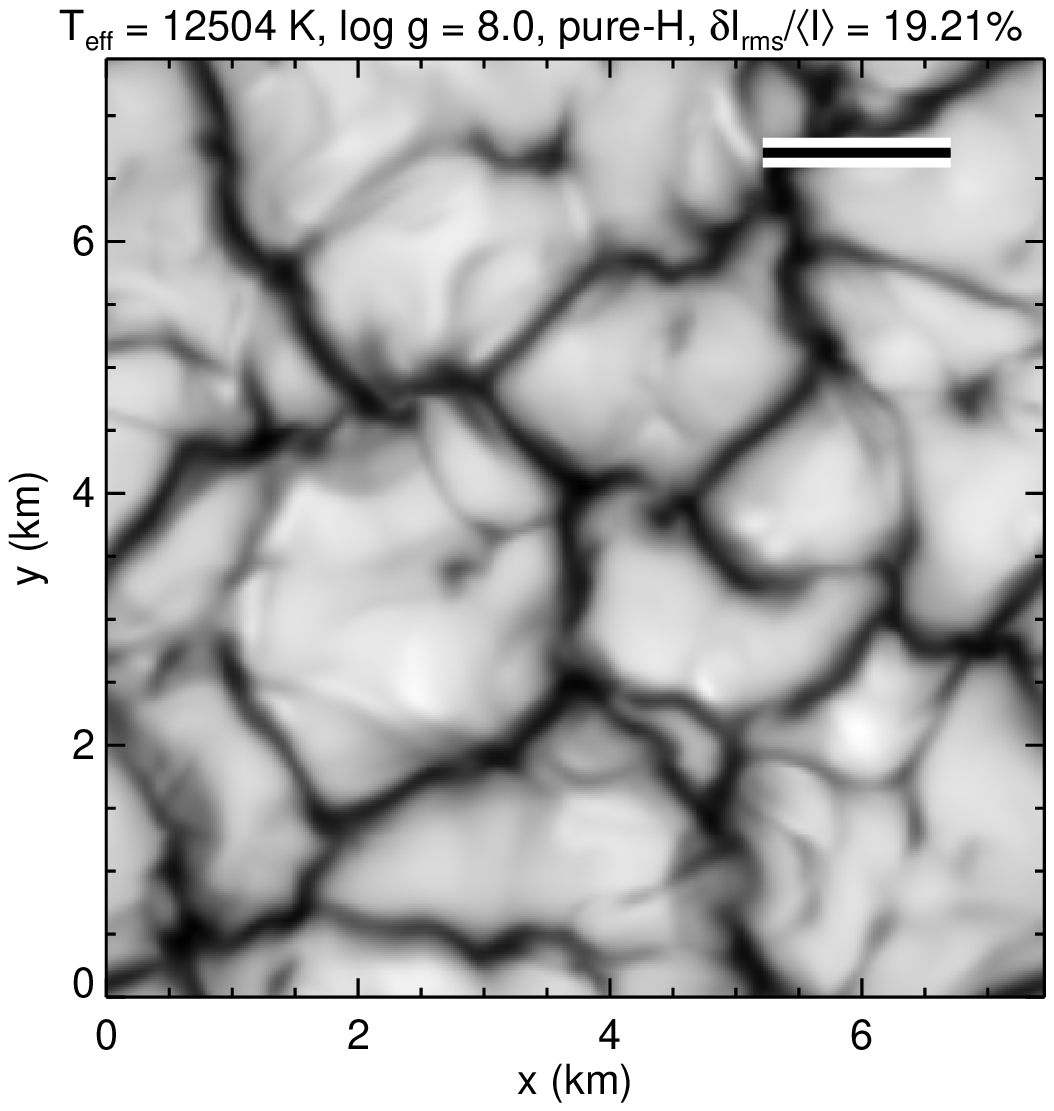}}
\subfloat[]{
\includegraphics[bb=130 400 442 675,width=2.20in]{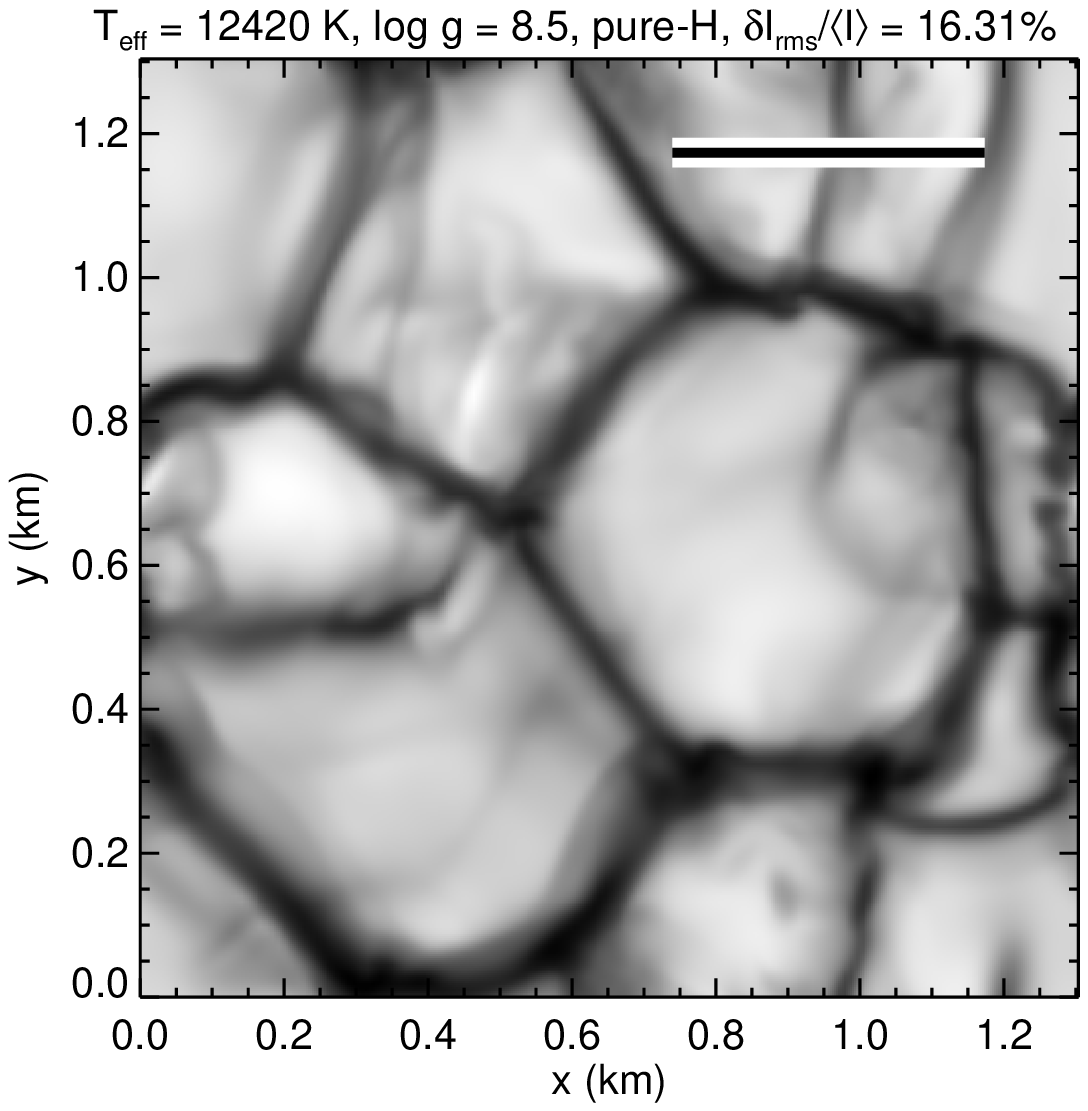}}
\caption{Emergent bolometric intensity for white dwarfs at $T_{\rm eff}$ =
  6000 (1st row), 9000 (2nd row), 11,000 (3rd row), and 12,500~K (4th row) and
  $\log g$ = 7.5, 8.0 and 8.5 (columns from left to right).
\label{fg:f8}}
\end{center}
\clearpage
\end{figure*}

\subsection{White dwarfs}

We rely on 60 CO$^5$BOLD simulations for pure-hydrogen atmosphere white dwarfs (spectral type DA)
in the range $13,000 > T_{\rm eff}$ (K) $> 6000$ K. The 12 models at $\log g =
8.0$ are described in detail in \citet{tremblay13}. We have now extended the
grid of models to $\log g$ = 7.0, 7.5, 8.5 and 9.0. The CO$^5$BOLD setup for
these additional models is identical to the $\log g = 8$ case.  In brief, the
resolution is of 150$\times$150$\times$150 grid points, and the simulations
extend well below and above the photosphere. We rely on an equation-of-state
and opacity tables which have the same microphysics as the standard 1D models
\citep{TB11}. We opted for a 8 opacity bins setup for the radiative transfer
scheme.

Compared to the dwarf and giant models described above, one important
difference is that hotter white dwarfs have a convection zone that is thinner
than the typical vertical dimension of the atmosphere. As a consequence, the
6, 4, 3, and 1 hottest models at $\log g$ = 7.0, 7.5, 8.0 and 8.5,
respectively, were computed with a bottom layer that is closed to convective
flows (zero vertical velocities). In that case, we imposed the radiative flux
at the bottom based on the diffusion approximation. Other than the slightly
different resolution, white dwarf models have been computed with a more recent
version of CO$^5$BOLD. The main difference is that we have now switched to a
less dissipative 2nd-order reconstruction method to solve the hydrodynamical
equations \citep{freytag13}, and no artificial viscosity was
used. Furthermore, regarding the time integration scheme, the corner-transport
upwind (CTU) method \citep{colella90} was adopted. However, we have verified
that these different numerical parameters have almost no impact on the mean
properties of white dwarfs \citep{tremblay13}, hence we believe that white
dwarf models can be compared directly to dwarf and giant simulations.

\subsection{Mean quantities}

In Fig.~\ref{fg:f1}, we present our set of 3D simulations in a HR-type diagram
with $\log g$ vs. $T_{\rm eff}$. The colour- ($T_{\rm eff}$) and size-coding
($\log g$) introduced in this figure is used throughout this paper for dwarfs
and giants (filled symbols) and white dwarfs (open triangles).

We derived a large number of mean thermodynamic and dynamic quantities from
our sequence of 3D simulations that are representative of the photosphere. All
quantities, unless otherwise noted, are spatial and temporal averages over the
constant geometrical depth which corresponds to $\langle\tau_{\rm R}\rangle_{\rm x,y} = 1$.
This layer approximates the region of the photosphere where the intensity is formed, although 3D convection is 
a non-local phenomenon and surface granulation is also influenced by deeper layers. 
We initially restrict our analysis to one characteristic layer in order to compare granulation properties to local parameters, and Sect.~3.2 further
describes the effect of deeper layers. The temporal average was performed over 250 or more snapshots in the last half of the
simulations, where they are all relaxed. Furthermore, we determined the
$T_{\rm eff}$ values identified in Table~1 from the mean emergent flux of the
same snapshots. Finally, the emergent intensity from these snapshots is the
input to study the granulation properties in Sect.~3.

\subsection{Independent variables}

Our aim is to cover a large part of the HR diagram with our simulations, yet
to find similarities between them. For describing the properties of
granulation it is not clear whether $T_{\rm eff}$ and $\log g$ are the most
suitable control variables. Nevertheless, we note that the 
hydrostatic pressure scale height

\begin{equation}
H_{\rm p} = \frac{\langle P \rangle}{\langle \rho \rangle g} ~,
\end{equation}

{\noindent}where $P$ is pressure and $\rho$ the density, is inversely proportional
to $g$ and roughly proportional to $T_{\rm eff}$ if we neglect 
molecular mass variations. To first order, we can therefore scale characteristic sizes as a
function of the atmospheric parameters. Furthermore, the turnover timescale in
the convective zone, or the advective timescale, is defined as $t_{\rm adv} =
H_{\rm _p} / v_{\rm rms}$ where $v_{\rm rms}$ is the convective
velocity

\begin{equation}
v^2_{\rm rms} = \langle v^2 \rangle - \frac{[\langle \rho v_{\rm x} \rangle^2 + \langle \rho v_{\rm y} \rangle^2 + \langle \rho v_{\rm z} \rangle^2]}{\langle \rho \rangle^2}~.
\end{equation}

{\noindent}The mass flux weighted mean velocity is removed from the rms value since it is sensitive
  to the setup of the numerical parameters and oscillations. Fig.~\ref{fg:f15} presents the advective timescale with $H_{\rm p}$ 
and $v_{\rm rms}$ derived from our 3D simulations and averaged over $x$, $y$, and $t$ at $\langle \tau_\mathrm{R} \rangle_{\rm x,y}
  = 1 $. Clearly, the characteristic turnover
time is anticorrelated with gravity, as are characteristic sizes. 

The atmospheric densities are show in Fig.~\ref{fg:f2} as a function of
  $T_{\rm eff}$. It demonstrates that stars and white dwarfs share a common range of photospheric densities ($-8
  < \log \rho < -4$) despite the different surface gravities. The opacities typically increase with $T_{\rm eff}$ and
metallicity, hence the photosphere is pushed towards lower densities. On the
other hand, higher gravities imply higher densities, although because of
different opacities, the relation is not strict. However, it is seen that by
increasing both $T_{\rm eff}$ and $\log g$, it is possible to keep the
photospheric density constant.

It is demonstrated in Fig.~\ref{fg:f3} that in the photosphere, the density
correlates relatively well with the Mach number, the ratio of flow and
sound speeds

\begin{equation}
{\rm Mach} = \frac{v_{\rm rms}}{c_s} = \sqrt{\frac{\langle\rho\rangle v^2_{\rm rms}}{\langle \Gamma_1 \rangle \langle P \rangle}} ,
\end{equation}

{\noindent}where $\Gamma_1$ is the first adiabatic exponent. We observe a
small offset between white dwarfs, dwarfs and to a lesser degree giants for
which we only have a few models. This is not entirely surprising since for a
given density, the higher gravity objects have larger temperatures. The energy density 
is then higher, but the vertical velocity must also increase  
to obtain the desired convective flux, which is confirmed by the MLT equations. The hot white dwarfs are one exception,
and they do not follow the trend since the convective to radiative flux ratio
becomes increasingly small in these atmospheres. The CIFIST grid does not
cover main-sequence A-stars where a similar downturn in the Mach number is
observed from other CO$^{5}$BOLD simulations \citep{astars,freytag12}.

The P\'eclet number is the ratio of the radiative and advective timescales

\begin{equation}
{\rm Pe} = \frac{t_{\rm rad}}{t_{\rm adv}} = \frac{ \langle \rho \rangle \langle c_p \rangle 
  v_{\rm rms} \tau_{\rm e}}{16 \sigma \langle T \rangle^3} \left( 1 +
\frac{2}{\tau_{\rm e}^{2}} \right) ,
\end{equation}

{\noindent}where $c_{p}$ is the specific heat per gram, $T$ the temperature, $\sigma$ the
Stefan-Boltzmann constant, and $\tau_e = \langle \kappa_{\rm R} \rangle \langle \rho \rangle H_{\rm p}$ the
characteristic optical depth of a disturbance of a size $H_{\rm
  p}$, with $\kappa_{\rm R}$ the Rosseland mean opacity per gram. This number indicates which will be the dominant energy transfer process for the formation and evolution of
convective cells. This characteristic number is also proportional to the
convective efficiency in the 1D MLT. The P\'eclet and Mach numbers have a
  tight relation in the photosphere according to Fig.~\ref{fg:f4}, except for the hottest white dwarfs. This implies that 
  P\'eclet number and density are also closely correlated. For
  $\log \rho \lesssim -6$, the convection at $\tau_\mathrm{R} \sim 1$ becomes
  relatively inefficient and granules lose a significant part of their energy through
  radiation. The P\'eclet number is rapidly varying as a function of optical
depth, and values in Fig.~\ref{fg:f4} should be taken as an order of magnitude
estimate only.

In the following, we rely on the logarithm of the Mach number as the reference variable to
characterize most granulation properties. This quantity is not changing
rapidly as a function of $\tau_{\rm R}$, hence the value at $\tau_{\rm R} = 1$ is a suitable control variable. Since Mach and P\'eclet numbers
correlate well, and represent significantly different characteristics of the
plasma, it is difficult to separate the effects of both characteristic
numbers. We note that both numbers are derived from the simulated convective
velocities, which is a quantity that can only be roughly approximated by
1D models. We will discuss further in Sect.~4.3 about how it is possible to
overcome this issue.

\section{Granulation properties}

We derive in this section characteristic quantities computed from
snapshots of the 3D simulations such as those presented in Fig.~\ref{fg:f5}
(giants, solar metallicity), Fig.~\ref{fg:f6} (dwarfs, solar metallicity),
Fig.~\ref{fg:f7} (dwarfs, different metallicities), and Fig.~\ref{fg:f8}
(white dwarfs, pure-hydrogen). All snapshots are also available in the online
  Appendix~A. It is immediately clear from the scales of the plots that
gravity is the main factor in determining the size of convective cells,
although granulation is otherwise visually very similar in most simulations
across the HR diagram. The intensity contrast values, given above the snapshots,
are however varying substantially with temperature. The following subsections
aim at describing more quantitatively those observations.

The emergent intensity maps do depend on a certain range of the photosphere
and {\it proper} (solar-type) granulation occurs when the transition from
convectively unstable to stable is rather sharp. When the entropy minimum is
rather wide, the layers where the overturning from up to down has to occur are
not so well defined. In addition, the region where the visible light comes
from is wider. This is likely the explanation why in cool white dwarfs
(Fig. \ref{fg:f8}), where the entropy minimum is very wide and the P\'eclet
number is large, granulation appears fuzzier than in dwarfs
(Fig. \ref{fg:f6}).

In hotter white dwarfs, a different granulation pattern stands out with
narrower cool downdrafts and hot cells with smoother edges. A similar effect
is observed for red supergiants and main-sequence F- and A-stars \citep[][see
  Fig.~15]{freytag12}, both not included in this work.

In objects with a low intensity contrast, such as the white dwarfs at 6000~K
in Fig.~\ref{fg:f8} but also main-sequence M-stars \citep{ludwig06}, vortical
and knot structures are clearly seen in inter-granular lanes. \citet{ludwig06}
suggest that the reduced level of horizontal shearing at low Mach number may
be the explanation. Vortices are also observed and simulated in the Sun
\citep{nature12}, although they are less prominent in low resolution models.

\subsection{Intensity contrast}

The intensity contrast of surface granulation is a measure of the deviation
from the plane-parallel approximation. The root-mean-square (rms) relative
intensity contrast $\delta I_{\rm rms}$ defined as

\begin{equation}
\frac{\delta I_{\rm rms}}{\langle I \rangle} = \left\langle \frac{\sqrt{\langle I(x,y,t)^2 \rangle_{\rm x,y}
    - \langle I(x,y,t) \rangle_{\rm x,y}^2}}{\langle I(x,y,t)\rangle_{\rm
    x,y}}\right\rangle_{\rm t}
\end{equation}

\begin{figure}[h!]
\captionsetup[subfigure]{labelformat=empty}
\begin{center}
\includegraphics[bb=40 40 552 765,width=2.3in,angle=90]{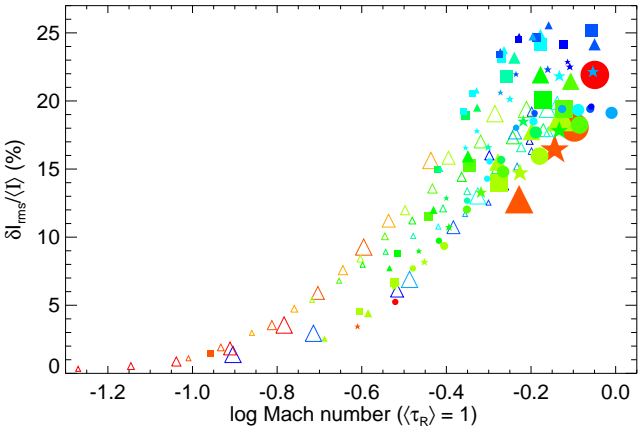}
\caption{Rms relative intensity contrast as a function of the logarithm of the
  Mach number (evaluated at $\langle\tau_\mathrm{R}\rangle = 1$).
\label{fg:f9}}
\end{center}
\end{figure}

{\noindent}is shown in Fig. \ref{fg:f9} as a function of the Mach number in
the photosphere. It is seen that the intensity contrast correlates well with
Mach number over the full range of the HR diagram. This might not be entirely
surprising since both the Mach number and the intensity contrast are a measure
of the strength of convection. The low metallicity dwarfs appear to have a
higher maximum intensity contrast than solar-like dwarfs and white dwarfs.
This will be investigated further in Sect.~4. The white dwarf sequence has two
branches at low Mach number, with cool white dwarfs under the quasi-adiabatic
convection regime, and hot white dwarfs with small convective to radiative
flux ratios. In both cases, the lower intensity contrast can be explained by
a smaller amount of energy to be transported by convection. Given the rather different physical
conditions (e.g. density and P\'eclet number) in those atmospheres, it is
expected that the Mach number will not uniquely describe the intensity
contrast.

\begin{figure*}[!]
\captionsetup[subfigure]{labelformat=empty}
\begin{center}
\subfloat[]{
\includegraphics[bb=100 10 532 735,width=2.2in,angle=90]{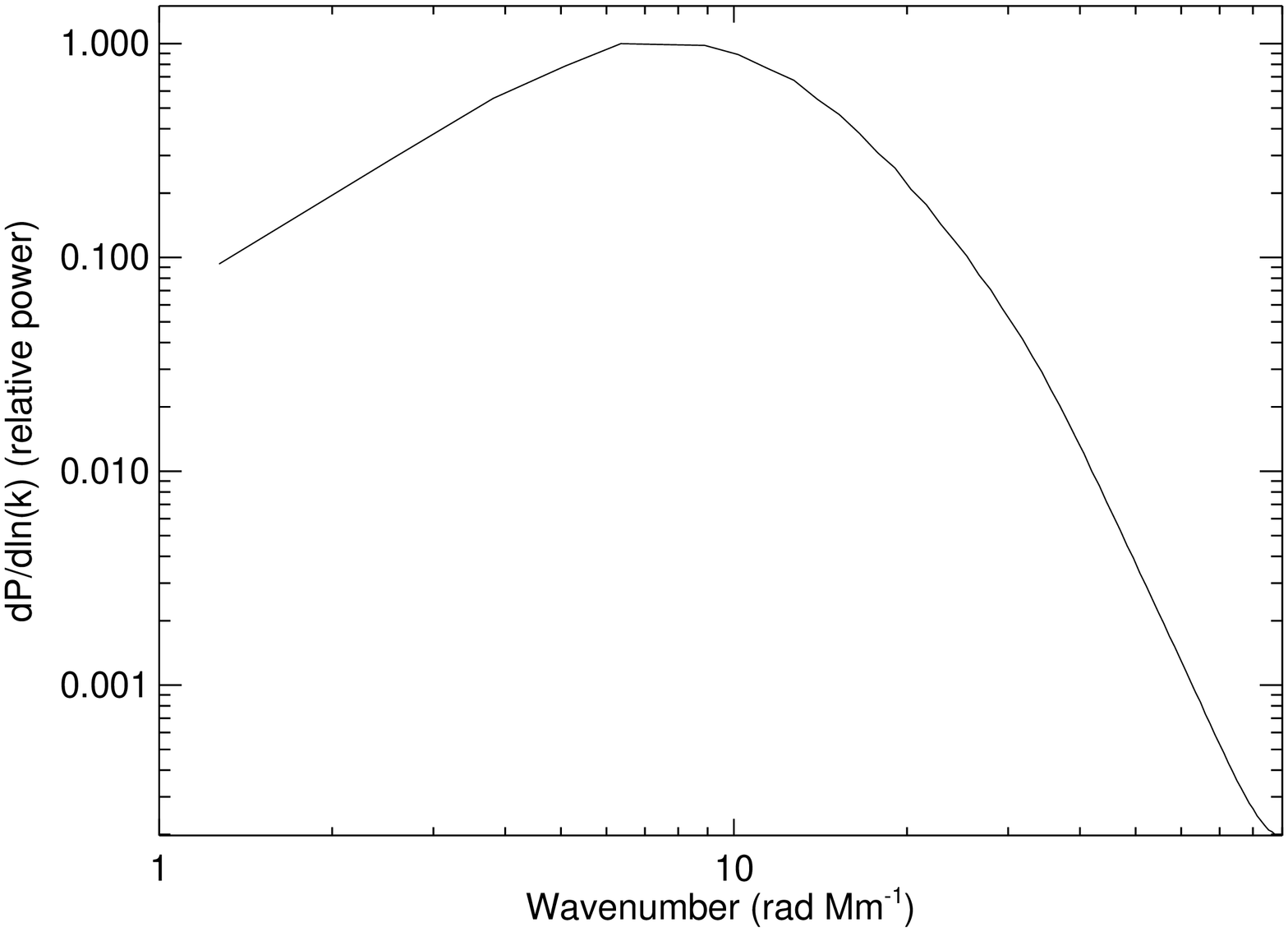}}
\subfloat[]{
\includegraphics[bb=100 10 532 735,width=2.2in,angle=90]{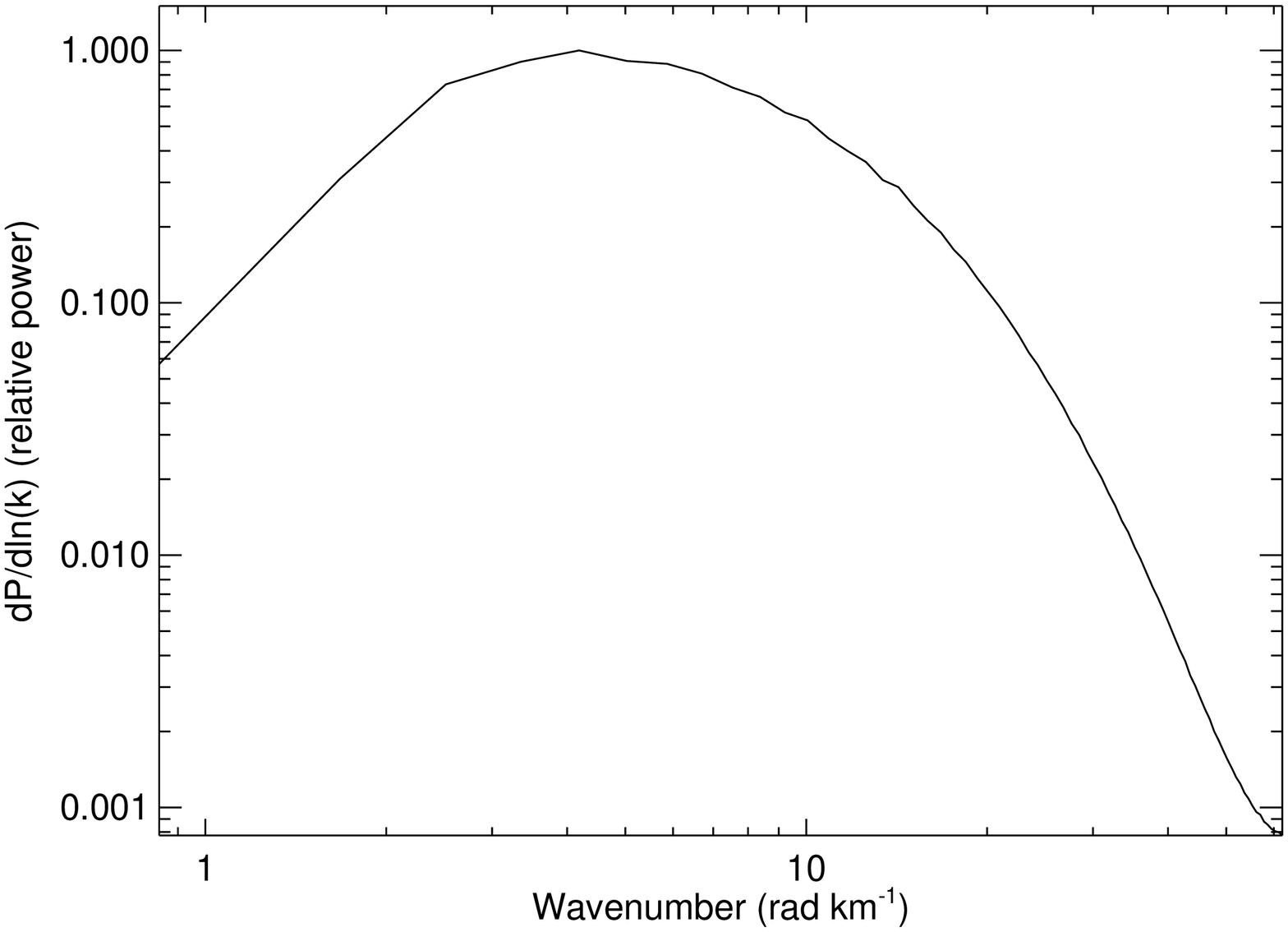}}
\caption{Mean power spectra as a function of the horizontal wavenumber
  (2$\pi$/$\lambda$) averaged over snapshots of the (bolometric) intensity
  of the $T_{\rm eff}$ = 5000~K, $\log g = 4.5$, [Fe/H] = 0 (left panel) and
  $T_{\rm eff}$ = 12,500~K, $\log g = 8.0$ (right panel) simulations.
\label{fg:f11}}
\end{center}
\end{figure*}

The behaviour observed in Fig.~\ref{fg:f9} is also predicted qualitatively 
by the mixing-length theory. The temperature fluctuation between a convective
element and its surroundings is related to the convective velocity according to

\begin{equation}
\frac{\delta T_{\rm MLT}}{T} = \frac{b}{a}\frac{1}{ g Q \Lambda}{v_{\rm MLT}^2}~,
\end{equation}

{\noindent}where $Q$ is the isobaric expansion coefficient, $a$ and $b$ are commonly used free parameters defining the convective flux and the velocity, respectively, and $\Lambda$ is the mixing-length \citep{mihalas78,tassoul90,ludwig99}. A relation
can also be found for temperature fluctuations as a function of P\'eclet number (derived from the MLT convective 
efficiency). We have solved the MLT equations for the mean 3D structures and found that inside the convective zones, temperature
fluctuations are indeed increasing with Mach number. However, it is difficult to compare directly
the results to Fig.~\ref{fg:f9} since the $\tau_{\rm R} = 1$ region is sometimes convectively stable according to the MLT, 
and also because of the free parameters in Eq.~6. Nevertheless, MLT supports
the view that the intensity contrast is a function of the Mach or the P\'eclet numbers.

On the other hand, we can also study the relative temporal variation of the
spatially averaged intensity defined as

\begin{equation}
\frac{\sigma_{\rm I}}{\langle I \rangle} =  \frac{\sqrt{\langle \langle I(x,y,t)\rangle_{\rm
    x,y}^2\rangle_{\rm t}-\langle I(x,y,t)\rangle_{\rm
    x,y,t}^2}}{\langle I(x,y,t)\rangle_{\rm
    x,y,t}} .
\end{equation}

{\noindent}For a fixed point on the stellar disk, i.e. without averaging, the
temporal intensity variation is the same as the spatial variation
$\delta I_{\rm rms}$ from the principle of ergodicity. However, the temporal
intensity variation decreases as more convective cells are included in the
spatial average, hence it is a function of the horizontal extent of a
simulation. \citet{l06} derived the following relation between the temporal
and geometrical intensity variation

\begin{equation}
\sigma_{\rm I} \sim f \frac{l_{\rm gran}}{l_{\rm sim}}
\delta I_{\rm rms} 
\end{equation}

{\noindent}where $l_{\rm gran}$ is the characteristic size of granules and
$l_{\rm sim} = \sqrt{A}$ the linear size of the box. The factor $f$ of order
unity arises in part from the fact that the intensities are spatially
correlated in granules and \citet{l06} approximate the value to 0.4 for solar
granulation. We have computed $\sigma_{\rm I}$ and characteristic granulation
sizes (see Sect. 3.2) for the CIFIST grid and find that $f$ covers the full
range $0.2-0.6$. While it is out of the scope of this work, these
results could be expanded to the full observable disk of stars to predict the
photometric variability (or granulation power), especially in the case of giants with fewer and
larger granules \citep{s75}. 

The temporal $T_{\rm eff}$ variation $\sigma_{\rm Teff}$ is expected to scale
approximately as one-fourth of the intensity variation. We note in Table~1
that for the CIFIST grid, $\sigma_{\rm Teff} < 0.6\%$ which is indeed
verified by Eq.~8 given a maximum observed value of $\delta I_{\rm rms} \sim
25 \%$, and the requirement that at least 3$\times$3 granules are resolved in
the simulations. This result is of practical interest for spectral synthesis
applications. Typically, 3D spectra calculations rely on the order of 20
snapshots, although in many cases the snapshots are specifically chosen so
that $T_{\rm eff}$ does not deviate substantially from the average
\citep{allende12}. Our results suggest that that even a random selection of about
20 snapshots would provide a very good representation (i.e. with an accuracy
much better than 1$\%$) of the mean atmospheric structure of any stellar simulation.

\begin{figure}[!]
\captionsetup[subfigure]{labelformat=empty}
\begin{center}
\includegraphics[bb=40 40 552 735,width=2.6in,angle=90]{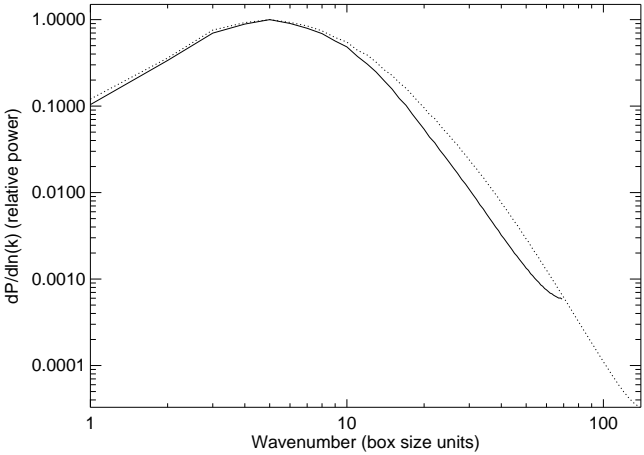}
\caption{Mean power spectra as a function of the horizontal wavenumber for
simulations at $T_{\rm eff}$ = 5750~K, $\log g = 3.75$, and [Fe/H]~=~$-3$ with a resolution
of 140$\times$140$\times$150 (solid line) and 220$\times$220$\times$150 (dotted line).
\label{fg:f12}}
\end{center}
\end{figure}

\subsection{Characteristic size}

We have performed a study of the characteristic power carrying lengths by
computing a power spectrum as a function of wavenumber for all available
emergent intensity maps. The mean power spectrum over all snapshots of one
simulation is shown in Fig.~\ref{fg:f11} for two typical cases. All power
spectra are available in online Appendix~A. We display power per logarithmic
wavenumber interval for a more direct identification of the power carrying
scales \citep{ludwig02}. The CIFIST grid was computed with the qualitative
requirement that at least of the order of 3$\times$3 cells are simulated. In
other words, the power spectrum peak should be relatively well resolved for
all simulations. We have derived the characteristic horizontal granulation
size, i.e. the wavelength of the peak of the power spectra, with third-order
polynomial fits (25 points around the maximum).  The dwarf and white
dwarf in Fig.~\ref{fg:f11} have rather similar power spectra. Over all
simulations, the mean full-width at half-maximum (FWHM) is 2.1 in
characteristic size unit, with a very small dependence on the atmospheric
parameters.

\begin{figure*}[!]
\captionsetup[subfigure]{labelformat=empty}
\begin{center}
\subfloat[]{
\includegraphics[bb=70 10 532 775,width=2.1in,angle=90]{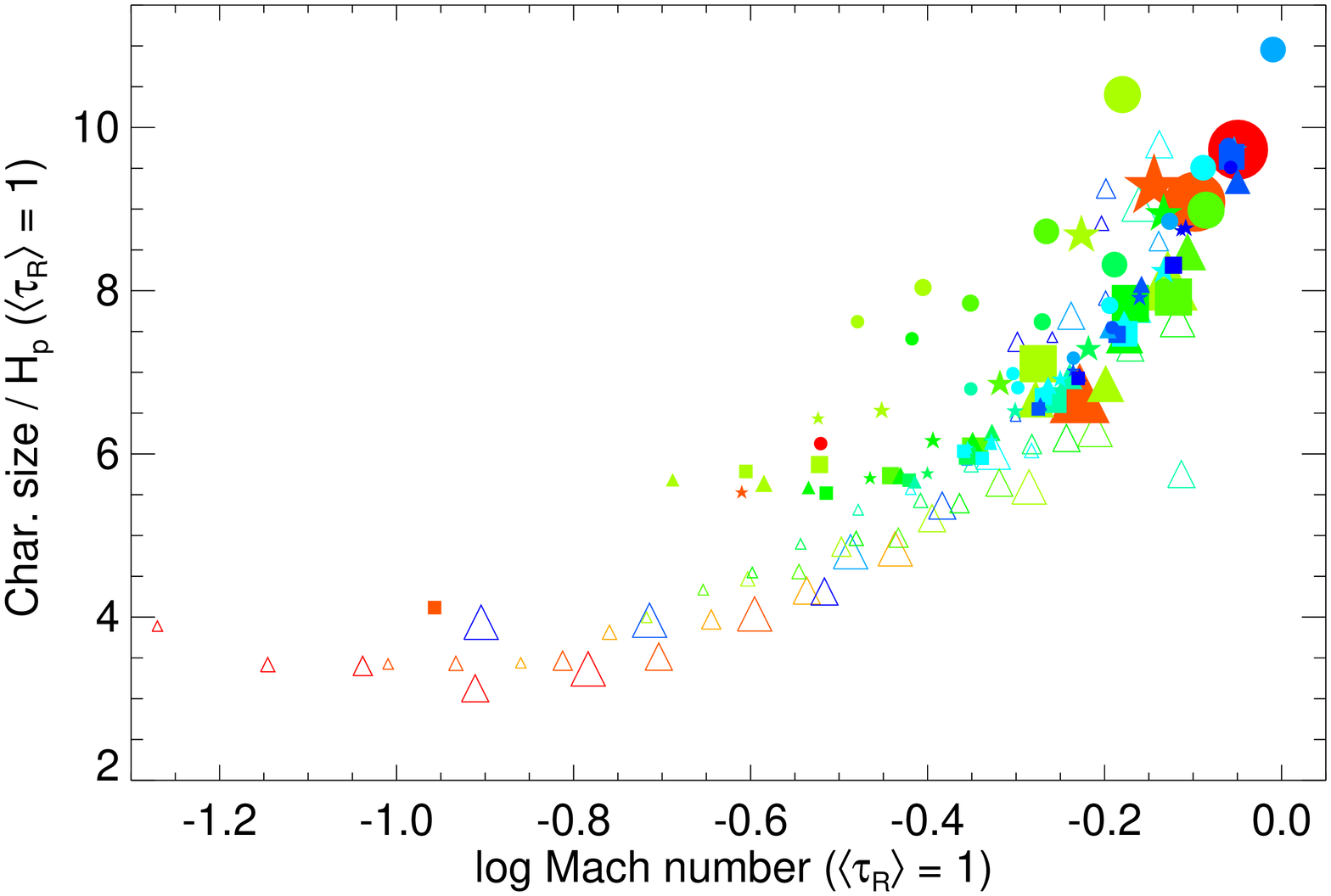}}
\subfloat[]{
\includegraphics[bb=70 10 532 775,width=2.1in,angle=90]{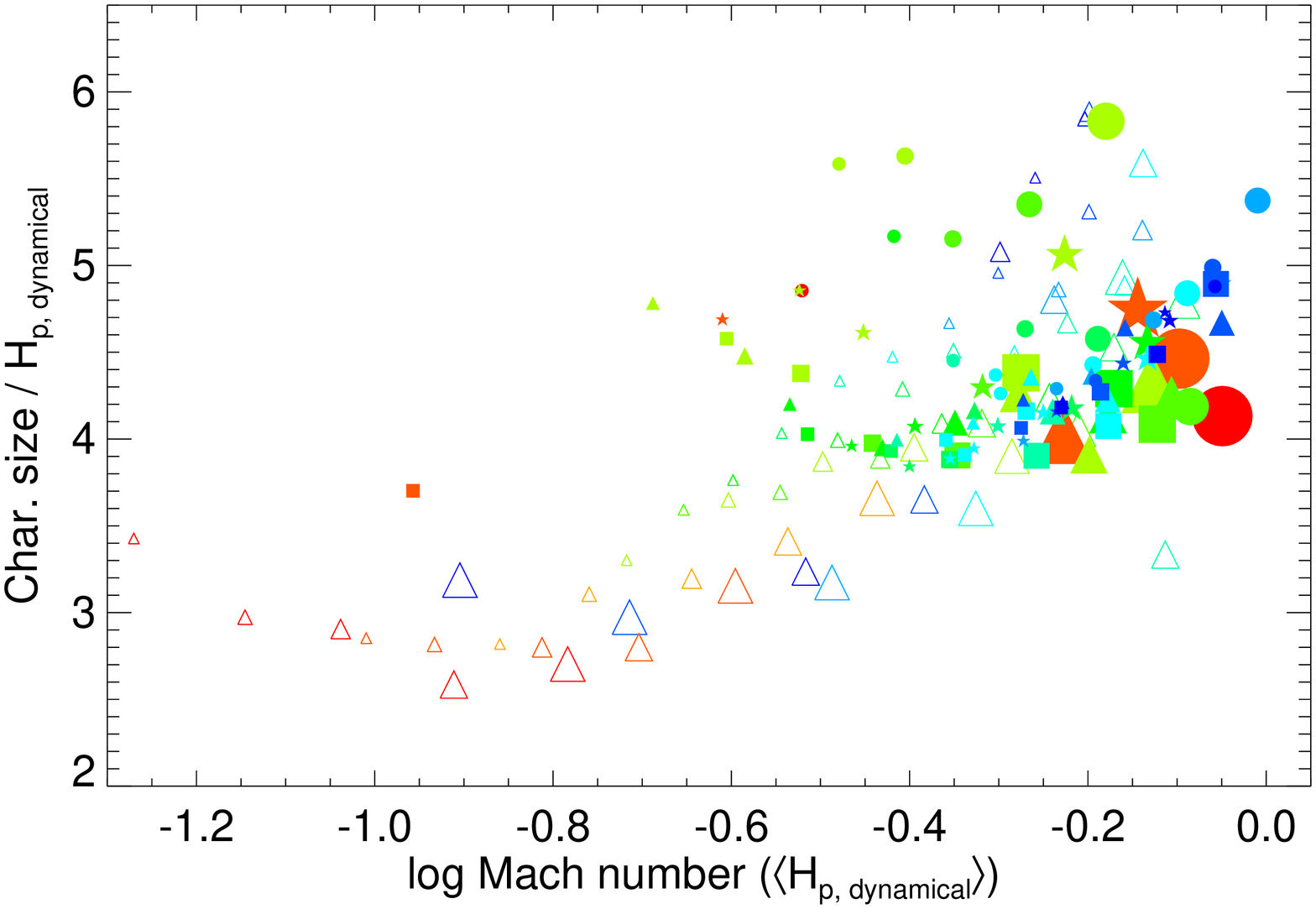}}
\caption{{\it Left:} Ratio of the characteristic granular size to the pressure
  scale height as a function of the logarithm of the Mach
  number (both quantities evaluated at $\langle\tau_\mathrm{R}\rangle = 1$). {\it Right:} Same as left panel but with
  the actual pressure scale height below the photosphere that is found in the
  3D simulations (see Eq.~11).
\label{fg:f13}}
\end{center}
\end{figure*}

It is well known that the shape of the granules can be influenced by either
explicit or implicit (finite resolution) viscosity. Fig.~\ref{fg:f12} illustrates 
that the power peak remains at the same wavenumber when horizontal resolution is increased by a factor of
two. On the other hand, the shape of the high wavenumber tail is significantly
different, with more fine structures in the high resolution case. For the
purpose of this work, it demonstrates that the characteristic granulation size
is rather insensitive to numerical parameters.

Early calculations of 2D model atmospheres have shown that granules
typically have an horizontal size about 10 times the local pressure scale
height, which is also a measure of the vertical extent of the granules
\citep{freytag97}.  In Fig.~\ref{fg:f13} (left panel), we present the ratio of
the characteristic granulation size to the hydrostatic pressure scale height
at $\langle\tau_\mathrm{R}\rangle = 1$. We conclude that the characteristic size is far from
a constant fraction of $H_{\rm p}$. In the objects with the most vigorous
convection, the ratio is typically around 10, although for objects with very
low Mach numbers, the ratio reaches a plateau around a value of $\sim$3-4. It
is shown that dwarfs with solar metallicity have larger sizes and deviate
slightly from the relation traced by sub-solar metallicity dwarfs and white
dwarfs.

To understand the characteristic size variation, we consider a simple analytical model of convection \citep{steffen89} with an
horizontal arrangement of square cells of size $L$ such that the mass flux
is approximated by

\begin{equation}
\rho v = \rho_{\rm o} v_{\rm o} \cos(2\pi x/L)\cos(2\pi y/L){\rm e}^{-(z-z_{o})/H} ,
\end{equation}

{\noindent}where $H$ is the vertical scale height of the momentum density and $z_{\rm o}$ a reference layer. 
For such a pattern, the continuity equation implies that 

\begin{equation}
\frac{v_{\rm h, rms}}{v_{\rm z, rms}} \propto \frac{L}{H} .
\end{equation}

\begin{figure}[!]
\captionsetup[subfigure]{labelformat=empty}
\begin{center}
\includegraphics[bb=40 40 552 765,width=2.3in,angle=90]{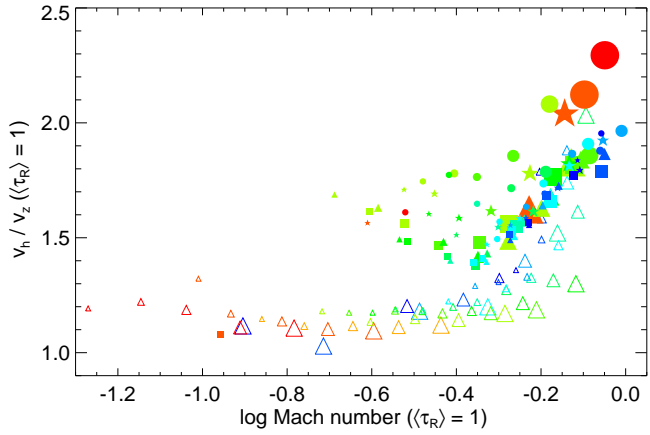}
\caption{Ratio of the horizontal to vertical rms velocity as a
  function of the Mach number (both quantities evaluated at $\langle\tau_\mathrm{R}\rangle = 1$).
\label{fg:f28}}
\end{center}
\end{figure}

{\noindent}where $v_{\rm h}^2 = v_{\rm x}^2 + v_{\rm y}^2$. We have verified
that the momentum density and pressure scale heights are the same within a 
few percent, except for hot models with small convective
fluxes. Hence, to first order, the variation of ${\rm Char.~size}/{H_{\rm p}}$ on Fig. \ref{fg:f13} could simply be related to an
increasing horizontal versus vertical velocity ratio \citep{steffen89,pmodes},
which is presented in Fig. \ref{fg:f28}. We indeed observe that the horizontal
velocities become proportionally larger for Mach numbers of order unity,
although the slope of the ${\rm Char.~size}/{H_{\rm p}}$ relation as a function of Mach number is steeper than the $v_{\rm h}/v_{\rm z}$ relation.
 We have found that the culprit for this discrepancy is that the $H_{\rm
  p}$ value at $\tau_\mathrm{R} \sim 1$, assuming hydrostatic equilibrium (Eq.~1), does
not represent particularly well the actual geometrical dimension
of one (dynamical) pressure scale height in the 3D simulations. In the right
panel of Fig. \ref{fg:f13}, we relied on the actual value of the pressure
scale height below the photosphere that is found in the simulations

\begin{equation}
H_{\rm p, dynamical} = z(\langle\tau_\mathrm{R}\rangle = 1)-z(\ln[P/P_{\langle\tau_\mathrm{R}\rangle = 1}] = 1)~,
\end{equation}

{\noindent}as well as the mean Mach number over the same range. The dynamical pressure scale height can be significantly larger
than the local hydrostatic estimate, to an extent that ${\rm Char.~size}/H_{\rm p, dynamical}$ is now varying by a factor of $\sim$2 across
the CIFIST grid, in very good agreement with our simple analytical model of mass conservation (Eq.~10) and Fig. \ref{fg:f28}. 
A further study of the depths of the granules, in addition to their widths derived here, could reveal more information
about which layers are relevant for the granulation formation.

\subsection{Characteristic lifetime}

We turn our attention in this section to the time evolution of
granulation. The power spectra of the local temporal variation of
intensity $I_{x,y}(t)$ reach a nearly constant value at low frequencies, hence
it is difficult to define a characteristic lifetime from the peak of the mean power
spectra, averaged over all pixels, as we did for characteristic lengths. In the field of
asteroseismology, the spectral power density $P$ is often characterised by a
generalised Harvey model \citep{harvey85,astero2}

\begin{equation}
P(\nu) = \frac{b}{1+(\nu/\nu{_{\rm o}})^{\alpha}}
\end{equation}

{\noindent}where $\nu$ is the cyclic temporal frequency. All other variables
are fitting parameters and in particular $\nu{_{\rm o}}$ could be interpreted
as the characteristic frequency of the granulation cycle. The Wiener-Khinchin
theorem implies that we can derive the related autocorrelation function

\begin{equation}
R(\tau) = \int_{-\infty}^{\infty}P(\nu)\cos(2\pi\nu\tau){\rm d}\nu ,
\end{equation}

{\noindent}where $\tau$ is the time lag. As an example, for an exponent $\alpha=2$ in Eq.~12, it is found
that

\begin{equation}
R(\tau) = b \nu_{\rm 0} \sqrt{\frac{\pi}{2}} {\rm e}^{-2\pi\nu_{\rm o}\tau}.
\end{equation}

{\noindent}It implies that the characteristic cycle timescale $\nu_{\rm
  o}^{-1}$ is $2\pi$ times larger than the e-folding decay time (${t_{\rm
    e\textrm{-}fold}}$) of the autocorrelation function. In Table~2, we
provide the characteristic ${t_{\rm e\textrm{-}fold}}$ for three different
power fitting functions. It shows that by varying the power spectrum fitting
function, the related decay time can be significantly different. Since it is
difficult to choose a single $\alpha$ parameter in Eq.~12 for all 3D
simulations, we have decided to compute only the e-folding decay time of
autocorrelation, averaged over all pixels. An estimate of the corresponding 
cycle timescale $\nu_{\rm o}^{-1}$ can be computed from Table 2.

 \begin{table*}[h!]
 \caption{Grid of CIFIST 3D model atmospheres}
 \label{tab1}
 \begin{center}
 \begin{tabular}{lllllllllll}
\hline
\hline
$T_{\rm eff}$ & $\log g$ & [Fe/H] & log(Char. size) & $A(I^{+})/A$ & $\delta I_{\rm rms}/\langle I \rangle$ & $\log$ Pe\tablefootmark{a} & Mach\tablefootmark{a} & $\log \rho$\tablefootmark{a} & $\log H_{\rm p}$\tablefootmark{a} & $\sigma_{\rm Teff}/T_{\rm eff}$ \\
(K) & & & [cm] & (\%) & (\%) & & & [g cm$^{-3}$] & [cm] & (\%) \\
\hline
     3714 &   1.00 &     0 &  11.37 &   56.0 &   21.9 &  -0.97 &  0.893 &  -8.01 &  10.39 &   0.42 \\
     4016 &   1.50 &     0 &  10.92 &   53.7 &   18.0 &  -1.20 &  0.799 &  -7.95 &   9.96 &   0.56 \\
     4039 &   1.50 &    -1 &  10.93 &   52.1 &   16.5 &  -0.78 &  0.718 &  -7.62 &   9.96 &   0.40 \\
     3988 &   1.50 &    -3 &  10.75 &   47.6 &   12.8 &   0.40 &  0.591 &  -6.91 &   9.92 &   0.14 \\
     4475 &   1.50 &    -3 &  10.85 &   49.2 &   18.8 &   0.05 &  0.744 &  -7.39 &   9.94 &   0.14 \\
     4487 &   2.00 &    -3 &  10.28 &   47.1 &   17.8 &   0.30 &  0.633 &  -7.00 &   9.45 &   0.18 \\
     4779 &   2.00 &    -2 &  10.36 &   48.2 &   19.4 &  -0.20 &  0.755 &  -7.42 &   9.46 &   0.23 \\
     4880 &   2.00 &    -3 &  10.38 &   49.9 &   21.5 &  -0.05 &  0.783 &  -7.37 &   9.45 &   0.18 \\
     4475 &   2.50 &     0 &  10.03 &   50.3 &   15.9 &  -1.04 &  0.661 &  -7.48 &   9.01 &   0.25 \\
     4492 &   2.50 &    -1 &   9.95 &   46.1 &   14.8 &  -0.59 &  0.594 &  -7.18 &   9.01 &   0.21 \\
     4477 &   2.50 &    -2 &   9.84 &   45.2 &   14.0 &  -0.00 &  0.531 &  -6.81 &   8.99 &   0.11 \\
     4522 &   2.50 &    -3 &   9.80 &   48.4 &   15.5 &   0.38 &  0.527 &  -6.64 &   8.97 &   0.07 \\
     4964 &   2.50 &     0 &   9.98 &   52.5 &   18.2 &  -0.93 &  0.822 &  -7.69 &   9.03 &   0.35 \\
     4990 &   2.50 &    -1 &   9.97 &   48.7 &   17.9 &  -0.50 &  0.736 &  -7.43 &   9.02 &   0.31 \\
     5018 &   2.50 &    -2 &   9.89 &   48.7 &   20.1 &  -0.17 &  0.675 &  -7.21 &   8.99 &   0.25 \\
     5018 &   2.50 &    -3 &   9.86 &   48.7 &   21.9 &   0.01 &  0.665 &  -7.11 &   8.99 &   0.17 \\
     4924 &   3.50 &     0 &   8.99 &   46.8 &   14.8 &  -0.75 &  0.542 &  -6.98 &   8.05 &   0.28 \\
     4928 &   3.50 &    -1 &   8.88 &   44.5 &   13.3 &  -0.34 &  0.481 &  -6.69 &   8.04 &   0.21 \\
     4976 &   3.50 &    -2 &   8.81 &   46.8 &   15.3 &   0.18 &  0.450 &  -6.42 &   8.03 &   0.12 \\
     4978 &   3.50 &    -3 &   8.81 &   48.8 &   15.9 &   0.42 &  0.448 &  -6.28 &   8.02 &   0.09 \\
     5427 &   3.50 &     0 &   8.99 &   48.1 &   17.7 &  -0.63 &  0.647 &  -7.16 &   8.07 &   0.41 \\
     5479 &   3.50 &    -1 &   8.92 &   46.5 &   18.5 &  -0.27 &  0.605 &  -6.95 &   8.06 &   0.29 \\
     5502 &   3.50 &    -2 &   8.87 &   48.0 &   21.8 &   0.01 &  0.553 &  -6.76 &   8.04 &   0.29 \\
     5533 &   3.50 &    -3 &   8.88 &   47.9 &   23.2 &   0.09 &  0.576 &  -6.73 &   8.03 &   0.22 \\
     5885 &   3.50 &     0 &   9.06 &   49.7 &   19.3 &  -0.74 &  0.815 &  -7.40 &   8.08 &   0.26 \\
     5889 &   3.50 &    -1 &   8.99 &   49.8 &   21.9 &  -0.47 &  0.736 &  -7.19 &   8.07 &   0.20 \\
     5864 &   3.50 &    -2 &   8.93 &   49.5 &   24.2 &  -0.23 &  0.665 &  -7.02 &   8.06 &   0.27 \\
     5873 &   3.50 &    -3 &   8.93 &   49.6 &   24.9 &  -0.18 &  0.664 &  -6.99 &   8.05 &   0.26 \\
     6143 &   3.50 &     0 &   9.13 &   51.8 &   19.1 &  -0.88 &  0.978 &  -7.59 &   8.09 &   0.43 \\
     6210 &   3.50 &    -1 &   9.07 &   51.8 &   22.2 &  -0.70 &  0.884 &  -7.43 &   8.08 &   0.25 \\
     6308 &   3.50 &    -2 &   9.07 &   53.1 &   25.2 &  -0.65 &  0.878 &  -7.40 &   8.08 &   0.24 \\
     6307 &   3.50 &    -3 &   9.05 &   52.7 &   24.2 &  -0.60 &  0.892 &  -7.38 &   8.07 &   0.23 \\
     4479 &   4.00 &     0 &   8.42 &   45.4 &    9.4 &  -0.34 &  0.393 &  -6.55 &   7.51 &   0.15 \\
     4527 &   4.00 &    -1 &   8.33 &   45.1 &    8.2 &  -0.05 &  0.353 &  -6.24 &   7.52 &   0.11 \\
     4505 &   4.00 &    -2 &   8.28 &   49.3 &    6.7 &   0.50 &  0.301 &  -5.83 &   7.51 &   0.07 \\
     4493 &   4.00 &    -3 &   8.24 &   55.0 &    4.4 &   0.95 &  0.260 &  -5.56 &   7.49 &   0.03 \\
     4953 &   4.00 &     0 &   8.45 &   45.1 &   12.0 &  -0.57 &  0.445 &  -6.68 &   7.55 &   0.26 \\
     4984 &   4.00 &    -1 &   8.35 &   43.8 &   10.8 &  -0.21 &  0.404 &  -6.39 &   7.55 &   0.15 \\
     4956 &   4.00 &    -2 &   8.30 &   47.5 &   11.5 &   0.32 &  0.362 &  -6.07 &   7.54 &   0.09 \\
     4990 &   4.00 &    -3 &   8.30 &   47.7 &   12.0 &   0.32 &  0.371 &  -6.09 &   7.54 &   0.10 \\
     5475 &   4.00 &     0 &   8.47 &   46.2 &   15.7 &  -0.56 &  0.536 &  -6.83 &   7.58 &   0.22 \\
     5532 &   4.00 &    -1 &   8.39 &   46.2 &   16.6 &  -0.18 &  0.500 &  -6.62 &   7.58 &   0.19 \\
     5472 &   4.00 &    -2 &   8.34 &   47.7 &   18.9 &   0.16 &  0.442 &  -6.37 &   7.56 &   0.11 \\
     5474 &   4.00 &    -3 &   8.35 &   48.2 &   19.5 &   0.32 &  0.471 &  -6.30 &   7.55 &   0.11 \\
     5930 &   4.00 &     0 &   8.49 &   47.7 &   18.5 &  -0.53 &  0.639 &  -7.02 &   7.60 &   0.19 \\
     5852 &   4.00 &    -1 &   8.43 &   48.2 &   20.2 &  -0.23 &  0.562 &  -6.77 &   7.59 &   0.19 \\
     5855 &   4.00 &    -2 &   8.40 &   48.9 &   23.1 &   0.01 &  0.538 &  -6.61 &   7.57 &   0.12 \\
     5846 &   4.00 &    -3 &   8.40 &   48.7 &   23.7 &   0.09 &  0.545 &  -6.57 &   7.56 &   0.13 \\
     6237 &   4.00 &     0 &   8.55 &   49.8 &   19.4 &  -0.62 &  0.747 &  -7.18 &   7.61 &   0.28 \\
     6260 &   4.00 &    -1 &   8.50 &   49.5 &   22.3 &  -0.42 &  0.691 &  -7.02 &   7.60 &   0.17 \\
     6278 &   4.00 &    -2 &   8.46 &   50.3 &   24.6 &  -0.28 &  0.652 &  -6.91 &   7.59 &   0.17 \\
     6241 &   4.00 &    -3 &   8.47 &   49.7 &   24.7 &  -0.24 &  0.636 &  -6.88 &   7.59 &   0.14 \\
     6487 &   4.00 &     0 &   8.60 &   51.0 &   19.4 &  -0.79 &  0.871 &  -7.35 &   7.61 &   0.26 \\
     6502 &   4.00 &    -1 &   8.55 &   51.5 &   22.5 &  -0.60 &  0.779 &  -7.19 &   7.61 &   0.22 \\
     6532 &   4.00 &    -2 &   8.52 &   51.5 &   24.2 &  -0.48 &  0.755 &  -7.11 &   7.60 &   0.16 \\
     6411 &   4.00 &    -3 &   8.50 &   51.2 &   25.6 &  -0.36 &  0.695 &  -6.99 &   7.59 &   0.18 \\
     6726 &   4.25 &     0 &   8.36 &   51.7 &   19.6 &  -0.75 &  0.876 &  -7.27 &   7.38 &   0.25 \\
     6730 &   4.25 &    -1 &   8.32 &   51.9 &   22.9 &  -0.57 &  0.770 &  -7.11 &   7.37 &   0.20 \\
     5785 &   4.44 &     0 &   8.00 &   46.0 &   14.3 &  -0.45 &  0.497 &  -6.65 &   7.15 &   0.21 \\
     5770 &   4.44 &    -1 &   7.94 &   46.8 &   16.6 &  -0.11 &  0.442 &  -6.42 &   7.16 &   0.18 \\
     5757 &   4.44 &    -2 &   7.92 &   47.6 &   19.2 &   0.21 &  0.437 &  -6.23 &   7.14 &   0.18 \\
\hline
\end{tabular} 
\end{center} 
\tablefoottext{a}{Evaluated at $\langle\tau_\mathrm{R}\rangle = 1$.}
\end{table*}

\setcounter{table}{0}
 \renewcommand*\thetable{\arabic{table}}
 \begin{table*}
 \caption{continued}
 \label{tab1}
 \begin{center}
 \begin{tabular}{lllllllllll}
\hline
\hline
$T_{\rm eff}$ & $\log g$ & [Fe/H] & log(Char. size) & $A(I^+)/A$ & $\delta I_{\rm rms}/\langle I \rangle$ & $\log$ Pe\tablefootmark{a} & Mach\tablefootmark{a} & $\log \rho$\tablefootmark{a} & $\log H_{\rm p}$\tablefootmark{a} & $\sigma_{\rm Teff}/T_{\rm eff}$ \\
(K) & & & [cm] & (\%) & (\%) & & & [g cm$^{-3}$] & [cm] & (\%) \\
\hline
     3964 &   4.50 &     0 &   7.73 &   54.7 &    5.2 &   0.53 &  0.301 &  -6.03 &   6.94 &   0.06 \\
     4000 &   4.50 &    -1 &   7.68 &   56.2 &    3.5 &   0.93 &  0.245 &  -5.71 &   6.93 &   0.03 \\
     4000 &   4.50 &    -2 &   7.50 &   55.5 &    1.5 &   1.28 &  0.110 &  -5.21 &   6.89 &   0.02 \\
     4510 &   4.50 &     0 &   7.89 &   45.8 &    7.7 &  -0.09 &  0.332 &  -6.25 &   7.01 &   0.15 \\
     4500 &   4.50 &    -1 &   7.82 &   46.8 &    6.3 &   0.26 &  0.299 &  -5.93 &   7.01 &   0.09 \\
     4539 &   4.50 &    -2 &   7.77 &   52.4 &    4.5 &   0.77 &  0.248 &  -5.53 &   7.00 &   0.06 \\
     4521 &   4.50 &    -3 &   7.74 &   57.1 &    2.6 &   1.23 &  0.205 &  -5.25 &   6.98 &   0.02 \\
     4982 &   4.50 &     0 &   7.92 &   43.9 &    9.7 &  -0.35 &  0.382 &  -6.39 &   7.05 &   0.23 \\
     5060 &   4.50 &    -1 &   7.82 &   44.3 &    9.0 &  -0.03 &  0.343 &  -6.11 &   7.06 &   0.16 \\
     5013 &   4.50 &    -2 &   7.79 &   49.3 &    8.8 &   0.49 &  0.306 &  -5.77 &   7.05 &   0.08 \\
     4992 &   4.50 &    -3 &   7.79 &   52.8 &    7.7 &   0.75 &  0.292 &  -5.62 &   7.04 &   0.06 \\
     5488 &   4.50 &     0 &   7.92 &   44.6 &   12.7 &  -0.45 &  0.446 &  -6.52 &   7.09 &   0.26 \\
     5473 &   4.50 &    -1 &   7.85 &   45.6 &   12.9 &  -0.07 &  0.398 &  -6.26 &   7.09 &   0.17 \\
     5479 &   4.50 &    -2 &   7.83 &   48.6 &   15.0 &   0.33 &  0.379 &  -6.03 &   7.08 &   0.10 \\
     5486 &   4.50 &    -3 &   7.82 &   49.1 &   15.0 &   0.47 &  0.385 &  -5.95 &   7.07 &   0.08 \\
     5866 &   4.50 &     0 &   7.95 &   45.8 &   15.6 &  -0.41 &  0.503 &  -6.64 &   7.11 &   0.24 \\
     5898 &   4.50 &    -1 &   7.89 &   46.7 &   17.8 &  -0.08 &  0.470 &  -6.44 &   7.10 &   0.61 \\
     5923 &   4.50 &    -2 &   7.86 &   47.7 &   20.6 &   0.17 &  0.459 &  -6.29 &   7.09 &   0.15 \\
     5924 &   4.50 &    -3 &   7.88 &   50.1 &   20.8 &   0.26 &  0.469 &  -6.25 &   7.09 &   0.14 \\
     6232 &   4.50 &     0 &   7.98 &   47.3 &   18.0 &  -0.41 &  0.582 &  -6.79 &   7.12 &   0.24 \\
     6239 &   4.50 &    -1 &   7.93 &   48.2 &   20.6 &  -0.17 &  0.534 &  -6.61 &   7.12 &   0.18 \\
     6321 &   4.50 &    -2 &   7.92 &   48.7 &   23.4 &  -0.02 &  0.531 &  -6.53 &   7.11 &   0.18 \\
     6270 &   4.50 &    -3 &   7.92 &   48.1 &   23.6 &   0.06 &  0.534 &  -6.48 &   7.10 &   0.18 \\
     6458 &   4.50 &     0 &   8.01 &   48.5 &   19.1 &  -0.47 &  0.643 &  -6.90 &   7.13 &   0.19 \\
     6458 &   4.50 &    -1 &   7.97 &   49.3 &   22.0 &  -0.26 &  0.581 &  -6.73 &   7.12 &   0.16 \\
     6535 &   4.50 &    -2 &   7.96 &   49.1 &   24.5 &  -0.15 &  0.589 &  -6.68 &   7.12 &   0.15 \\
     6561 &   4.50 &    -3 &   7.96 &   49.3 &   24.8 &  -0.14 &  0.591 &  -6.67 &   7.11 &   0.18 \\
     6112 &   7.00 &   ... &   5.25 &   50.0 &    3.5 &   1.03 &  0.165 &  -4.90 &   4.72 &   0.03 \\
     7046 &   7.00 &   ... &   5.39 &   46.5 &    9.3 &   0.57 &  0.254 &  -5.35 &   4.79 &   0.08 \\
     8027 &   7.00 &   ... &   5.52 &   47.0 &   15.6 &   0.25 &  0.366 &  -5.78 &   4.84 &   0.16 \\
     9025 &   7.00 &   ... &   5.64 &   48.8 &   19.1 &  -0.03 &  0.518 &  -6.24 &   4.89 &   0.20 \\
     9521 &   7.00 &   ... &   5.72 &   50.9 &   19.4 &  -0.15 &  0.615 &  -6.48 &   4.92 &   0.17 \\
    10018 &   7.00 &   ... &   5.85 &   53.3 &   19.3 &  -0.24 &  0.763 &  -6.73 &   4.97 &   0.22 \\
    10540 &   7.00 &   ... &   6.00 &   62.9 &   18.9 &  -0.35 &  0.806 &  -6.95 &   5.02 &   0.09 \\
    11000 &   7.00 &   ... &   6.03 &   63.0 &   19.4 &  -0.44 &  0.690 &  -7.06 &   5.07 &   0.06 \\
    11501 &   7.00 &   ... &   5.91 &   65.2 &   13.0 &  -0.62 &  0.472 &  -7.18 &   5.13 &   0.03 \\
    12001 &   7.00 &   ... &   5.88 &   61.4 &    6.9 &  -0.86 &  0.326 &  -7.32 &   5.20 &   0.02 \\
    12501 &   7.00 &   ... &   5.87 &   54.2 &    2.9 &  -1.31 &  0.193 &  -7.46 &   5.27 &   0.02 \\
    13003 &   7.00 &   ... &   5.91 &   51.0 &    1.4 &  -1.75 &  0.125 &  -7.56 &   5.31 &   0.03 \\
     6065 &   7.50 &   ... &   4.70 &   51.6 &    1.8 &   1.32 &  0.123 &  -4.56 &   4.21 &   0.01 \\
     7033 &   7.50 &   ... &   4.83 &   47.2 &    5.9 &   0.78 &  0.198 &  -5.01 &   4.29 &   0.05 \\
     8017 &   7.50 &   ... &   4.98 &   46.7 &   11.2 &   0.43 &  0.291 &  -5.42 &   4.34 &   0.12 \\
     9015 &   7.50 &   ... &   5.11 &   48.5 &   15.8 &   0.17 &  0.403 &  -5.84 &   4.39 &   0.16 \\
     9549 &   7.50 &   ... &   5.18 &   50.5 &   17.0 &   0.06 &  0.480 &  -6.07 &   4.42 &   0.16 \\
    10007 &   7.50 &   ... &   5.25 &   51.9 &   17.3 &  -0.01 &  0.571 &  -6.28 &   4.46 &   0.22 \\
    10500 &   7.50 &   ... &   5.36 &   53.9 &   17.9 &  -0.09 &  0.675 &  -6.51 &   4.50 &   0.23 \\
    10938 &   7.50 &   ... &   5.30 &   56.6 &   19.2 &  -0.16 &  0.770 &  -6.68 &   4.54 &   0.12 \\
    11498 &   7.50 &   ... &   5.59 &   61.7 &   19.9 &  -0.27 &  0.727 &  -6.84 &   4.60 &   0.13 \\
    11999 &   7.50 &   ... &   5.54 &   64.9 &   17.7 &  -0.42 &  0.578 &  -6.96 &   4.66 &   0.05 \\
    12500 &   7.50 &   ... &   5.44 &   65.1 &   10.7 &  -0.63 &  0.414 &  -7.07 &   4.71 &   0.02 \\
    13002 &   7.50 &   ... &   5.41 &   59.4 &    6.1 &  -0.91 &  0.304 &  -7.18 &   4.77 &   0.01 \\
     5997 &   8.00 &   ... &   4.22 &   53.6 &    0.9 &   1.67 &  0.092 &  -4.20 &   3.69 &   0.01 \\
     7011 &   8.00 &   ... &   4.32 &   48.4 &    3.5 &   1.01 &  0.154 &  -4.68 &   3.78 &   0.03 \\
     8034 &   8.00 &   ... &   4.45 &   46.9 &    7.6 &   0.62 &  0.227 &  -5.08 &   3.85 &   0.06 \\
     9036 &   8.00 &   ... &   4.59 &   48.5 &   12.0 &   0.35 &  0.318 &  -5.47 &   3.90 &   0.12 \\
     9518 &   8.00 &   ... &   4.62 &   49.9 &   13.6 &   0.25 &  0.369 &  -5.66 &   3.92 &   0.13 \\
    10025 &   8.00 &   ... &   4.69 &   51.6 &   14.4 &   0.17 &  0.433 &  -5.87 &   3.95 &   0.14 \\
    10532 &   8.00 &   ... &   4.78 &   52.8 &   15.0 &   0.10 &  0.522 &  -6.09 &   3.99 &   0.15 \\
    11005 &   8.00 &   ... &   4.86 &   53.8 &   16.6 &   0.06 &  0.598 &  -6.27 &   4.03 &   0.17 \\
    11529 &   8.00 &   ... &   4.96 &   55.8 &   17.7 &  -0.01 &  0.693 &  -6.47 &   4.07 &   0.18 \\
    11980 &   8.00 &   ... &   5.05 &   60.4 &   18.7 &  -0.06 &  0.726 &  -6.58 &   4.12 &   0.13 \\
    12504 &   8.00 &   ... &   5.14 &   61.9 &   19.2 &  -0.22 &  0.633 &  -6.71 &   4.17 &   0.11 \\
    13000 &   8.00 &   ... &   5.09 &   65.1 &   16.0 &  -0.39 &  0.503 &  -6.80 &   4.22 &   0.05 \\ 
\hline
\end{tabular} 
\end{center} 
\end{table*}

\setcounter{table}{0}
 \renewcommand*\thetable{\arabic{table}}
 \begin{table*}
 \caption{continued}
 \label{tab1}
 \begin{center}
 \begin{tabular}{lllllllllll}
\hline
\hline
$T_{\rm eff}$ & $\log g$ & [Fe/H] & log(Char. size) & $A(I^+)/A$ & $\delta I_{\rm rms}/\langle I \rangle$ & $\log$ Pe\tablefootmark{a} & Mach\tablefootmark{a} & $\log \rho$\tablefootmark{a} & $\log H_{\rm p}$\tablefootmark{a} & $\sigma_{\rm Teff}/T_{\rm eff}$ \\
(K) & & & [cm] & (\%) & (\%) & & & [g cm$^{-3}$] & [cm] & (\%) \\
\hline
     6024 &   8.50 &   ... &   3.70 &   53.9 &    0.5 &   1.93 &  0.072 &  -3.89 &   3.17 &   0.01 \\
     6925 &   8.50 &   ... &   3.81 &   50.5 &    1.9 &   1.30 &  0.117 &  -4.33 &   3.27 &   0.01 \\
     8004 &   8.50 &   ... &   3.93 &   46.9 &    4.7 &   0.84 &  0.174 &  -4.74 &   3.35 &   0.04 \\
     9068 &   8.50 &   ... &   4.05 &   48.0 &    8.4 &   0.54 &  0.249 &  -5.12 &   3.40 &   0.09 \\
     9522 &   8.50 &   ... &   4.08 &   50.1 &   10.1 &   0.44 &  0.285 &  -5.28 &   3.43 &   0.08 \\
     9972 &   8.50 &   ... &   4.14 &   50.8 &   11.2 &   0.36 &  0.331 &  -5.45 &   3.45 &   0.14 \\
    10496 &   8.50 &   ... &   4.21 &   51.9 &   12.1 &   0.28 &  0.391 &  -5.65 &   3.48 &   0.15 \\
    10997 &   8.50 &   ... &   4.28 &   52.8 &   13.4 &   0.22 &  0.446 &  -5.84 &   3.51 &   0.13 \\
    11490 &   8.50 &   ... &   4.33 &   53.4 &   14.3 &   0.18 &  0.522 &  -6.02 &   3.55 &   0.12 \\
    11979 &   8.50 &   ... &   4.43 &   54.5 &   15.2 &   0.13 &  0.585 &  -6.19 &   3.59 &   0.14 \\
    12420 &   8.50 &   ... &   4.54 &   56.8 &   16.3 &   0.08 &  0.633 &  -6.32 &   3.64 &   0.24 \\
    12909 &   8.50 &   ... &   4.64 &   61.1 &   17.1 &  -0.01 &  0.626 &  -6.44 &   3.69 &   0.31 \\
     6028 &   9.00 &   ... &   3.23 &   54.8 &    0.3 &   2.16 &  0.054 &  -3.57 &   2.64 &   0.01 \\
     6960 &   9.00 &   ... &   3.30 &   52.5 &    1.1 &   1.57 &  0.098 &  -4.02 &   2.77 &   0.01 \\
     8041 &   9.00 &   ... &   3.38 &   47.6 &    3.0 &   1.04 &  0.138 &  -4.42 &   2.85 &   0.03 \\
     8999 &   9.00 &   ... &   3.50 &   47.8 &    5.4 &   0.76 &  0.192 &  -4.74 &   2.90 &   0.06 \\
     9507 &   9.00 &   ... &   3.56 &   48.8 &    6.8 &   0.64 &  0.222 &  -4.90 &   2.92 &   0.08 \\
     9962 &   9.00 &   ... &   3.60 &   49.9 &    8.0 &   0.54 &  0.252 &  -5.06 &   2.95 &   0.07 \\
    10403 &   9.00 &   ... &   3.66 &   51.5 &    8.9 &   0.46 &  0.286 &  -5.22 &   2.97 &   0.10 \\
    10948 &   9.00 &   ... &   3.73 &   51.9 &   10.1 &   0.38 &  0.332 &  -5.41 &   3.00 &   0.10 \\
    11415 &   9.00 &   ... &   3.77 &   53.0 &   10.8 &   0.34 &  0.381 &  -5.57 &   3.03 &   0.13 \\
    11915 &   9.00 &   ... &   3.84 &   53.7 &   11.7 &   0.31 &  0.441 &  -5.74 &   3.06 &   0.10 \\
    12436 &   9.00 &   ... &   3.92 &   53.8 &   12.5 &   0.26 &  0.500 &  -5.91 &   3.11 &   0.13 \\
    12969 &   9.00 &   ... &   4.03 &   55.1 &   13.7 &   0.21 &  0.551 &  -6.06 &   3.16 &   0.15 \\
\hline
\end{tabular} 
\end{center} 
\end{table*}

\begin{table}
 \caption{Relation between temporal frequency and decay time}
 \label{tab2}
 \begin{center}
 \begin{tabular}{ll}
\hline
\hline
P($\nu$) & $\nu_{\rm o}^{-1}$/${t_{\rm e\textrm{-}fold}}$ \\
\hline
$ b/[1+(\nu/\nu{_{\rm o}})^{2}]$ & 6.28 \\
$ b/[1+(\nu/\nu{_{\rm o}})^{4}]$ & 3.84 \\
$ b{\rm e}^{-\nu/\nu_{\rm o}}$ & 4.78 \\
\hline
\end{tabular} 
\end{center} 
\end{table}

\begin{figure}[!]
\captionsetup[subfigure]{labelformat=empty}
\begin{center}
\includegraphics[bb=40 40 552 765,width=2.3in,angle=90]{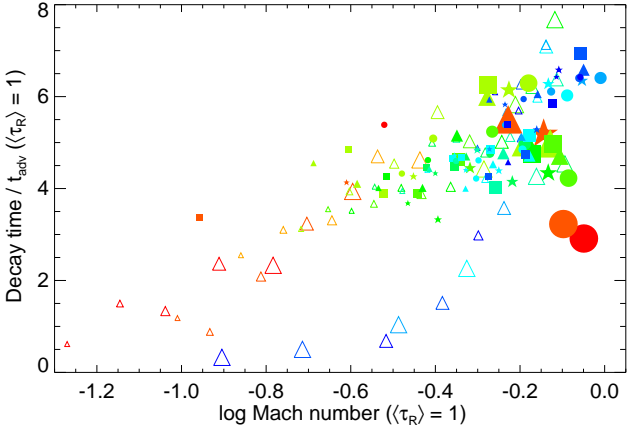}
\caption{Ratio of the characteristic decay time to
  the advective timescale as a function of the logarithm of the Mach number
  (both quantities evaluated at $\langle\tau_\mathrm{R}\rangle = 1$).
\label{fg:f17}}
\end{center}
\end{figure}

For the CIFIST grid, we present in Fig.~\ref{fg:f17} the ratio of our derived
$t_{\rm e\textrm{-}fold}$ decay time, and the semi-analytical advective timescale $H_{\rm
  p} / v_{\rm rms}$, where $v_{\rm rms}$ is also drawn from the 3D simulations. We
find that the decay time correlates well with Mach number, and unsurprisingly,
simulations with a large characteristic size to $H_{\rm p}$ ratio have longer
lifetimes. The observed decay times would be roughly in agreement with the
advective timescales by relying on characteristic sizes instead of $H_{\rm
  p}$. 
  
  We observe that hot white dwarfs feature very short lifetimes, even
given their relatively small granule sizes. According to the P\'eclet
  number values of Fig.~\ref{fg:f4}, the characteristic radiative timescale is
  much smaller than the advective timescale in those atmospheres which could 
  explain a faster evolution of granules. However, hot stars in the CIFIST grid
  with the same P\'eclet numbers still scale well with the advective
  timescale. The main difference between those objects is that the stars are
  fully convective below the photosphere while hot white dwarfs have a thin
  convective zone with the maximum convective to radiative flux ratio reaching a small value. The
  enhanced radiative energy transport below the photosphere in white dwarfs
  may have an effect on the evolution of granulation. It would be interesting to compare
  hot white dwarfs and A-stars with thin convective zones. The convection in main-sequence stars is also driven
  by helium ionization, hence these objects may behave differently than hot pure-hydrogen white dwarfs. 
  Finally, our decay times could be compared to the granulation timescales
  derived from power spectra of {\it Kepler} or {\it CoRoT} observations \citep[see, e.g.,][]{mathur11}.

\begin{figure*}[!]
\captionsetup[subfigure]{labelformat=empty}
\begin{center}
\subfloat[]{
\includegraphics[bb=85 65 400 355,width=2.15in]{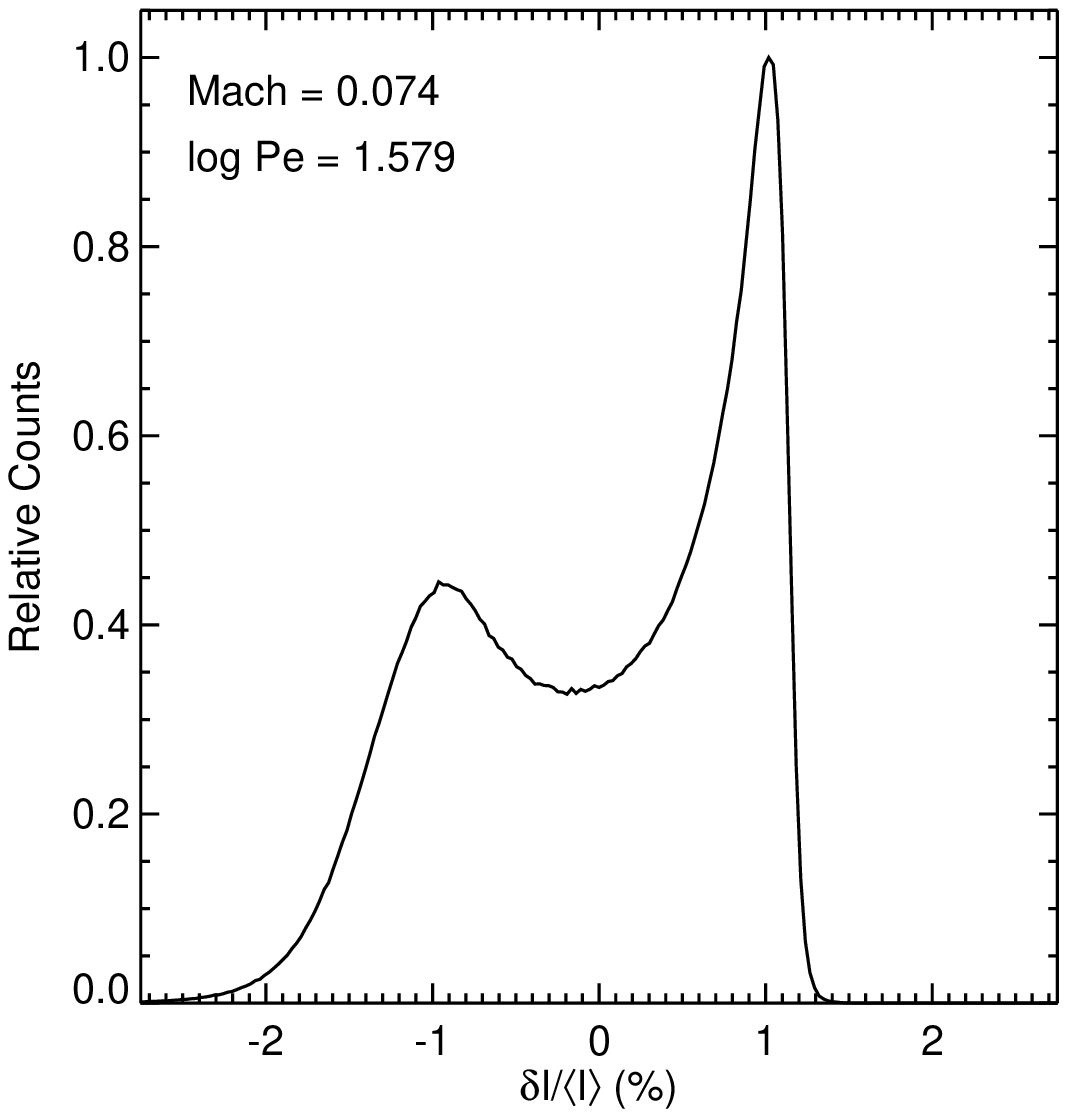}}
\subfloat[]{
\includegraphics[bb=85 65 400 355,width=2.15in]{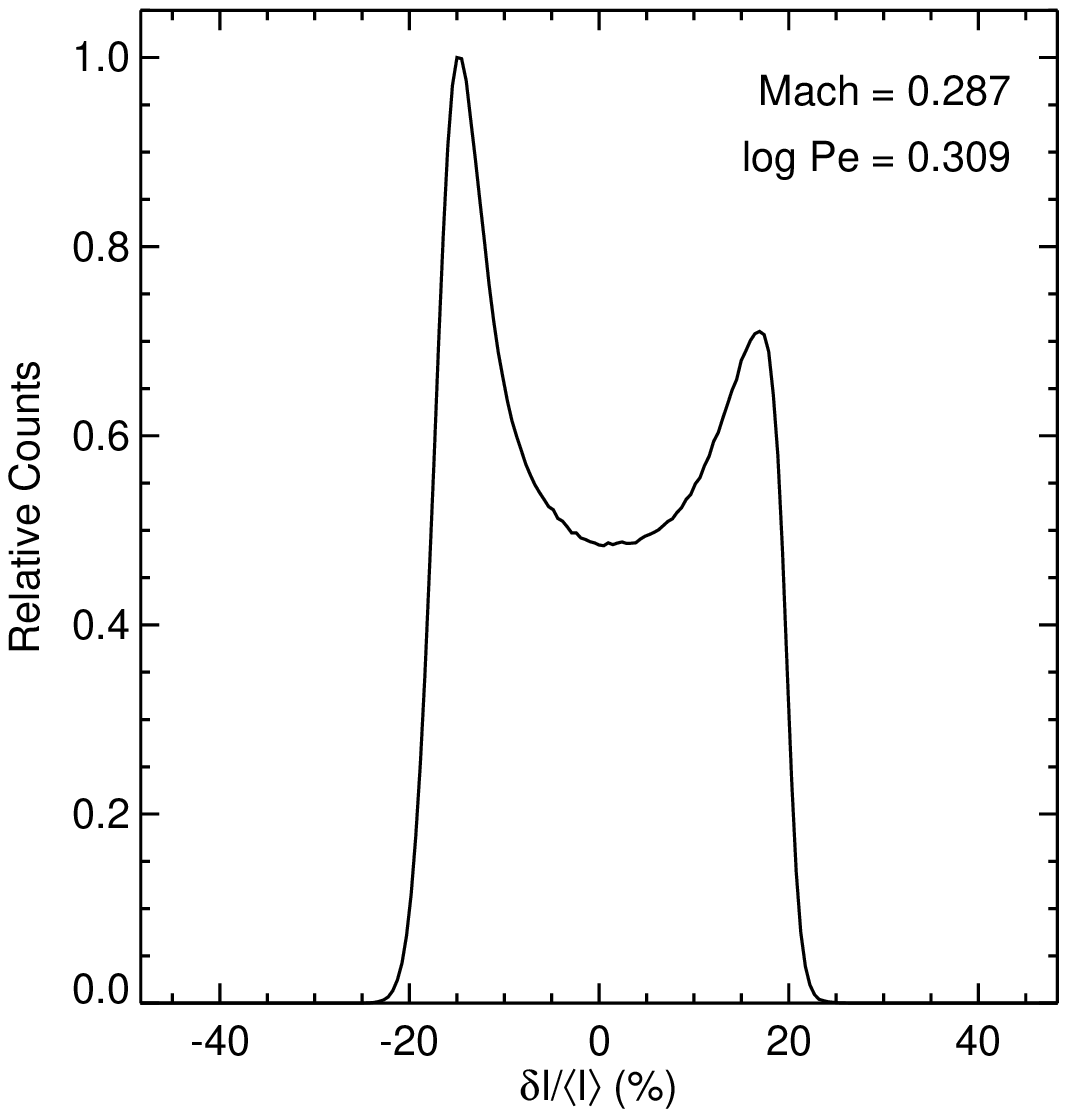}}
\subfloat[]{
\includegraphics[bb=85 65 400 355,width=2.15in]{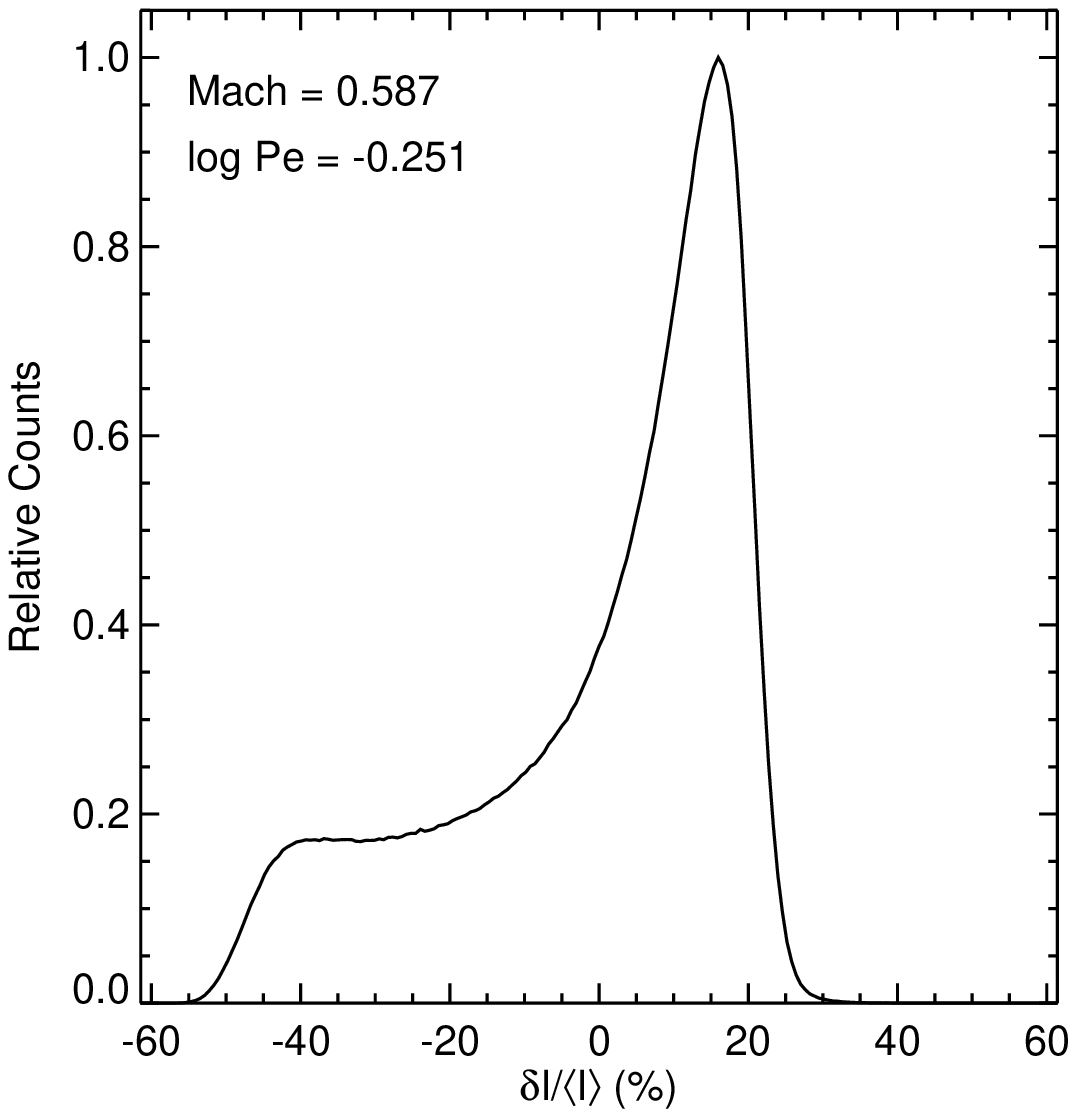}}
\caption{Relative intensity distribution for white dwarfs at $T_{\rm eff}$ =
  6000 (left), 9000 (middle) and 12,500~K (right panel) and $\log g$ =
  8.0. The Mach and P\'eclet numbers are identified on each panel.
\label{fg:fint}}
\end{center}
\clearpage
\end{figure*}

\section{Discussion}

We have demonstrated that the characteristic granulation size, and to a lesser
degree the relative intensity contrast and the characteristic lifetime, scale
well with the Mach number in the intensity forming region. The Mach number is
itself closely anti-correlated with the local density (except for hot white dwarfs), since a lower density
implies higher vertical velocities to transport the same amount of 
energy by convection. To a lesser degree, the Mach number is also sensitive to the
temperature of the photosphere, since a higher effective temperature implies a
larger total energy to transport.

One significant surprise of our analysis is that for low Mach numbers,
properties are very similar for cool white dwarfs with quasi-adiabatic
photospheric convection and hot white dwarfs with low P\'eclet numbers and
inefficient convection. In other words, the convective efficiency does
  not seem to be a dominant factor in the determination of the mean size and
  contrast of granules across the HR diagram.

\subsection{Metallicity}

The effects of metallicity on the granulation are subtle \citep[see
  also][]{allende12,magic13}. For a fixed $T_{\rm eff}$ and $\log g$, the density in the photosphere is generally increasing
for lower metallicities, hence the Mach number is lower, as are the
characteristic granulation size and the intensity contrast \citep{houdek99,samadi10}. This can be observed qualitatively in
Fig.~\ref{fg:f7} since the simulations at different metallicities have
  the same geometrical dimensions. However, this is only one aspect since for
a given density or Mach number, the intensity contrast is generally higher for
low metallicity stars (see Fig.~\ref{fg:f9}). The source of this behaviour is
likely the fact that for a given Mach number, the convective efficiency is
slightly higher for sub-solar metallicities according to
Fig.~\ref{fg:f4}. Hence, the rising cells in low metallicity dwarfs loose less
energy by radiation at $\tau_\mathrm{R} \sim 1$, and produce a slightly higher
intensity contrast. The same effect would cause the hot white dwarfs with
inefficient convection to have a slightly lower intensity contrast than cool
white dwarfs with quasi-adiabatic convection, but the same Mach number. 

\subsection{Convective efficiency}

Our analysis has so far looked at the mean characteristics of
  granulation, disregarding the differences between upflows and downflows seen
  in the intensity snapshots of Figs.~\ref{fg:f5}-\ref{fg:f8}. In
  Fig.~\ref{fg:fint}, we present the relative intensity distribution for three
  white dwarfs from the cool to hot end of the sequence. All intensity
  distributions are available in the online Appendix~A.  The rms relative
  intensity contrast previously discussed in this work is directly evaluated
  from the intensity distribution of individual snapshots. Fig.~\ref{fg:fint}
  demonstrates that both upflows and downflows have a characteristic
  distribution which results in a double-peaked relative intensity
  distribution. The sign of the local intensity contrast does not necessarily
  imply a direction for the convective flow. For instance, some of the
  downflows may be very bright \citep{ludwig12}. Therefore, it is not easily
  possible from the intensity snapshots to characterise the rms contrast of
  upflows and downflows. Nevertheless, we can still look at the position of the peak of the respective distributions.
  Fig.~\ref{fg:fint} shows that in some cases there is a significant asymmetry between upflows and downflows. For the 12,500~K
  white dwarf model, the downflows have an intensity peak further away from the mean intensity than
  the upflows, especially in comparison to the 9000~K case where upflows and
  downflows appear to have a similar distribution.

\begin{figure}[!]
\captionsetup[subfigure]{labelformat=empty}
\begin{center}
\includegraphics[bb=40 40 552 765,width=2.3in,angle=90]{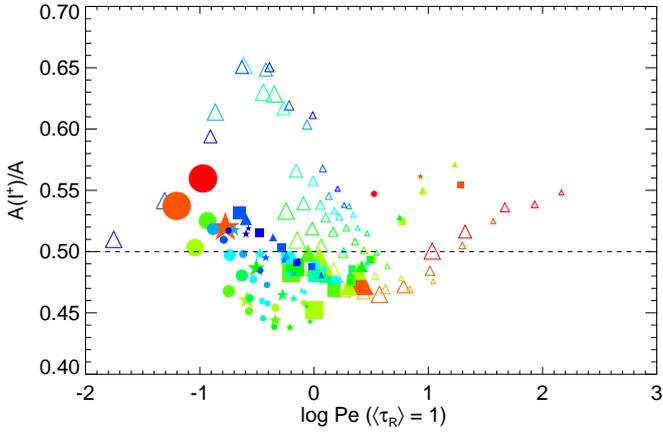}
\caption{Fraction of the geometrical surface (A) where the intensity is higher
  than the average as a function of the logarithm of the P\'eclet number
  evaluated at $\langle\tau_\mathrm{R}\rangle = 1$. The dashed line represents an even
  distribution of bright and dark cells.
\label{fg:f14}}
\end{center}
\end{figure}

In Fig.~\ref{fg:f14} we derive the fraction of the geometrical surface where
the intensity is higher than the average, as a function of P\'eclet number. It
confirms the earlier observation made on intensity snapshots that for hot
objects and especially white dwarfs, the fraction of the area occupied by
bright granules can be considerably larger than for cooler objects. The
minimum seen in Fig.~\ref{fg:f14} at about Pe = 1 suggests that variation of
the P\'eclet number could be a significant factor for this asymmetry, although
the Mach number certainly has a role as well given the second inflection in
the white dwarf sequence at low P\'eclet numbers. In summary, P\'eclet
  number appears to have a noticeable effect on the shape of granules and the
  asymmetry between bright cells and dark lanes (roughly approximated as upflows and downflows), 
  although it does not translate into a strong effect on the characteristic sizes.

\subsection{Parameterisation of 3D convective properties}

We have mentioned that the Mach number is a quantity that
can not easily be predicted a priori of a 3D simulation. The MLT predicts 
velocities but they are rather depth dependent in the photosphere and the free parameters in the 
theory needs to be adjusted for this application, e.g. by comparing 1D and 3D velocities. We have therefore
attempted to characterise granulation properties as a function of the
atmospheric parameters, similarly to \citet{trampe}. 
We find that the following parameterisations provide
characteristic sizes and decay times with a standard deviation\footnote{We exclude hot white dwarfs with a thin convective zone and short decay times in Fig.~\ref{fg:f17} since they are poorly represented by our parameterisation.} of $\sim$30\%
which we believe is a reasonable error given that the characteristic values 
are estimates.

\begin{equation}
 \frac{\rm Char.~size}{\rm [km]} = 13.5 g^{-1} [T_{\rm eff}-300\log g]^{1.75} 10^{0.05 \rm [Fe/H]},
\end{equation}

\begin{equation}
 \frac{\rm Decay~time}{\rm [s.]} = 2.08 g^{-1} [T_{\rm eff}-300\log g]^{1.75} 10^{0.05 \rm [Fe/H]},
\end{equation}

{\noindent}where $T_{\rm eff}$ is in K units, $g$ in cgs units, and white dwarfs are assumed to have [Fe/H] = $-4$. Eq.~15 and 16 are only valid within the range
  of the HR diagram studied in this work, i.e. for convective objects with
  $T_{\rm eff} \gtrsim 600\log g$. 

Compared to the parameterisation given in \citet{trampe}, we find sizes
  that are $\sim$20\% larger for solar-metallicity dwarfs. Furthermore, our
  rms intensity contrasts are on average $\sim$10\% lower. We have currently
  no explanation for this discrepancy. The comparison with observed granulation properties
  \citep{title,rieutord10,mathur11,samadi13} could help in further constraining 
  the different predictions. However, the different codes are currently
  in fairly good agreement \citep{beeck12} while the
  comparison with observations for stars other than the Sun show larger deviations \citep{astero2,mathur11}.
  
It is hoped that our findings that granulation is very similar across the HR
diagram will help in improving the 1D models of convection. Given the
completely different nature of 1D and 3D models, it is difficult to find
any direct implications for the 1D MLT parameterisation. However, our results could be 
useful for more sophisticated models, such as the 2-column (up- and downflow) model of \citet{2col}. 
This treatment requires the geometry of both columns, and therefore would also require
a more detailed study of the asymmetry between upflows and downdrafts.   

\section{Conclusions}

The characteristic granulation properties were investigated for the 148
  3D model atmospheres of dwarfs, giants, and white dwarfs in the CIFIST
  grid. We have derived the ratio of the characteristic horizontal size to the pressure
  scale height in the intensity forming region and showed that it is strongly
  correlated with the Mach number. The ratio increases from values of 3 (Mach
  = 0.1) to 10 (Mach $\sim$ 1). This variation is caused by the increase of the horizontal to vertical velocity ratio
  as a function of Mach number and the resulting constraint from the mass 
  conservation. A more quantitative explanation would
  require the computation of the characteristic depths of the granules and the 
  dynamical pressure scale height over the same geometrical range.
 
The decay time and relative intensity contrast are also shown to be correlated
with Mach number, but it appears that the convective efficiency, or P\'eclet
number, which is rapidly changing as a function of depth, also has a
considerable effect. The intensity contrast increases with the efficiency of
convection for a constant Mach number. Finally, we provided fitting functions
for the characteristic horizontal size and decay time of granulation across
the HR diagram as a function of $T_{\rm eff}$, $\log g$ and [Fe/H] that can be
useful for further hydrodynamical calculations and asteroseismic applications.

Our characteristic values certainly do not represent the full picture. A close
examination of the intensity maps presented in this work shows that the shape
of the individual convective cells, e.g. the brightness profile, the fuzziness
of the edges and the apparition of vortices, is a function of the atmospheric
parameters and does not necessarily scale well with the Mach number. These substructures may also be
  impacted by numerics and neglected magnetic fields. We
  observe an asymmetry between the size of the bright cells and dark narrow lanes which
  appears to be correlated with both the Mach and P\'eclet numbers in a
  complex way. Future works may study this aspect in more
  detail, for instance by combining intensity maps with a set of velocity maps
  at certain depths. The velocity maps may also provide characteristic depths for the granules 
  reaching the photosphere, which would improve our understanding of surface granulation.

\begin{acknowledgements}
P.-E. T. is supported by the Alexander von Humboldt Foundation. 3D model
calculations of white dwarfs have been performed on CALYS, a mini-cluster of
320 nodes built at Universit\'e de Montr\'eal with the financial help of the
Fondation Canadienne pour l'Innovation. This work was supported by Sonderforschungsbereich SFB 881
"The Milky Way System" (Subproject A4) of the German Research Foundation
(DFG). B.F.\ acknowledges financial support from the {\sl Agence Nationale de
  la Recherche} (ANR), and the {\sl ``Programme Nationale de Physique
  Stellaire''} (PNPS) of CNRS/INSU, France.
\end{acknowledgements}

\bibliographystyle{aa} 

\end{document}